\documentclass[fleqn,usenatbib]{mnras}

\usepackage{newtxtext,newtxmath}

\usepackage[T1]{fontenc}
\usepackage{ae,aecompl}

\usepackage{graphicx}	
\usepackage{amsmath}	

\newcommand*{\swap}[2]{#2#1}

\title[Sub-stellar Companions of Intermediate-mass Stars with CoRoT]{Sub-stellar Companions of Intermediate-mass Stars with CoRoT: \\CoRoT--34b, CoRoT--35b, and CoRoT--36b.}

\author[D. Sebastian, E.W.Guenther, M. Deleuil et al.]{
D. Sebastian,$^{1}$\thanks{E-mail:D.Sebastian.1@bham.ac.uk }
E.W.Guenther,$^{2}$
M. Deleuil,$^{3}$
M. Dorsch,$^{4}$
U. Heber,$^{4}$
C. Heuser,$^{4}$
\newauthor
D. Gandolfi,$^{5}$
S. Grziwa,$^{6}$
H.J. Deeg,$^{7,8}$
R. Alonso,$^{7,8}$
F. Bouchy,$^{9}$
Sz. Csizmadia,$^{10}$
\newauthor
F. Cusano,$^{11}$
M. Fridlund,$^{12,13}$
S. Geier,$^{14}$
A. Irrgang,$^{4}$
J. Korth,$^{15}$
D. Nespral,$^{7,8}$
\newauthor
H. Rauer,$^{16}$
L. Tal-Or,$^{17}$
and the CoRoT-team \\
$^{1}$ School of Physics and Astronomy University of Birmingham, Birmingham University, Edgbaston Park Rd, Birmingham B15 2TT, UK \\
$^{2}$ Th\"uringer Landessternwarte Tautenburg, Sternwarte 5, 07778 Tautenburg, Germany\\
$^{3}$ Laboratoire d'Astrophysique de Marseille, 38 rue Fr\'ed\'eric  Joliot-Curie, F-13388 Marseille Cedex 13, France \\
$^{4}$ Astronomisches Institut , der Universit\"at Erlangen--N\"urnberg, Dr. Remeis--Sternwarte, Sternwartstr. 7, D-96049 Bamberg, Germany    \\
$^{5}$ INAF, Osservatorio Astrofisico di Torino, via Osservatorio 20, 10025 Pino Torinese, Italy \\
$^{6}$ Rheinisches Institut f\"ur Umweltforschung an der Universi\"at zu K\"oln, Aachener Stra\ss e 209, D-50931 K\"oln, Germany  \\
$^{7}$  Instituto de Astrof\'\i sica de Canarias, C. V\'\i a L\'actea S/N, E-38205 La Laguna, Tenerife, Spain \\
$^{8}$ Universidad de La Laguna, Dept. de Astrof\'\i sica, E-38206 La Laguna, Tenerife, Spain \\
$^{9}$  Observatoire Astronomique, Universit\'e de Gen\`eve, 51 Ch. des Maillettes, 1290 Versoix, Switzerland \\
$^{10}$ Centre for Astronomy and Astrophysics, TU Berlin, Hardenbergstra\ss e 36, D-10623 Berlin, Germany \\
$^{11}$ INAF-Osservatorio di Astrofisica e Scienza dello Spazio, Via Gobetti 93/3, 40129 Bologna, Italy \\
$^{12}$ Leiden Observatory, Leiden University, NL-2333 CA Leiden, The Netherlands \\
$^{13}$ Department of Space, Earth and Environment, Chalmers University of Technology, Onsala Space Observatory, SE-439 92 Onsala, Sweden \\
$^{14}$ Institute for Physics and Astronomy, University of Potsdam, Karl-Liebknecht-Str. 24/25, 14476 Potsdam, Germany\\
$^{15}$ Department of Space, Earth and Environment, Astronomy and Plasma Physics, Chalmers University of Technology, SE-412 96 Gothenburg, Sweden\\
$^{16}$ Institute of Planetary Research, DLR, Berlin and Free University, Berlin \\
$^{17}$ Department of Physics, Ariel University, Ariel 40700, Israel \\
}

\date{Accepted. Received ; in original form 2021}

\pubyear{2021}

\begin{document}
\label{firstpage}
\pagerange{\pageref{firstpage}--\pageref{lastpage}}
\maketitle

\begin{abstract}
Theories of planet formation give contradicting results of how frequent
close-in giant planets of intermediate mass stars (IMSs; $\rm 1.3\leq M_{\star}\leq 3.2\,M_{\rm \odot}$) are.
Some theories predict a high rate of IMSs with close-in gas giants, while others
predict a very low rate.
Thus, determining the frequency of close-in giant planets of IMSs is an important test for theories of planet formation.
We use the CoRoT survey to determine the absolute frequency of IMSs that
harbour at least one close-in giant planet and compare it to that 
of solar-like stars. The CoRoT transit 
survey is ideal for this purpose, because of its completeness for gas-giant planets with orbital periods of less than 10 days and its large sample of main-sequence IMSs.
We present a high precision radial velocity follow-up programme and conclude on 17 promising transit candidates of IMSs, observed with CoRoT.

\noindent We report the detection of CoRoT--34b, a brown dwarf close to the hydrogen burning limit, orbiting a 1.1\,Gyr A-type main-sequence star. We also confirm two inflated giant planets, CoRoT--35b, part of a possible planetary system around a metal-poor star, and CoRoT--36b on a misaligned orbit.
We find that $0.12 \pm 0.10\,\%$ of IMSs between $1.3\leq M_{\star}\leq 1.6 M_{\rm \odot}$ observed by CoRoT do harbour at least one close-in giant planet. 
This is significantly lower than the frequency ($0.70 \pm 0.16\,\%$) for solar-mass stars, as well as the frequency of IMSs harbouring long-period planets ($\rm \sim 8\,\%$).
\end{abstract}




\section{Introduction}\label{ch1}


Up to now most surveys of extra-solar planets have concentrated on stars
that have one solar-mass, or less (e.g. \citealt{naef05, cumming08, dong13, wittenmyer20b}) Thus, our knowledge of planets orbiting
stars more massive than the Sun is currently quite limited. This is very
unfortunate, because planets orbiting intermediate-mass stars give
us important clues to how planets form and evolve. As intermediate-mass 
stars (IMSs) we denote main-sequence stars with 
$\rm M_{\star}=1.3-3.2\,M_{\rm \odot}$.

Planets orbiting such stars on close orbits (with $\rm a \lesssim$ 0.1\,AU) are heated up enormously making them ideal 
laboratories to study the inflation of planetary radii and the 
evaporation of  planetary atmospheres 
\citep{shporer11, mazeh12, vonessen15}. 
For example, KELT-9b, which orbits a $M = \rm 2.5\,M_{\odot}$ star, has a day-side temperature of 4600 K, similar to 
a K-star \citep{gaudi17}. Many of them also have high obliquity's
\citep{winn10}. Giant planets (GP) of IMSs also give us important clues of how planets
form. The life-time of the disks of IMSs is on average about half as 
long but more massive than that of solar-like stars \citep{mamajek09}. 
By studying planets of IMSs we can, thus, constrain the time-scale of 
planet formation. 


\subsection{Radial-velocity surveys of intermediate-mass stars}

Using Radial-velocity (RV) for this purpose is not easy. Many IMSs rotate
fast with $\rm v\,sin(i_\star) > 100\,$km\,s$\rm ^{-1}$ \citep{royer07,pribulla14} and many are
located within the instability strip of the Hertzsprung–Russell diagram. Furthermore, the
number of spectral-lines is much smaller than that of later-type stars.
This limits the sensitivity for detecting IMSs planets
\citep{galland06a,galland06b,desort09a,desort09b,guenther09,borgniet14}. 
The most recent result was published by \cite{borgniet19} who derived 
a frequency of $\rm 3.7^{-1}_{+3}\,\%$ for A \& F stars with masses of less than $\rm 1.5 M_{\sun}$ to harbour GPs closer than 2-3 AU. This frequency is consistent with that of FGK-stars.

Another approach is to search for planets of giant stars that
had $\rm 1.3\leq M_{\star}\leq 3.2 M_{\rm \odot}$ when they
were on the main-sequence \citep[e.g.][]{Lovis07}. However, the disadvantage is that
giant stars may not have short period planets. \citet{johnson10, johnson10b} conclude that the frequency of GPs ($\rm M_{Planet} \geq 0.5\,M_{Jup}$) of giant 
and sub-giant stars at distances $\rm a<2.5\,AU$ increases proportional to the mass of the host star. \cite{reffert15} found that the GP frequency increases in the stellar mass interval from 1 to 1.9\,$M_{\odot}$ but then decreases for higher-mass stars again in a similar study. \cite{wittenmyer20a} derived a GP frequency of $7.8^{+9.1}_{-3.3}\%$ for a sample
of long-period (< 5 yr) planets orbiting evolved IMSs. 
This frequency is not significantly higher than that for solar-like stars. The surveys of giants stars, thus, indicate that the frequency of GPs of IMSs is equal, or higher than that of FGK-stars.

However, the cases of the K-giants Aldebaran \citep{hatzes15b,reichert19} and $\gamma$ Draconis \citep{Hatzes18} and others similar \citep[e.g.][]{Delgado18} are worrying. Those show long-period RV variations and for a long time and it was believed  that these are the
signatures of planets, because they passed all the 
classical tests. However, after decades of observations it was 
discovered that the RV signal
had changed, which rules out the planet 
hypothesis. Thus, it is possible that surveys of giant stars over-estimate the planet frequency.

All RV surveys of giant stars show a lack of close-in 
($ <$ 0.1\,AU) planets. There are some hypothesis that explain
that lack of close-in planets of giant stars: 
The first hypothesis assumes that 
IMSs originally had many more planets than FGK-stars,
but most of them migrated inwards, for example by
planet-planet scattering, or disc migration. Most of these
planets ended up in orbits close to the host stars and were
then engulfed when they became giants. That means we 
see in giant stars only those planets that were not
engulfed, and these are only GPs at large distances.
In the framework of this model, \citet{hasegawa13} 
estimates that 5-11\,\% of main-sequence IMSs
should have a close-in GP. This is one order of magnitude higher 
than for solar-mass stars \citep{naef05, cumming08, wittenmyer20b}. 

The second hypothesis is that the lack of close-in planets is an effect of planet evolution. In this hypothesis it is assumed that
the migration processes are less effective for IMSs than for FGK stars.
The migration time-scale for IMSs is then too long for the planets
to migrate inwards. In the framework of this model it is
not expected to observe a large population of close-in GPs
of IMSs. \cite{currie09} predicts the frequency 
of close-in planets of main-sequence IMSs should be
smaller than 1.5\,\%.

A possible third hypothesis would be that close-in planets
evaporate due to the intense radiation of the IMSs. Nevertheless, massive Giant planets orbiting A-stars closer than 0.1 AU have been detected \citep[e.g.][]{cameron10, gaudi17}. 
Their existence rule out that atmospheric evaporation plays a mayor role for the lack of close-in planets orbiting such IMSs.

\subsection{Transit surveys of intermediate-mass stars}

As outlined in the previous section, the frequency of IMSs with close-in GPs tells us a lot about planet formation, evolution
and migration, but classical RV surveys are less than ideal for determining this number and close-in planets are not detected in surveys of giant stars. 

Transit surveys have the advantage that they are not biased against
rapidly rotating stars. Furthermore, they are excellent for 
detecting short-period planets. The first detection of 
a transiting GP orbiting a bright A-type
star was WASP33 b \citep{cameron10}. Since this star 
is a Delta Scuti pulsator, the mass of this planet could 
only be measured by analysing a large amount of spectra 
\citep{lehm15}. For transit surveys, the geometric
probability to observe a transit of a planet orbiting an 
A5 main-sequence star with $\rm 1.7\,R_{\odot}$ 
\citep{gray05} in a 6\,d orbit (a = 0.08\,AU) is about
10\,\%. It is thus not surprising that many GP orbiting IMSs 
were detected in ground based transit surveys. Examples are
WASP-189b (0.05\,AU; \cite{anderson18}), KELT-9b (0.03\,AU; \cite{gaudi17}), 
KELT-21b (0.05\,AU; \cite{johnson18}), or MASCARA-4b (0.05\,AU; \cite{dorval19}).
Short-period brown dwarf companions can also be detected in 
this way. The discovery of the first
transiting close-in brown dwarfs orbiting IMSs,
HATS 70b (0.04\,AU; \cite{zhou19a}) and TOI-503 b (0.06\,AU; \cite{subjak2020}) gave us important insights how sub-stellar objects can form. As outlined by \cite{subjak2020}, detecting such objects is particularly interesting, because it is generally thought that high-mass brown dwarfs ($\rm M > 40\,M_{Jup}$) must
form in-situ via core fragmentation, whereas low-mass brown 
dwarfs could also form in disks, just like planets.

Space-based transit surveys are perhaps the best way to 
determine the statistics of close-in GPs, because of their high sensitivity 
and the long monitoring time. The case of Kepler-13Ab shows that space-based data allow to detect planetary ephemeries from photometric variability induced by the companion \citep{shporer11}. The discovery of Kepler-448\,b \citep{bourrier15} shows that it is even possible to detect planets of IMSs with an orbital distance of more than 0.15\,AU.  

Using planets detected by TESS \citep{ricker15} and confirmed 
by ground-based transit surveys, like HATNet (Hungarian-made Automated Telescope Network), \cite{zhou19b} derived a
frequency of close-in GPs of $\rm 0.71\pm0.31\%$ for G-stars, 
$\rm 0.43\pm0.15\%$ for F-stars, and $\rm 0.26\pm0.11\%$ for A-stars. 
This study also indicates that the frequency for IMSs 
is not significantly higher than that of solar-like stars as found for planets in wider orbits with radial velocity methods. 

From the discovery of a peculiar features in the periodogrammes 
of A-stars in the Kepler data, \cite{balona14} concluded that about 
8\% of the A-type stars could have massive planets, or brown dwarf
companions with orbital periods of about six days or less.  
However, a subsequent study by \cite{sabotta19} showed that 
these features are not caused by GPs and that the
true frequency must be smaller than 0.75\%.

\subsection{A survey for giant planets and brown dwarfs orbiting IMSs with CoRoT}

CoRoT was the first space survey dedicated to the photometric
search for exoplanets \citep{baglin06}. The detection of CoRoT--7b \citep{leger09} with a transit depth of
0.03\% showed that space-based transit surveys are much more efficient
for detecting shallow transits than ground based surveys. 
According to \cite{deleuil18b}, the completeness
of the CoRoT survey is about 90\% for GPs with orbital
periods of less than 10 days. However, GPs orbiting A-stars show shallower transits than those orbiting FGK-stars, due to the larger size of the star.
After several years in space, the noise limit of
CoRoT in 2h, was still about 0.1\,\% for a 15-mag star \citep{moutou13}. By phase-folding the light curve to the transit period the noise limit will further decrease, such that transits of GPs
around IMSs can be detected with at least 5$\,\sigma$ even for the
faintest stars in the CoRoT sample which is about R=16\,mag. The completeness for
GPs of A-stars, thus, is similar to that of FGK-stars.


CoRoT has observed 101\,083 main-sequence stars \citep{deleuil18b} 
of which about 30\% are IMSs. Amongst the 36 sub-stellar companions, 
already published in the literature, only four are orbiting IMSs. If the frequency of GPs of IMSs were the same as that for solar-like FGK stars, we expect to find in the order of 10 GPs in this sample. 

Thus, we have initiated a survey to search for close-in companions 
of IMSs based on the results obtained from the CoRoT--space observatory.
The low RV accuracy for IMSs is not an obstacle, if we are just 
interested in the statistics of  planets. An upper limit
in the planetary regime combined with an analysis that
shows that the object is not a false-positive (FP) is
sufficient. Most of the known planets have been confirmed by excluding FPs, For most of them we do not have a mass determination via the RV method.
An important aspect of the CoRoT mission was that all stars were
searched for transits with the same method. The methods, as well as a complete 
list of detections and candidates are given in \cite{deleuil18}. 

The preliminary results of our survey for giant planets of IMSs
has been given in \cite{guenther16} (Hereafter G16). In that article, we pointed out 
that the stars CoRoT 110756834 (LRa02\,E1\,1475), CoRoT 659460079 (LRc09\,E2\,3322), 
and CoRoT 652345526 (LRc07\,E2\,0307) most likely harbour planets or brown dwarf companions. 

In this article, we conclude our survey by presenting a detailed radial velocity analysis of our candidates, as listed in G16, including the detailed transit analysis and RV confirmation of those three objects. Furthermore, we discuss the candidate CoRoT 659668516 (LRc08\,E2\,4203) which was mentioned in G16 and close the cases for the remaining candidates CoRoT 110660135 (LRa02\,E2\,4150), CoRoT 310204242 (LRc03\,E2\,2657), CoRoT 102850921 (IRa01\,E2\,2721), CoRoT 102605773 (LRa01\,E2\,0963), and CoRoT 659721996 (LRc10\,E2\,3265).

In line with the main CoRoT survey, we will use the nomenclature for confirmed planets and brown dwarfs: CoRoT--34 (LRa02\,E1\,1475), CoRoT--35 (LRc09\,E2\,3322), and CoRoT--36 (LRc07\,E2\,0307), respectively for these three stars.
From our results, and given the known number of IMSs that CoRoT has observed, we can now calculate the frequency of close-in GPs for IMSs and compare it with the frequency of close-in GPs for lower mass stars.

This article is structured as follows. In Sect.~\ref{sec_stellar} we give an overview on the stellar 
sample, observed with the CoRoT satellite. In Section~\ref{sec_cand} we explain the selection of our candidates, being followed up in more detail. The methods used for these follow-up observations are detailed in Sect.~\ref{sec_obs}. We introduce the sub-stellar objects discovered in our survey in Sect.~\ref{sec_planets} and summarise the outcome of the survey in Sect.~\ref{sec_candidates}. In Sect.~\ref{sec_occurence} we combine the observations from the CoRoT follow-up team with our survey to draw a conclusion on the frequency of close-in 
GP of IMSs and discuss our results in Sect.~\ref{sec_concls}.

\section{The stellar sample of the CoRoT survey}\label{sec_stellar}

In order to derive the frequency of planets for stars with 
different masses, we
have to know how many stars of which mass the sample
contains. CoRoT observed fields in two different directions in the sky. 
One set of fields is located close to $\rm RA= [18h 50m]$. The fields are called 
"Galactic center fields". Stars in these fields that were observed for
up to 152\,days are labeled LRc (Long-Run-Center), followed by E1/E2 for the
CCD used and a running number. The other fields are close to $\rm RA= [6h 50m]$.
These fields are called ``Galactic anticentre fields'' and long observing runs are labeled 
LRa. In preparation for the CoRoT mission, 
the spectral type, luminosity class, apparent brightness, and contamination factor 
of all stars in the fields of the CoRoT sample were determined by \cite{deleuil09} 
using multi-colour photometry. 
The CoRoT mission was primarily designed to detect transiting planets 
with spectral types $\rm F$, $\rm G$, and $\rm K$. When selecting the targets, higher preference was given to stars that were more likely to be main-sequence F,G,K-stars but other types of stars were not excluded.

A refined photometric classification, based on the same data was published by \cite{damiani16}. Furthermore, several spectroscopic surveys have been carried out 
that included a large number of these stars. Those offer an independent evaluation 
of the photometric classification of the sample by directly comparing the resulting 
spectral types. 

\cite{sarro13} carried out a spectroscopic survey that focused on a sub-sample 
that has been found to be photometrically variable stars with CoroT. 
A large survey, focusing on solar-mass stars was published by \cite{gazzano10}. 
The most comprehensive survey has been carried out by \cite{sebastian12} and \cite{guenther12} 
who used the multi-object spectrograph AAOmega at the AAT to determine the spectral types and 
luminosity  classes of a representative  sample of 11\,466 stars in 
the CoRoT anticentre direction. This study shows that 32.8\,\% of all 
main-sequence stars in the CoRoT anticentre fields are IMSs. The 
comparison of the photometric and spectroscopic results by \citet{damiani16} showed that the median deviation between the temperatures derived photometrically and 
spectroscopically are only 5\% for late-type and 
10\% for early-type stars, respectively. The accuracy of the luminosity
classes IV and V is also within the $\rm 1.4 \sigma$ confidence level\footnote{85\% agreement between photometric and spectroscopic classification}.
Another result from this comparison is, a systematic difference of the effective temperatures with of the adopted interstellar extinction. The photometrically derived temperatures are systematically lower for G-M type stars in the anticentre fields. 
\citet{damiani16} reports the average spectroscopic distances in the centre fields to be 1.6\,kpc for main-sequence stars. This is in good agreement to the distances derived by \cite{gazzano13}. Therefore, the systematics due to extinction, found in the anticentre fields are negligible for main-sequence stars in the centre fields. Thus, the determinations are reliable and can be used to determine how many main-sequence A,F,G,K, and M-stars the sample contains.

\begin{figure}
   \centering
   \includegraphics[width=\columnwidth]{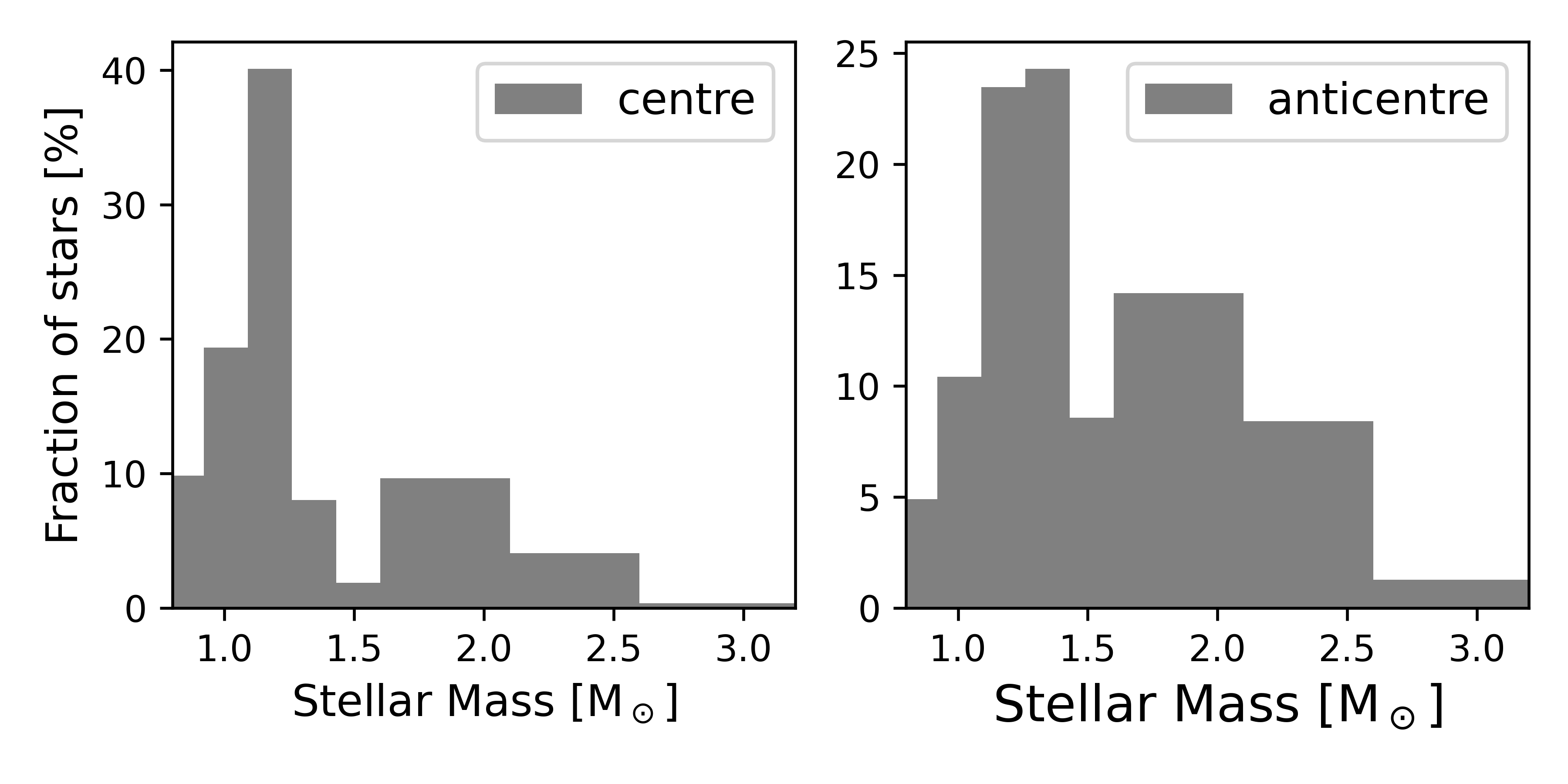}
   
    \caption{Mass distribution of main-sequence stars in the CoRoT fields. left - Galactic centre fields from photometric classification based on Exodat \citep{deleuil09, damiani16}, right - Galactic anticentre fields based on data from \citep{sebastian12, guenther12}.
              }
         \label{FigStarStat}
   \end{figure}

In the next step, we estimated the stellar masses distribution of this sample by converting the spectral types to average mass ranges. We can do this by assuming main-sequence stars of solar metallicity, which is in good agreement with the findings of \cite{gazzano13} for the CoRoT fields. Using the conversion factors given in \cite{gray05}\footnote{Appendix Table B.1., page 506} we derive following upper mass limits: B7 -- 3.2$\rm M_\odot$, B9.5 -- 2.6$\rm M_\odot$, A4 -- 2.1$\rm M_\odot$, F0 -- 1.6$\rm M_\odot$, F4 -- 1.43$\rm M_\odot$, F8 -- 1.26$\rm M_\odot$, G3 -- 1.09$\rm M_\odot$, K0 -- 0.92$\rm M_\odot$. The resulting mass-distribution for stars with luminosity classes IV and V in both viewing directions of CoRoT is shown in Fig.~\ref{FigStarStat}. These upper mass limits translate to a constant bin size of 0.17$\rm M_\odot$ for stars below 1.6$\rm M_\odot$ and a larger bin size of 0.5$\rm M_\odot$ for more massive stars. 

\section{Identifying the candidates}\label{sec_cand}

The survey strategy is explained in detail in \cite{guenther16}. 
We give a short summary on the target selection and the method, 
used to derive the radial velocity of the candidates.
Based on both, photometric and, if available, 
spectroscopic classification of all CoRoT targets, we selected stars
with spectral types earlier than F3 and luminosity class IV and V. 
Using this criterion, we identified 25\,243 CoRoT stars for which the light curves 
have been searched for signals of transiting planets.
For this analysis we used the \texttt{EXOTRANS} algorithm 
developed by \cite{grziwa12}. We also included all candidates 
identified by the CoRoT detection teams. 

As planet candidates we selected all objects were the transit
depth is smaller than 1.5\%. Since \texttt{EXOTRANS} is optimised to find periodic signals, after identifying the planet-candidates, we analysed their light curves in detail to derive all relevant parameters of the candidates. We removed first phase-folded the light curves and removed trends and stellar variability from each epoch using a polynomial fit. This allows to measure the transit depth, but also to investigate the light curve visually.
As detailed in \cite{ammler15} all candidates were observed using the Tautenburg low-resolution ($R \sim$ 1000) Nasmyth spectrograph. Similarly to \cite{sebastian12}, the spectral classification was done by comparing the spectra to catalogue reference spectra \citep{valdes04} using an automatic least-square minimisation pipeline. The accuracy of this method is about one to two sub-classes for IMSs \citep{ammler15}.
For a first estimate of the radii of the transiting objects, we estimated the stellar radii from conversion factors given in \cite{gray05}\footnote{Appendix Table B.1., page 506}. 
These radii are sufficient to remove many binary companions. However, because M-stars have roughly the same size as gas-giants, such false-positives can only be removed by RV measurements.

Since the photometric mask of CoRoT has a size of about $\rm 30\arcsec\times16\arcsec$, 
it is important to confirm that the object with the transit is the bright star, and not a faint eclipsing binary within the photometric mask. 
Removing such false-positives (FPs) is an important step for all planet 
search programs. To remove these we used archival data of the CoRoT fields 
from the the MegaPrime/MegaCam wide-field imager \citep{boulade03} at CFHT, Hawaii. That data has a resolution of 0.19"/pix and thus allowed us to 
exclude bright contamination stars with a resolution of $\leq $1 arcsec to 
the candidate host-star. 
The 17 candidates found in this survey are listed in Table~\ref{tab.061}. The coordinates and apparent magnitudes of all candidates can be found in Table~\ref{tab:obs_param}. Two of the candidates (CoRoT 632279463 and CoRoT 631423419) have been been ruled out as planet candidates from detailed light curve analyses, showing a shallow secondary eclipse which only allows a binary solution for the system. Both were, thus, not followed-up with high resolution spectroscopy. 

\section{Observations and Methods}\label{sec_obs}

Given the relative faintness of our targets ($\rm R=11-15.3\,mag$), intermediate and high-resolution spectrographs mounted on telescopes with apertures of 2m and larger were needed. Thus, we used a variety of instruments for the RV follow-up observations: 

\begin{itemize}
\item[i.)] SANDIFORD spectrograph mounted on the Cassegrain focus of the 2.1m Otto-Struve telescope at McDonald observatory \citep{mccarthy93}. The grating was adjusted to achieve a wavelength coverage from 400 to 440\,nm at a resolution of $\rm \lambda/\Delta \lambda = R\approx60\,000$. We took ThAr-frames for wavelength calibration directly before and after each spectrum to minimise radial velocity variations, introduced due to flexing of the instrument during pointings.

\item[ii.)] TWIN spectrograph at 3.5m telescope at
Calar Alto observatory, Spain. The TWIN spectrograph is a long-slit spectrograph designed for simultaneous observation in a red and blue arm\footnote{\url{http://www.caha.es/CAHA/Instruments/TWIN/}}. We observed using the T01 and T10 gratings, resulting in an intermediate resolution of $\rm R\approx7000$. Wavelength calibration was done, using Helium-Argon exposures before and after the science spectra. Despite the relative high signal to noise, achieved for our candidates, the instrumental stability does not allow to achieve a stability of less then 10\,km\,s$\rm ^{-1}$.

\item[iii.)] Calar Alto Fiber-fed Échelle spectrograph (CAFE) at the 2.2m telescope at Calar Alto observatory, Spain \citep{aceituno13}. It is a fibre-fed spectrograph, which is mechanically stabilised but only passively temperature stabilised. It provides a wavelength coverage from 400 to 950\,nm at a resolution of $\rm R\approx60\,000$.

\item[iv.)] FIES spectrograph mounted on the 2.6\,m Nordic Optical Telescope (NOT) at Roque de los Muchachos Observatory, La Palma, Spain \citep{telting14}. We used the $\rm 1\farcs3$ \,high-resolution fibre, which gives a resolving power of $\rm R\approx67\,000$ and covers the wavelength range of 360--740\,nm.

\item[v.)] UVES mounted on the ESO VLT UT-2 (KUEYEN) of the ESO Paranal observatory \cite{dekker00}. We used the standard 390+580 setting, which covers the wavelength region from 327\,nm to 682\,nm. Unfortunately, orders that are only partly covered by the detector, cannot be used for precise radial velocity measurements. Thus, the blue channel effectively covers only the spectral range from 329 to 452\,nm without gaps and the two detectors for the red channel, cover the range from 478 to 574\,nm, and from 582 to 676\,nm respectively. We used a slit-width of 0.8 arcsec, which provides a resolution of about $\rm R\approx61\,000$. 

\item[vi.)] HARPS at the ESO 3.6m telescope at La Silla \citep{pepe02,mayor03}. The spectra cover the region from
378\,nm to 691\,nm in 72 spectral orders. Depending on the brightness of the candidate, we have obtained spectra either with the standard high accuracy mode (HAM) with a resolving power $\rm R\approx115\,000$ or the high efficiency mode (EGGS) which allows a 1.75 times higher throughput compared to the HAM mode due to a larger projected aperture of the fibre (1.4 arcsec). Depending on the actual seeing this mode offers on average a reduced resolving power of $\rm R\approx80\,000$.

\item[vii.)] HIRES at the 10.0 Keck telescope (\citep{vogt94}. The spectra reach a resolving power of $\rm R\approx67\,000$. 

\end{itemize}

In the following, we will briefly describe the data reduction methods used for all
spectra. For HARPS, we used the HARPS pipeline to reduce and extract the spectra (bias subtraction, flat-field, scattered light subtraction, and wavelength
calibration). The sky-fibre was used to remove stray-light from the
moon if necessary. All other spectra were bias-subtracted, flat-fielded, the scattered light
removed, and extracted using standard \texttt{IRAF} routines \citep{tody93}. The wavelength calibration was done using calibration lamp spectra (mostly ThAr spectra for high resolution spectrographs), obtained close to the target spectrum.

Since most of these stars are relatively faint, the signal-to-ratio (SNR) obtained
was  between 20 and 80. Thus, we developed a special analysis
program to determine the radial velocities. In this program a
synthetic or an observed high SNR spectrum is used as a template, which is fitted to the observed spectrum using the least-square method to
all Echelle orders without merging them. 
Because the observed stellar flux in each Echelle order is larger in the centre of the blaze function, we used a weighted fit to avoid too noisy parts of the spectrum. We, thus, modelled the template using the blaze function of the Echelle spectrograph. Tests show that including the blaze
function into the template improves the precision significantly. 
This improvement is particularly important for spectra with a low SNR \citep{sebastian17}.
The output is simply the relative velocity to the used template spectrum, thus, we applied the barycentric as well as the telluric-line correction for all spectra that include telluric features.  

Synthetic templates were derived from line tables for different stellar parameters 
\citep{lehmann11} to match the best parameters of the stars. However, if a sufficient number of spectra were available to reach a combined SNR larger than 50, we constructed the templates for the stars themselves by combining all these spectra taking the relative RV of the spectra to a synthetic template into account. This combination is realised by first matching all spectra by linear interpolation to the wavelength scale of the synthetic template spectrum and, second, applying a simple median function. This interpolation is applied on the level of single orders of 2D spectra which allows to combine spectral orders without additional interpolation.  
For determining the RV, we always used all orders, except those that contain strong telluric absorption lines. If available, telluric lines were used to monitor instrumental drifts. 
To measure the semi-amplitude K, as well as the orbital phase, we used a least-square optimisation to fit a Keplerian orbit to the measurements, taking the measurement errors as well as instrumental offsets into account. The photometric period was kept as a fixed parameter. We determined the error of the fit by deriving the variance of the best fit solutions by randomly excluding data from the sample.
To test the accuracy of this method, we obtained SANDIFORD spectra of the two transiting F-type binary stars LRc07\,E2\,0482 (F7V, UNSW-TR-3; \citet{hidas05}) and LRc02\,E1\,0132 (F3V, CoRoT 105906206; \citet{silva14}). Both are IMS within the same brightness-range of the candidates in our survey. LRc02\,E1\,0132, is a fast rotating IMS with $\rm v_\mathrm{rot}\sin i_\star \approx 20$\,km\,s$^{-1}$. We obtained six RV measurements of the primary star, well distributed over the binary orbit, and confirmed the orbit with a residual error of 60\,m\,s$\rm ^{-1}$, which is in agreement with the expected photon noise of about 200\,m\,s$\rm ^{-1}$ per measurement and including an instrumental stability of 100\,m\,s$\rm ^{-1}$.

Table~\ref{tab.061} shows an overview of the measurements obtained per instrument. RV measurements obtained for all candidates and instruments are available as supplemental data\footnote{RV measurements are available in machine readable form as radial\_velocities.csv.}.

\begin{figure}
\centering
\includegraphics[width=\hsize]{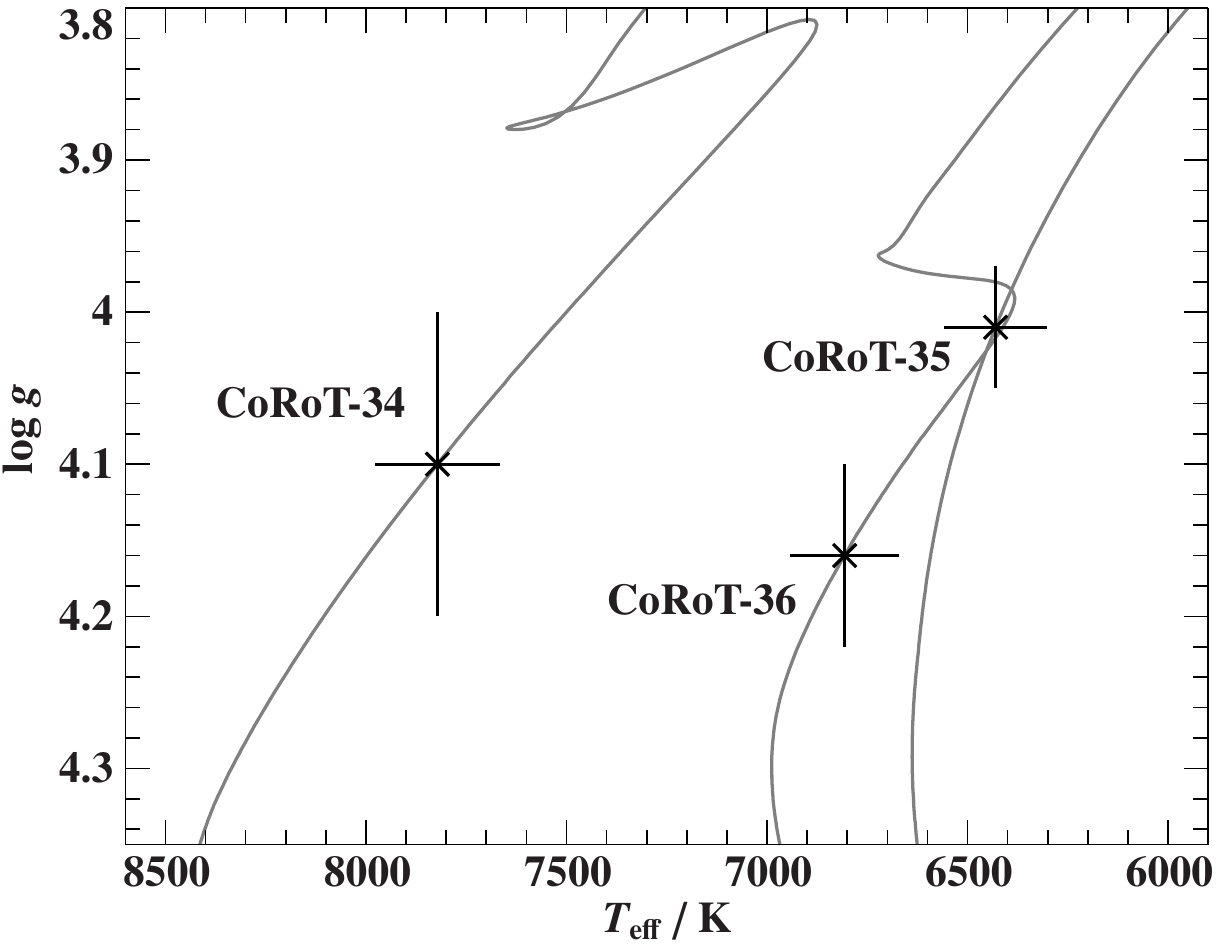}

\caption{Kiel diagram showing the positions of CoRoT--34, 35, \& 36.
The best-fitting MIST evolutionary track \citep{choi2016} is shown for each star.
}
\label{fig:kiel}
\end{figure}

We obtained atmospheric parameters using the high-resolution spectra. For our candidates with early spectral types: CoRoT--34, CoRoT 310204242, and CoRoT 110660135 we derived the effective temperature $T_\mathrm{eff}$, surface gravity $\log(g)_{sp}$, and metallicity [Fe/H] using grids of synthetic spectra, based on ATLAS12 model atmospheres \citep{Kurucz1996}.
The most important elements were considered using a hybrid non local thermal equilibrium (NLTE) approach with the DETAIL/SURFACE package \citep{Giddings1981}. 
Our model spectra, as well as the fitting method are described in detail in \citet{2014A&A...565A..63I} and \citet{heuser18}. For all other candidates, the synthetic spectra were calculated in LTE using the ATLAS12 \& SYNTHE codes \citep{Kurucz1993}.
To consider all sensitive features in the observed spectra, we performed global $\chi ^2$ fits. In the cases where combined spectra were available from several instruments, we fitted them simultaneously using the same model grid taking different resolution and radial velocity into account.
Regions that were not well reproduced were removed from the fit.
This includes the cores of hydrogen lines, lines that are not included in our model spectra, as well as lines with uncertain atomic data.
It is challenging to determine the surface gravity for F-type stars, because it is correlated with both the $T_\mathrm{eff}$ and the abundance of calcium and magnesium. 
We assumed systemic uncertainties of 0.2\,dex for $\log(g)_{sp}$, of 0.1\,dex for [Fe/H], and 2\,\% in $T_\mathrm{eff}$, which were added in quadrature to the much smaller statistical uncertainties. The atmospheric parameters from spectroscopy are listed in Table~\ref{tab:sub_obs}.
Second, we derived stellar age, mass, and radius by fitting the observed parameters to MESA evolutionary tracks \citep{choi2016} in the ($T_\textrm{eff}$, $\log\,g$, [Fe/H]) plane as described by \cite{2016A&A...591L...6I}. 
For CoRoT--34, 35, \& 36, we used the stellar mass and radius as input for the light curve fit, which we used to constrain $\log(g)_{lc}$ from the light curve directly (For details, see description of the light curve fit below). We derived the final stellar age, mass, and radius as listed in Table~\ref{tab:sub_obs} by repeating the MESA fit using $\log(g)_{lc}$ as prior. The best-fit evolutionary tracks are shown in Fig.~\ref{fig:kiel}. 

Reliable parallaxes from the early third Gaia release \citep[EDR3,][]{2021A&A...649A...1G,2021A&A...649A...2L,2021MNRAS.tmp..394E} allow us to derive precise distances. For CoRoT-34, however, the EDR3 quality parameters\footnote{The {\it re-normalised unit weight error}, RUWE \citep{RUWE}, as well as the \texttt{ipd\_gof\_harmonic} and \texttt{ipd\_frac\_multi} parameters are inspected.} indicate an inconsistent parallax measurement. Thus, we derived Gaia EDR3 distances for CoRoT--35 and CoRoT--36 only, by applying the parallax zero point offset following \cite{Lindegren2021}.
For CoRoT--34, we derived the spectrophotometric distance from the spectral energy distribution (SED) along with the interstellar reddening and using $\log(g)_{lc}$ as prior (see Appendix~\ref{sect:sed_fit}). We fitted the SED also for CoRoT--35 and 36 and found that the effective temperature derived from the SED consistently agrees with the spectroscopic effective temperature. 
The SED fitting also allows us to derive the angular diameters which are displayed in Table\,\ref{tab:sed_parameter}.

The CoRoT data were obtained from the IAS CoRoT Public Archive\footnote{\url{http://idoc-corot.ias.u-psud.fr}}. We used the Version 4 legacy data release, which had been extensively reprocessed over the original CoRoT releases \citep{2016cole.book...61C}. We used the BAR fluxes, which had been corrected from aliasing, offsets, backgrounds, the jitter of the satellite, differences in the flux due to the change of the mask, the change of the temperature set point, and the loss of long term efficiency. Furthermore, spurious points were replaced by interpolation. Of note is that the time-stamps in the legacy release are given in Barycentric Dynamical Time (BJD\_TDB), instead of the heliocentric UTC time of the previous Versions 1 to 3, which were used as the basis of all previous CoRoT planet discoveries. The difference between the two time-scales is about one minute --for details on this topic, please refer to  \citet{2010PASP..122..935E}. 

The CoRoT light curves were then corrected from stellar variability, as well as residual instrumental effects. For this fit we masked the transit times, which were known from the detection ephemeris and periods. To allow for slight changes in these parameters, we included each 30min out of transit time before and after the predicted transit time into the mask. We modelled the light curve using a polynomial fit by applying the \textit{Chebyshev} function, implemented in the \texttt(numpy.polynomial)\footnote{\url{https://numpy.org/doc/stable/reference/routines.polynomials.html}} package. The polynomial order was determined visually to only model variability on longer timescales that the expected transit duration. In this way, we were able to account for intensity variations due to pulsations that would affect the transit shape. 
To finally model the transit light curve, we used the software package \texttt{ALLESFITTER} \citep{guenther19} which uses the \texttt{ellc} \citep{maxted16} eclipse model to fit the transit light curve. The main fitting parameters are the orbital period P, the transit epoch, the dimensionless planet-star radius ratio $\rm b_{rr} = R_b / R_\star$, as well as $\rm b_{rsuma} = (R_\star + R_b) / a$, which both parameterise the scaled semi-major axis $\rm a/R_\star = (1+b_{rr})/b_{rsuma}$. $\rm R_\star$ is the stellar radius, $\rm R_b$ the planetary radius, and a the semi-major axis.
Further orbit parameters are $\rm cos(i_b)$, with the inclination $\rm i_b$ which can be used to derive the impact factor $\rm b_{tra;b}$, as well as the parameters $\rm f_c = \sqrt{e} cos(\omega)$ and $\rm f_s = \sqrt{e} sin(\omega)$, with the eccentricity $\rm e$ and the longitude of periastron $\rm \omega$. The limb darkening is parameterised using the quadratic parameters u1 and u2 which were sampled using the parameters q1 and q2 as defined in \cite{Kipping13}. 
We sampled the posterior probability distribution (PPD) of the model parameters using nested sampling \citep{speagle19}. In order to optimise the run-time we used wide uniform priors based on transit parameters, derived from a preliminary solution obtained from a least-squares fit. We used a stellar density prior from stellar mass and radius estimates, derived from spectroscopy and took the dilution within the CoRoT mask into account.
We then used the function \texttt{massradius} from the software package \texttt{pycheops} \citep{maxted21} which applies a Monte Carlo approach to derive physical system parameters from the sampled parameter $\rm period$, $\rm a/R_\star$, $\rm b_{rr}$, $\rm cos(i_b)$, the radial velocity semi-amplitude K, as well as the stellar mass and radius estimates. For CoRoT--34, 35, and 36, we derive the stellar density directly from Keplers third law and the before mentioned light curve parameters \citep{seager03}, which we used to derive the surface gravity $\log(g)_{lc}$ of the host star. We used this to optimise the mass and radius of the host stars, but also of the planets and brown dwarf. As a comparison, we listed in Table~\ref{tab:sub_obs} the surface gravity derived from spectroscopy ($\log(g)_{sp}$), as well as that from the light curve fit ($\log(g)_{lc}$).

\begin{table*}

\centering
\begin{tabular}{lccc|cccccccc|l}
CoRoT ID & Win ID& Spt & $\rm M_\star$ & Period[d] & 	SANDIFORD & CAFE & TWIN & UVES & FIES & HIRES & HARPS &  Status\\\hline\hline
102806520 & IRa01\_E1\_4591 & A5V & 1.9$^\star$ & 4.29 &   &  6 &  &   &   &   &    &     EB  \\
102850921 & IRa01\_E2\_2721 & A6V & 2.0$^\star$ & 0.61 & 4 &  8 &  &10 &   &   &    &     V  \\
102584409 & LRa01\_E2\_0203 & F1V & 1.3$^\star$ & 1.86 & 16& 8  &  &5  &   &   &    &     BEB  \\\hline 
102605773 & LRa01\_E2\_0963 & F0V & 1.7$^\star$ & 4.65 &  3& 16 &  &   & 4 &   &    &      FD \\
102627709 & LRa01\_E2\_1578 & F6V & 1.3$^\star$ & 16.06 &  4&    &  &   &   &   &    &     EB  \\
110853363 & LRa02\_E1\_0725 & A5IV & 2.1$^\star$ & 9.09 &  5&    &  &   &   &   &    &     EB  \\\hline
110756834 & LRa02\_E1\_1475 & A7V & 1.66 & 2.12 &  &   &  & 15&   & 5 & 17 &     BD  \\
110858446 & LRa02\_E2\_1023 & F3V & 1.4$^\star$ & 0.78 &  4&    &  &   &   &   &    &     BEB\\
110660135 & LRa02\_E2\_4150 & B4V & 2.26 & 8.17 & 14& 7  &  &   &10  &   & 2  &    EB \\\hline
310204242 & LRc03\_E2\_2657 & A7III & 6.49 & 10.3 &   &    &6 & 4 &   &   &  2  &     BEB \\
659719532 & LRc07\_E2\_0108 & A9IV & 1.8$^\star$ & 14.45 &  7&    &13&   &   &   &    &     EB \\
652345526 & LRc07\_E2\_0307 & F3V & 1.32 & 5.62 &  3&    &9& 11&28 &   &  6 &     planet \\\hline
659668516 & LRc08\_E2\_4203 & F3V & 2.27 & 3.29 &   &    & 6& 4 &   &   & 2   &    candidate \\
659460079 & LRc09\_E2\_3322 & F3V & 1.01 & 3.23 &   &    &  &   &   &   &   8 &   planet \\
659721996 & LRc10\_E2\_3265 & F6V & 2.1$^\star$ & 4.83 &   &    & 6&   &   &   &    &     BEB \\\hline
632279463 & LRc07\,E2\,0146 & F0V & 1.2$^\star$ & 0.49 &   &    &  &   &   &   &    &     EB$^1$ \\
631423419 & LRc07\,E2\,0187 & F8IV & 1.2$^\star$ & 3.88 &   &    &  &   &   &   &    &     EB$^1$ \\
\end{tabular}
\caption{Summary of the 17 candidates. WIN ID is the Window ID of the CoRoT data set, which has been used as reference in G16. The third to fifth columns show the spectral type and stellar mass obtained from low resolution spectroscopy, as well as the photometric detection period, respectively. The number of radial velocity measurements obtained in this study is given for each instrument. The last column shows the final status of the candidate (Special abbreviations: FD: False detection, V: stellar variability).
$^1$ Not followed-up with high resolution spectroscopy, $^\star$ Mass estimate from spectral type.} 
\label{tab.061}
\end{table*}

\begin{table*}
	\centering
	\begin{tabular}{lccc} 
		\hline
		Stellar parameters & CoRoT--34 & CoRoT--35 & CoRoT--36 \\\hline
        RA (J2015.5) & 06 51 29.01 & 19 17 15.43 & 18 31 00.24 \\
        DEC (J2015.5) & -03 49 03.46 & -02 46 28.82 & +07 11 00.05 \\
        CoRoT ID & 110756834 & 659460079 & 652345526 \\
        CoRoT Win ID & LRa02\,E1\,1475 & LRc09\,E2\,3322 & LRc07\,E2\,0307 \\
        Gaia EDR3 id & 3102398059230983296 & 4213081309275729792 & 4477340334766250496 \\
        2MASS id & J06512900-0349034 & J19171544-0246288 & J18310024+0711001 \\
        G [mag] & $\rm 14.10\pm0.02$ & $\rm 15.248\pm0.001$ & $\rm 12.941\pm0.03$ \\
        $\rm K_{S}$ [mag] & $\rm 12.68\pm0.04$ & $\rm 12.94\pm0.03$ & $\rm 11.59\pm0.02$ \\
        $^1$distance [pc] & $\rm 1540^{+210}_{-190}$ & $\rm 1140^{+50}_{-40}$ & $\rm 954\pm18$ \\
        Spectral type & A7 V & F6 V & F3 V \\
        $^2$$\rm T_{eff}$ [K] & $\rm 7820\pm160$ & $\rm 6390\pm130$ & $\rm 6730\pm140$ \\
        $^2$$\rm log(g)_{sp}$ & $\rm 3.88\pm0.20$ & $\rm 4.02\pm0.2$ & $\rm 3.92\pm0.2$ \\
        $^3$$\rm log(g)_{lc}$ & $\rm 4.10\pm0.10$ & $\rm 4.01\pm0.05$ & $\rm 4.16\pm0.06$ \\
        $^2$$\rm v_{rot}\sin i_\star$ [km\,s$\rm ^{-1}$] & $\rm 141.7\pm2.7$ & $8.8\pm0.3$ & $\rm 25.6\pm0.3$ \\
        $\mathrm{[Fe/H]}$$^2$&$\rm-0.2\pm0.2$&$\rm-0.5\pm0.1$&$\rm-0.1\pm0.1$\\
        $\rm Age [Gyr]$&$\rm 1.09^{+0.19}_{-0.21}$&$\rm 6.1^{+1.3}_{-1.3}$&$\rm 2.1^{+0.6}_{-0.5}$\\
        $\rm M_{\star} [M_{\sun}]$ & $\rm 1.66^{+0.08}_{-0.15}$ & $\rm 1.01^{+0.07}_{-0.06}$ & $\rm 1.32^{+0.09}_{-0.09}$ \\
        $\rm R_{\star} [R_{\sun}]$ & $\rm 1.85^{+0.29}_{-0.25}$ & $\rm 1.65^{+0.10}_{-0.11}$ & $\rm 1.52^{+0.20}_{-0.10}$ \\
        \hline
        Fitted parameters & CoRoT--34b & CoRoT--35b & CoRoT--36b\\\hline
        
        period [d] & $\rm 2.11853\pm0.00006$ & $\rm 3.22748\pm0.00008$ & $\rm 5.616531 \pm0.000023$ \\
        phot. epoch [$\rm BJD_{TDB}$] & $\rm 2454787.7411\pm0.0016$ & $\rm 2456074.2415 \pm 0.0018$ & $\rm 2455662.52236\pm0.00034$ \\
        $\rm b_{rr}$        & $\rm0.0609\pm0.0022$ & $\rm 0.1047\pm0.0017$ & $\rm 0.0953\pm0.0013$\\
        $\rm b_{rsuma}$       & $\rm0.244\pm0.025$   & $\rm 0.197\pm0.010$ & $\rm 0.121\pm0.008$\\
        K [km\,s$\rm ^{-1}$] & $\rm 7.68\pm0.85$ & $\rm 0.15\pm0.05$ & $\rm 0.065\pm0.045$ \\
        e (fixed) & 0 & 0 & 0 \\
        
        \hline
        Companion parameters & CoRoT--34b & CoRoT--35b & CoRoT--36b\\
        \hline
        $\rm a_\mathrm{b}$ [AU] & $\rm 0.03874 \pm 0.00081$ & $\rm 0.04290\pm0.00092$ & $\rm 0.066\pm0.007$ \\
        $\rm i_\mathrm{b}$ [deg] & $\rm 77.8_{-1.6}^{+1.4}$ & $\rm 84.1\pm0.1$ & $\rm 85.83\pm0.26$ \\
        $\rm b_\mathrm{tra;b}$ & $\rm 0.92\pm0.02$ & $\rm 0.58_{-0.07}^{+0.06}$ & $\rm 0.573_{-0.025}^{+0.023}$ \\
        transit depth [mmag] & $\rm 3.34_{-0.17}^{+0.21}$ & $\rm 11.47\pm0.31$ & $\rm 9.214_{-0.10}^{+0.099}$ \\
        T$_\mathrm{tot;b}$ [h] & $\rm 2.02_{-0.11}^{+0.12}$ & $\rm 4.19\pm0.08$ & $\rm 4.98\pm0.07$ \\
        quadr. limb u1 & $\rm 0.32\pm0.08$ & $\rm 0.25\pm0.15$ & $\rm 0.27_{-0.07}^{+0.06}$ \\
        quadr. limb u2 & $\rm 0.17\pm0.08$ & $\rm 0.28\pm0.17$ & $\rm 0.24_{-0.09}^{+0.12}$ \\
        $\rm M_\mathrm{b}$ [$\mathrm{M_{Jup}}$] & $\rm 71.4_{-8.6}^{+8.9}$ & $\rm 1.10_{-0.37}^{+0.37}$ & $^\star$$\rm 0.68_{-0.43}^{+0.47}$  \\
        $\rm R_\mathrm{b}$ [$\mathrm{R_{Jup}}$] & $\rm 1.09_{-0.16}^{+0.17}$ & $\rm 1.68\pm0.11$ & $\rm 1.41\pm0.14$ \\
        $\rm \rho_b [g\,cm ^{-3}]$ & $\rm 60\pm21$ & $\rm 0.29\pm0.11$ & $\rm 0.25\pm0.17$\\ 
         $\rm T_\mathrm{eq;b}$ [K] & $\rm 2425_{-128}^{+137}$ & $\rm 1747\pm62$ & $\rm 1567\pm35$ \\

        \hline
	\end{tabular}
	
	\caption{Parameters of the detected planets and brown dwarf companion and their host-stars. CoRoT WIN ID is the Window ID of the CoRoT data set, which has been used as reference in G16. $^1$ Displayed distances were determined by spectrophotometry for CoRoT--34, and from Gaia EDR3 parallaxes for CoRoT--35, and 36. $^2$Parameters derived from high-resolution spectroscopy, $^3$$\rm log(g)_{lc}$ derived from light curve fitting. $^\star$The reported mass of CoRoT--36b is an upper limit.}
	\label{tab:sub_obs}
\end{table*}

\section{Sub-stellar companions discovered} \label{sec_planets}

As we will show in this section, CoRoT-34 (CoRoT 110756834) is a mid A-type star with a companion that is just at the border between a very low-mass star and a brown dwarf, and CoRoT--35 (CoRoT 659460079), as well as CoRoT--36 (CoRoT 652345526) are mid F-type stars, with each harbour a giant planet.

\subsection{CoRoT--34}
 
\subsubsection{Exclusion of background sources for CoRoT--34}

This star was discovered as a transit candidate in the original CoRoT survey with a short period of about 2.12\,d. \cite{guenther13} obtained adaptive optic imaging and spectroscopy in the K-band, using CRIRES at the VLT at the Paranal Observatory. They excluded any background eclipsing binary (BEB) earlier than K3V within 0.8arcsec and ruled out any physical companion earlier than F6V. Furthermore, they excluded all stars in and close to the CoRoT mask being eclipsing binaries, using seeing-limited imaging obtained with CFH12K prime focus camera of the 3.6 m Canada France Hawaii Telescope (CFHT; located at Mauna Kea, Hawaii, USA). All these observations confirmed that the transit originates from the star itself and not from a background star within the PSF.

\subsubsection{Analysis of the CoRoT light curve of CoRoT--34}

\begin{figure}
   \centering
   \includegraphics[width=\hsize]{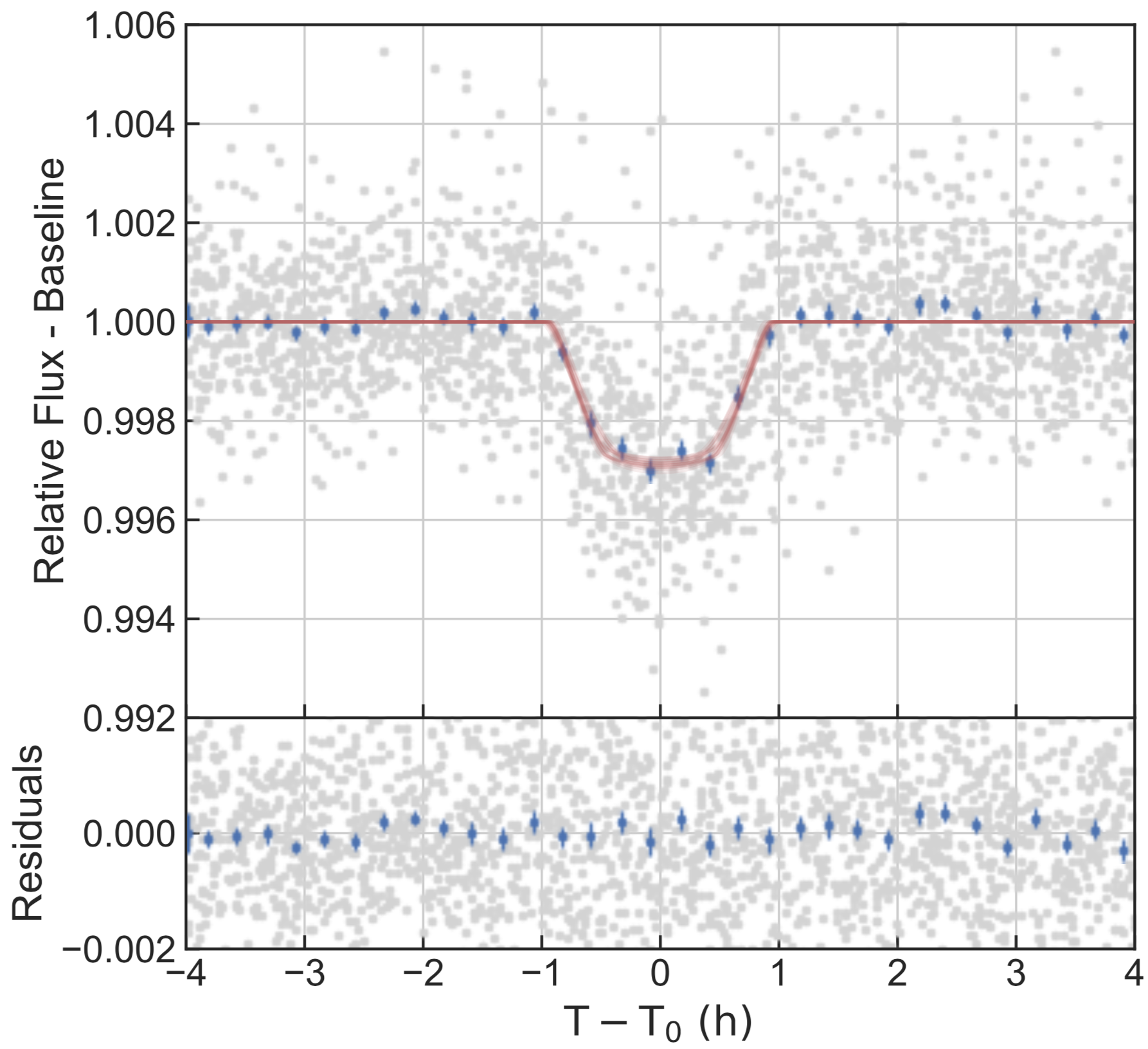}
      
   \caption{Upper panel: Phase folded, light curve of CoRoT--34. The red line is 
   the best fitting model, using a dilution factor of $F_{\text{cont}}/F_{\text{source}} =$ 0.014. The lower panel shows the residuals from the transit fit.}
         \label{Fig_lc_1475}
\end{figure}

The star was observed with CoRoT between November $\rm 16^{th}$ 2008 and March $\rm 11^{th}$ 2009. Thanks to the star's brightness, three colour light curves have been obtained. \cite{sarro13} used CoRoT light curves combined with stellar parameters derived from Giraffe spectra to automatically classify different pulsation pattern. They classified CoRoT--34 as a pulsating star with a period of 1.4\,d. By inspecting the light curve, we find, that the pulsation period is actually half this period (0.71\,d) with a variable amplitude of up to 1\%.
After detrending the light curve for these pulsations, we used the color information to determine the transit depth in all three bands. Despite increased noise in the blue and green CoRoT light curves, we cannot see any significant decrease in transit depth, and, thus, can exclude any late type BEB of K or M-type. This is also supported by the non detection of any infrared excess by fitting the SED of CoRoT--34 (see Fig.~\ref{fig:photometry_sed_corot34}). 
The phase folded light curve is shown in Fig.~\ref{Fig_lc_1475} and all derived transit parameters are listed in Table~\ref{tab:sub_obs}.

\subsubsection{High-resolution spectroscopy of CoRoT--34}

We obtained 5 spectra with HIRES, 17 spectra with HARPS (using the HAM mode, under ESO programme 184.C-0639) and 15 spectra with UVES (under ESO programme 092.C-0222) over a total time-span of 3.1\,yrs. 
We used the combined HARPS and UVES spectra to derive atmospheric parameters from our global spectral fit. The resulting parameters for $\rm T_{eff}$, $\rm log(g)_{sp}$, [Fe/H], and $\rm v_{rot}\sin i_\star$ are listed in Table~\ref{tab:sub_obs} which are consistent with the values derived by \cite{sarro13}. Unfortunately, these would put CoRoT--34 close to the termination of the main-sequence evolution, that is to the end of core hydrogen burning, making it difficult to derive the stellar mass and radius because of the changing shape of the evolutionary tracks. From the distribution of best-fit evolutionary tracks within the spectroscopic uncertainties, we derive a mass of  $1.76^{+0.28}_{-0.15}$\,$M_{\sun}$, which we used as a prior to fit the light curve.
$\rm log(g)_{lc}$, derived from the light curve, allowed us to constrain the stellar mass and radius of CoRoT-34 (see Fig.~\ref{fig:kiel}), which are listed in Table~\ref{tab:sub_obs}.

The star is fast rotating with $\rm v\,sin(i_{\star}) = 141.7\pm2.7$\,km\,s$\rm ^{-1}$, which complicates the mass determination of the companion. We used our least square fitting method, and got a good fit with a mean accuracy of 100\,m\,s$\rm ^{-1}$. Nevertheless, the radial velocity varies by several km\,s$\rm ^{-1}$ even for spectra taken in the same night. One possible explanation is the impact of stellar oscillations leading to distortions of the line profile. 
In order to measure this effect, we modelled the least-squares deconvolution (LSD) line profiles \citep{donati97} for each spectrum using a line list, with line weights optimised for an IMS with $\rm T_{eff} = 8000\,K$ and $\rm log(g) = 4.0$ \citep{lehmann11} based on data from the Vienna Atomic Line Database (VALD, \citealt{kupka00}).  
The LSD profile clearly shows the distortions of the line-profile caused by the pulsations. Since the magnitude of the observed distortions is only a fraction of the broadened line profile, a simple Gaussian least squares fit is used to obtain the mean velocity. In a second step, we calculated 
the integral of the profile to account for the asymmetry of the profile as a measure of the line distortions. The variance of both measurements is, thus, used as uncertainty. The UVES spectra showed an instrumental offset between the blue and red CCD. We decided to measure both spectra individually, correct for the constant offset and average the RV measurements. The RV results are listed in Table~\ref{tab:rv_1475}.

\begin{figure}
   \centering
   \includegraphics[width=\hsize]{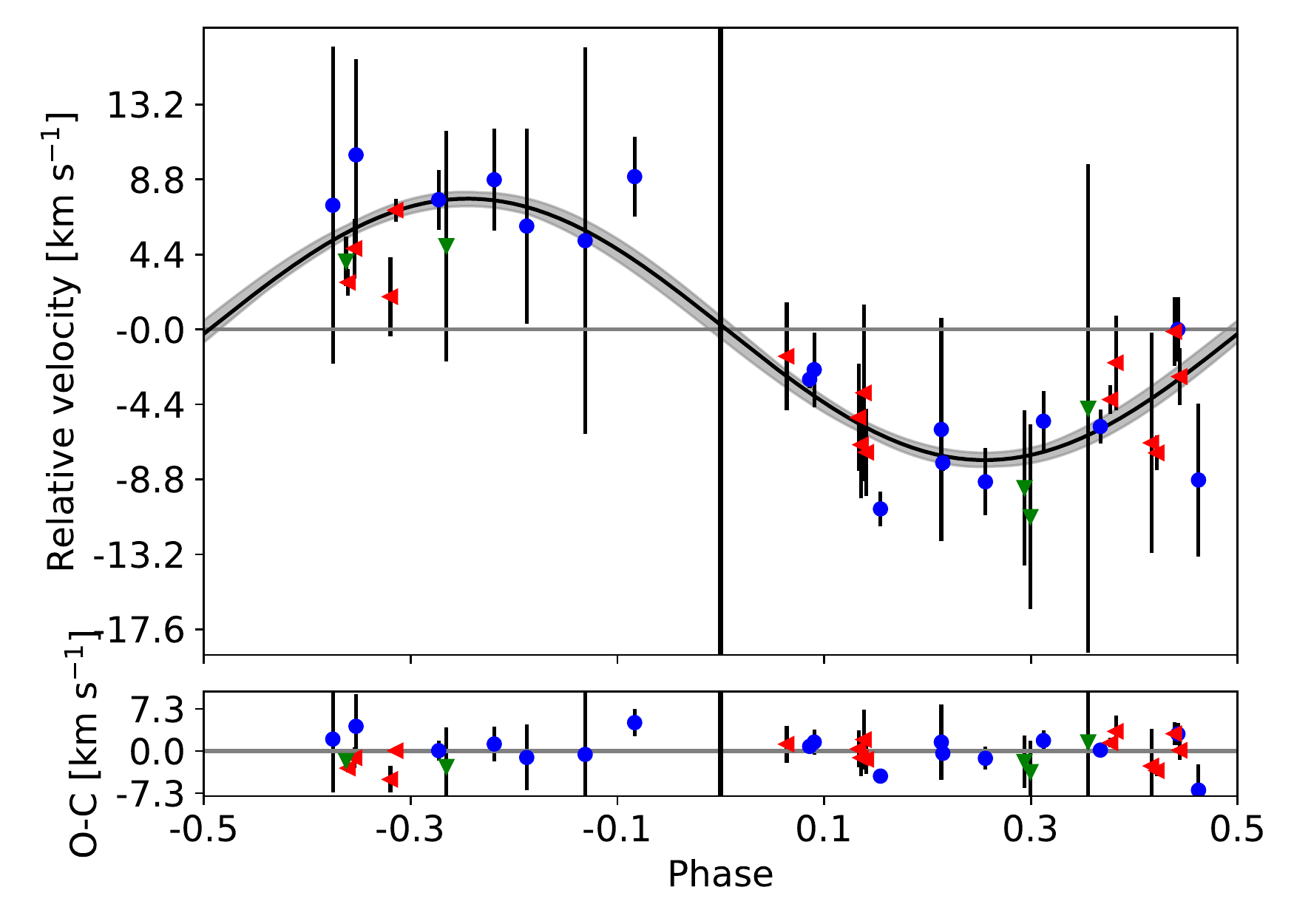}
      
   \caption{upper panel: Phase folded, radial-velocity curve of CoRoT--34. Blue points: HARPS measurements; green triangles: HIRES measurements; red triangles: UVES measurements. The photometric epoch of the transit is at phase\,=\,0. The black lines shows the best fitting orbital solution, with the grey area depicting the uncertainty of the fit. The lower panel shows the residuals from the orbital fit.}
\label{Fig_rv_1475}
\end{figure}

The radial velocity curve has been constructed from all high resolution spectra to determine the mass of the companion. The phase folded RV measurements for all instruments are shown in Fig.~\ref{Fig_rv_1475}. We found an orbital solution that is in perfect agreement with the photometric period and ephemeris. The residuals of the orbital fit follow a Gaussian distribution which we have verified using a Kolmogorov-Smirnov test by comparing it to a Gaussian distributed sample. The standard deviation of the residuals is $\rm 2.6\,km\,s\rm ^{-1}$, which is smaller than the average uncertainty of our RV measurements. Thus, we can conclude that the orbital solution is (i) stable  for data obtained with different instruments, (ii) stable over several years, and (iii) independent from the variable photometric amplitude of the stellar pulsations. 
Using the known orbit inclination, we derived the mass of the companion and found it to be a high-mass brown dwarf close to the hydrogen burning limit with an extreme mass ratio q = $\rm M_{BD}/M_{\star}$  = $\rm 0.0412\pm0.0048$. All parameters derived for this object are given in Table~\ref{tab:sub_obs}.


\subsection{CoRoT--35}

\subsubsection{Excluding background sources for CoRoT--35}
The transit was discovered during LRc09, the last Galactic center fields observation run of CoRoT, with a transit depth of $1\%$.

Using the CFH12K prime focus camera of the CFHT we obtained on and off transit images on August $\rm 27^{th}$ and $\rm 28^{th}$ 2012. These images show that CoRoT--35 is $\rm 0.0097\pm0.0026$ mag fainter during the transit. No other nearby star showed a significant change in brightness. We, thus, conclude that CoRoT--35 is the star with the
transit. The images show no stars in the direct vicinity and the dilution factor is $\rm 0.5\pm0.2\%$, which is small, compared to other targets in this survey (which is usually $\rm >1\%$). 

\subsubsection{Analysis of the CoRoT light curve of CoRoT--35}

\begin{figure}
  \includegraphics[height=.27\textheight,angle=0]{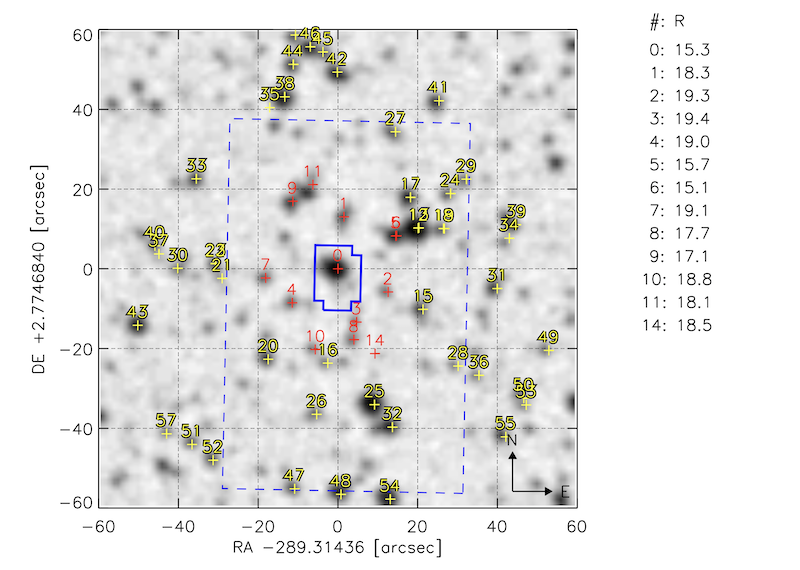}

    \caption{DSS image of CoRoT--35 with the photometric mask overlaid.
North is up, and east is on the right.}
  \label{mask35}
\end{figure}

\begin{figure}
   \centering
   \includegraphics[width=\hsize]{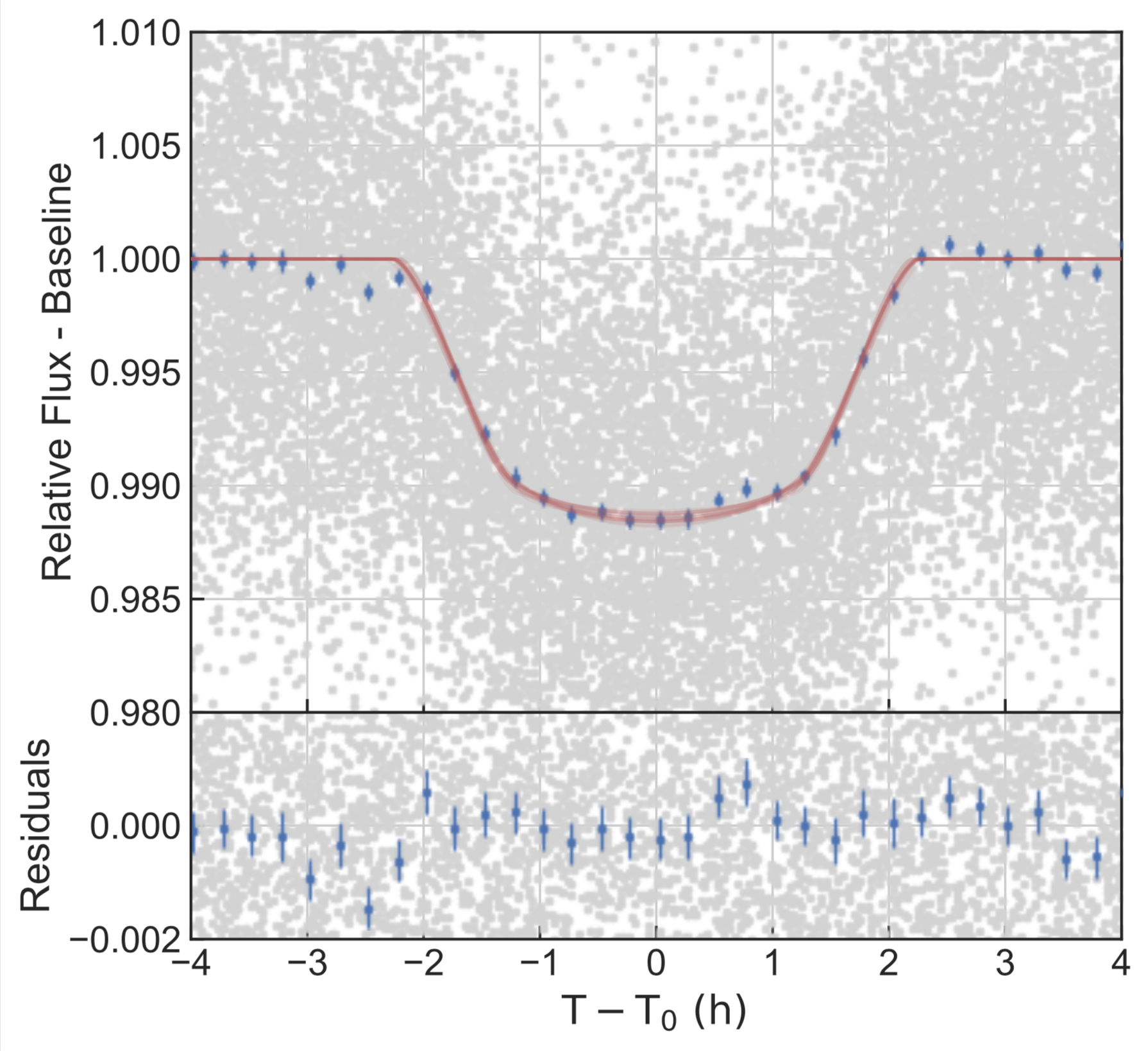}
      
   \caption{Phase folded, light curve of CoRoT--35. The red line is the best fitting model. The lower panel shows the residuals from the transit fit. Blue points represent the data binned to 12\,min.}
         \label{Fig_lc_3322}
\end{figure}

CoRoT--35 was observed in monochromatic mode for 84\,days in 2012 from April $\rm 12^{th}$ to July $\rm 7^{th}$, first with the low cadence mode (8.5\,min), and after the star has been found to be transiting with the high cadence mode (32\,sec). Fig.\,\ref{mask35} shows an DSS image taken with the CoRoT-mask superimposed.
Twenty six transits have been observed and we found that the transit mid-times vary by up to one hour during the CoRoT observations. This strongly hints for another massive, but possibly sub-stellar object orbiting CoRoT--35. To model the out-of-transit variations using a polynomial fit, we increased the transit mask to 1.5\,h before and after the predicted transit times to account for the mid-time variations. After the out-of-transit variations were corrected, we modelled the mid-time variations using \texttt{ALLESFITTER} and took these into account to model the transit parameters from the light curve and to determine the size and other parameters of the planet. The mid-time variations are listed in Table~\ref{tab:ttvs}, the phase folded light curve is shown in Fig.~\ref{Fig_lc_3322}. Despite the fit does agree well with the data, we note a small hump during the transit about 0.7h after the transit mid-time. Such humps generally could hint for stellar activity such as spot crossing events. Since our transit model does not include spot crossing events, our fit might underestimate the planet-to--star radius ratio \citep[e.g.][]{Oshagh14}. Given the magnitude of $\rm R=15.28$, the light curve of CoRoT--35 has larger uncertainties, compared to the other companions found in this survey, thus spot crossing events are not resolved in individual transits. We repeated our transit fit using a different binning of 36\,min, to increase the photometric precision of the data, and derived similar transit parameters compared to our initial fit, but find no sign of a hump. We conclude that it is more likely an artifact from white and residual red-noise of the light curve. All results of the transit fit are listed in Table~\ref{tab:sub_obs}.

\subsubsection{High-resolution spectroscopy of CoRoT--35}
We took 8 spectra of CoRoT--35 with the HARPS spectrograph using the EGGS mode (under ESO programme 188.C-0779). The stellar parameters from our global spectral fit are listed in Table~\ref{tab:sub_obs}. It shows CoRoT--35 to be a late F-type star that is slowly rotating. We found that the star is relatively metal-poor. This is in line with its comparatively low mass of $\rm M_\star = 1.01\pm0.13/,M_\odot$ and old age of $\rm \approx6\,Gyr$ as derived from MIST evolutionary tracks. Using the optimised $\rm log(g)_{lc}$ from the light curve fit, we again derived the stellar mass and radius from evolutionary tracks (see Fig.~\ref{fig:kiel}). In this case, the $\rm log(g)_{sp}$ from spectral fitting matches well the light curve fit, but given the improved accuracy of $\rm log(g)_{lc}$ we were able to better constrain the stellar radius which is listed in Table~\ref{tab:sub_obs}. 

\begin{figure}
    \centering
   \includegraphics[width=\hsize]{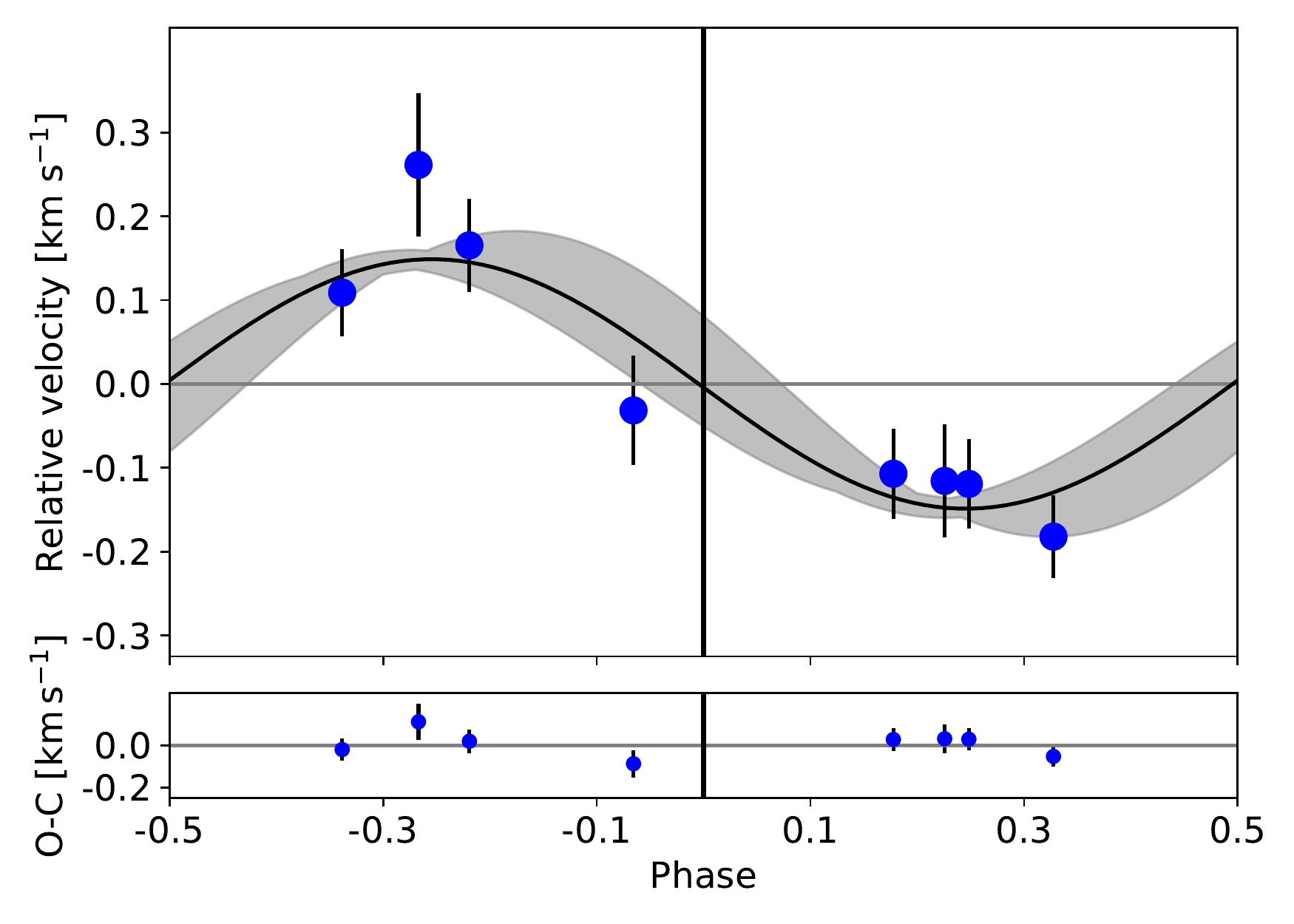}
   
   \caption{Phase folded radial-velocity measurements of CoRoT--35b from HARPS. The photometric epoch of the transit is at phase\,=\,0. The black curve represents the best fitting orbit model, with the grey shaded region depicting the uncertainty of the fit. The lower panel shows the residuals from the orbital fit.}
         \label{Fig_3322_RV}
 \end{figure}

Since the rotational velocity is with $\rm v_\mathrm{rot}\sin i_\star <10\,km\,s\rm ^{-1}$ relatively small, we determined the RVs using the classical cross-correlation method with a numerical mask that corresponds to a G2 star \citep{baranne96,pepe02}. The RV measurements were obtained by fitting a Gaussian function to the average cross-correlation function (CCF), after discarding the ten bluest and the two reddest orders of the spectra that were of too low SNR. The results are given in Table~\ref{tab:RV2}. One of the eight HARPS spectra, taken on September $\rm 9^{th}$ 2013, deviates slightly ($\rm <1.5\,\sigma$) from the orbital solution. This particular night suffered from strong wind ($\rm >15\,m\,s\rm ^{-1}$ ) and variable seeing conditions (1.1 -- 1.7\,arcsec), reducing the SNR to about 50\%, and increasing the relative error compared to the other data. Fig.~\ref{Fig_3322_RV} shows the best orbital fit to the HARPS measurements.

\subsection{CoRoT--36}

\begin{figure}
  \includegraphics[height=.27\textheight,angle=0]{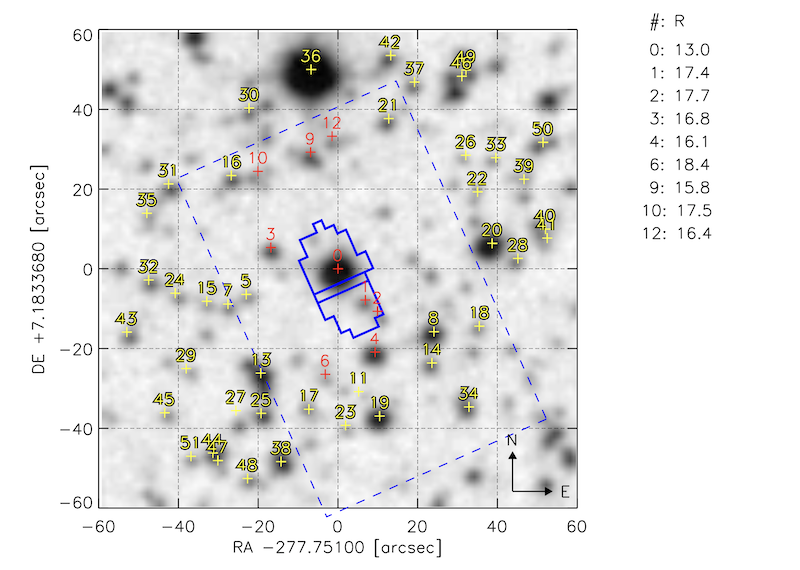}

    \caption{DSS image of  CoRoT-36 with the photometric mask of CoRoT 
   overlayed.  North is up, and east is on the right.}
  \label{mask36}
\end{figure} 

\begin{figure}
  \includegraphics[height=.25\textheight,angle=0]{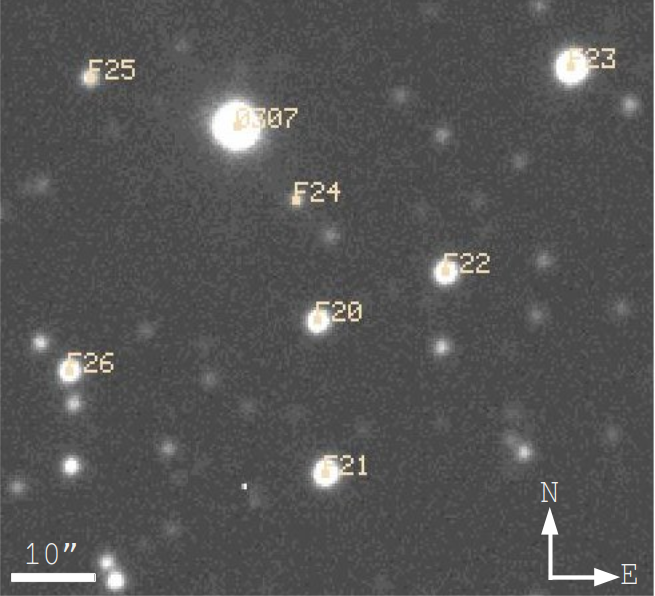}

 \caption{Image taken of CoRoT--36 (labelled as 0307) with the 0.8m telescope of the IAC. North is up, and east is on the right. The size of the image is about $\rm 60\times 60 \arcsec$, and the orientation is the same as in Fig.\,\ref{mask36}. The stars labeled with F20 and F22 are just outside the mask but star F24 is inside.}
  \label{IAC36}
\end{figure} 

\begin{figure}
   \centering
   \includegraphics[width=\hsize]{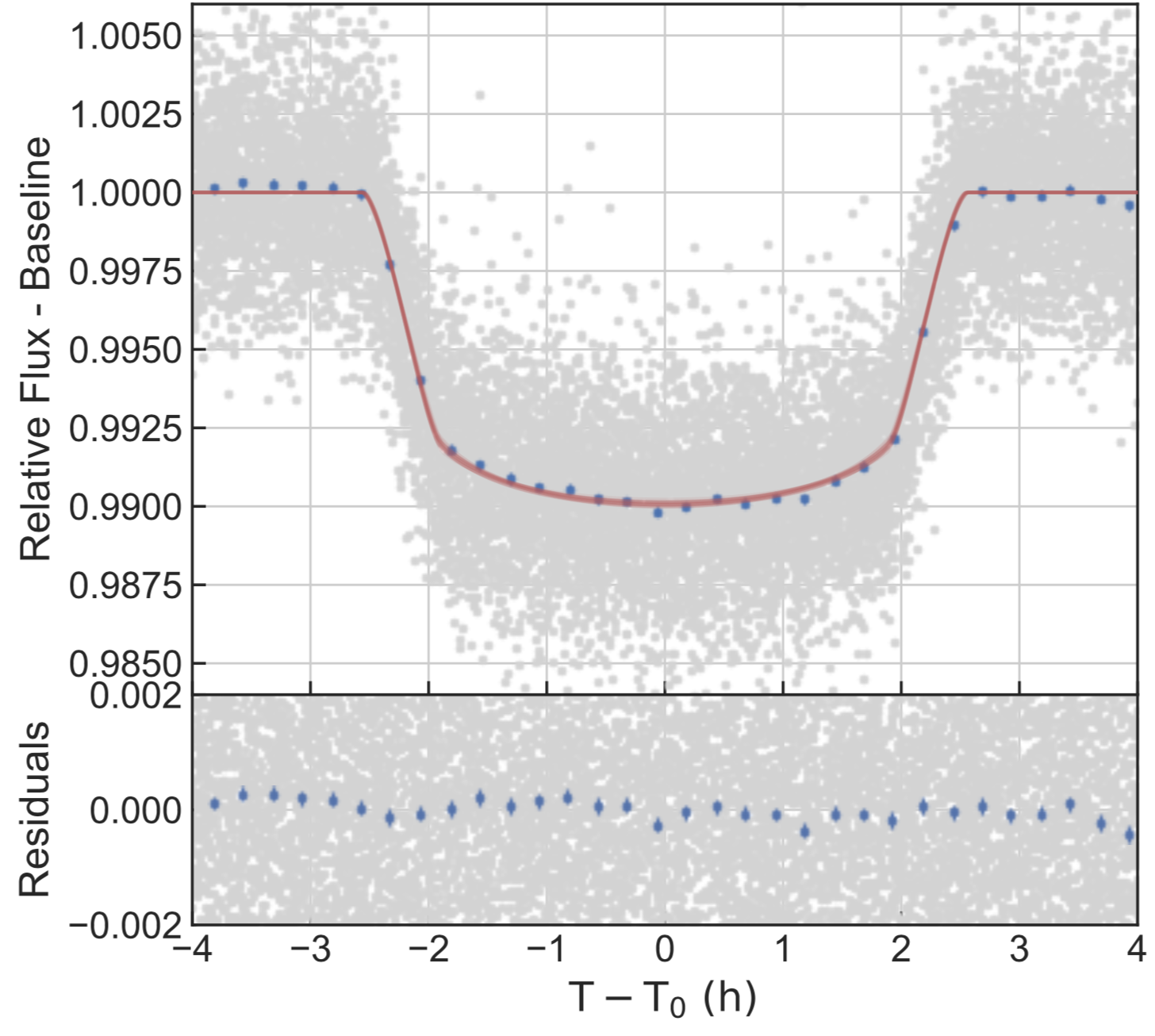}
   
   \caption{Phase folded, white light curve of CoRoT--36. The red line is the best fitting model. The lower panel show the residuals from the transit fit. Blue points represent the data binned to 12\,min.}
         \label{307_trans}
\end{figure}

\subsubsection{Excluding a background binary within the photometric mask of CoRoT--36} 
\label{Sec:onoff}

To exclude that CoRoT--36 is an FP, we used seeing limited imaging as well as adaptive imaging.
The rate and nature of FPs in the CoRoT exoplanets search, and the way how to detect and remove FPs is described in \cite{almenara09}.

\subsubsection*{Seeing limited imaging of CoRoT--36}
As part of the ground-based photometric follow-up for CoRoT candidates \citep[described in][]{2009A&A...506..343D}, we obtained time-series imaging of CoRoT--36 during two predicted transits. Both were acquired in R-band, one with the WISE-1-m telescope on  September $\rm 9^{th}$ 2011 and the other one with the IAC-80-cm telescope on July $\rm 25^{th}$ 2014  (Fig.\,\ref{IAC36}). The light curves and transit mid-times from both transits have been published in \citep{2020JAVSO..48..201D}. A comparison of Fig.\,\ref{mask36} with Fig.\,\ref{IAC36}
shows that the stars labeled with F20 and F22 are just outside the mask
but star F24 is inside. Both data-sets show that the target varied by an amount that would correspond to the depth of the
transit, whereas the variations of the other stars are at least one order of
magnitude less than what would be expected for an FP. 
The data obtained with the WISE
telescope were obtained 
73 days after the end of the CoRoT-observations and contained only an egress, whereas the IAC-80 observations were obtained over 3 years later, and contained a nearly complete transit. The IAC-80 timing given in \citet{2020JAVSO..48..201D} was, therefore, converted to the BJD\_TDB time-scale of the CoRoT data, and was used to derive an orbital period with a precision (see Table \ref{tab:sub_obs}) that is significantly improved over one based only on the CoRoT data.

\bigskip

\begin{figure}
  \includegraphics[height=.23\textheight,angle=0]{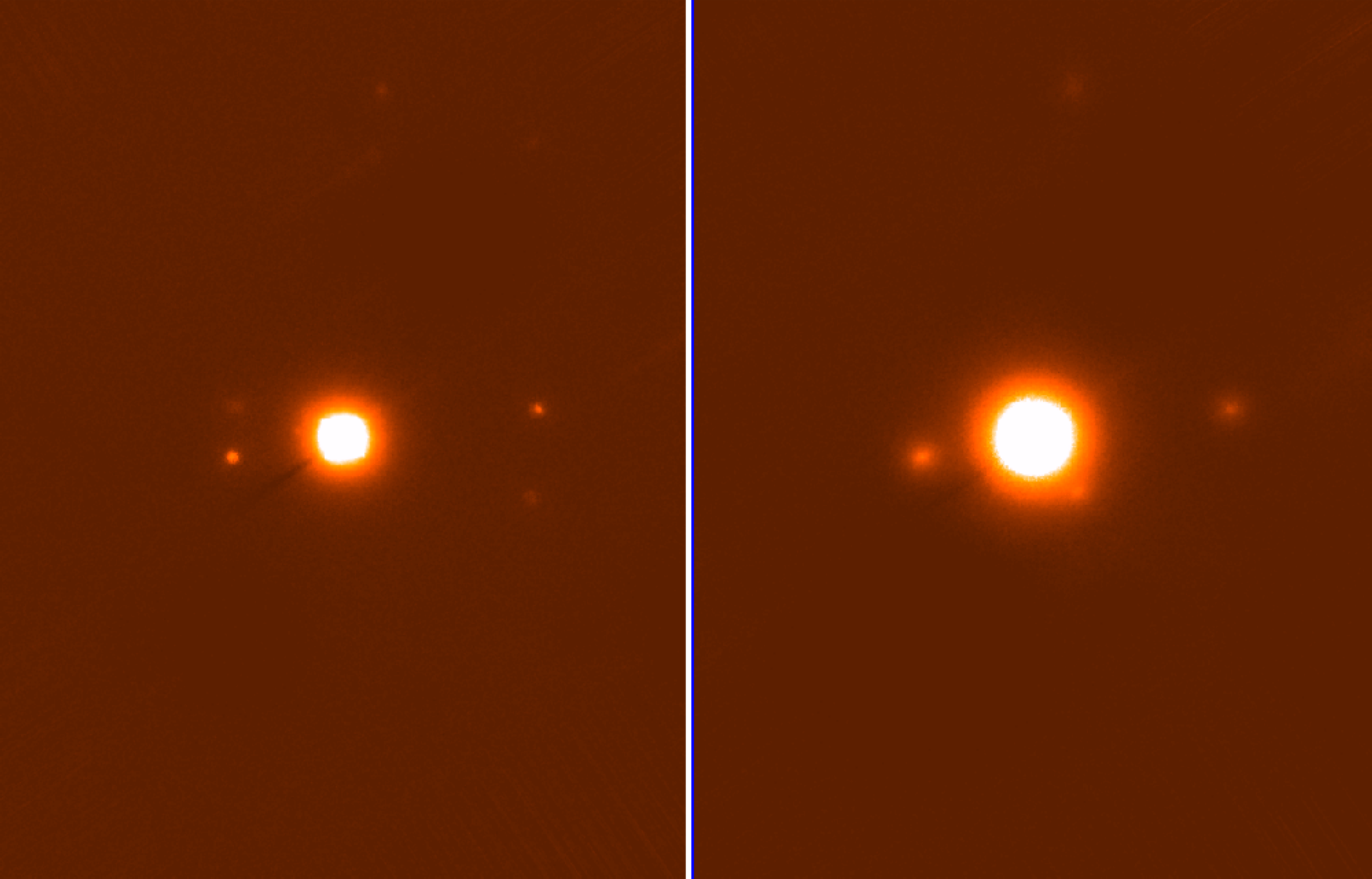}

    \caption{Images taken with the AO-system PISCES at the LBT in the K
  (left) and in the J-band (right). The size of the image is only
  about $6\times8\arcsec$, north is up and east is left. Two previously
  unknown stars at distances of $\rm 1\farcs96\pm0\farcs04$ and
  $\rm 3\farcs46\pm0\farcs04$ are visible. The orientation is as in Fig.~\ref{IAC36}.}
  \label{LBT}
\end{figure} 

\subsubsection*{Adaptive optics imaging of CoRoT--36}
In critical cases AO-imaging is essential for confirming transiting
planets, otherwise the probability for an FP is unacceptably 
high. We
consider CoRoT--36b a critical case, because it is very difficult
to confirm the planet by RV measurements due to the stellar rotation.

Observations were carried using PISCES \citep{Guerra11} together with
the  First Light Adaptive Optics (FLAO, \citealt{esposito10}) system mounted
on the Large Binocular Telescope (LBT). PISCES was the first 
adaptive optical imager of the LBT, but it is now  decommissioned. It was equipped with
a 1kx1k Hawaii-1 (HgCdTe) detector which provided an images scale of
0.019 arcsec/pixel.  For optimal sky-subtraction the images were
obtained by jittering. The final images were then obtained by
co-adding the overlapping region. Since the size of the image shown in
Fig.~\ref{LBT} is only $\rm 6\times 8\arcsec$, the star labeled F24 in
Fig.~\ref{IAC36} is far outside the field of view. A disadvantage of
adaptive optics imaging is that the observations are taken at infrared
wavelengths, whereas CoRoT observes within the optical. From the depth of
the transit, and the brightness of the star, we concluded that FPs have
to be brighter than V=$\rm 18.13\pm0.07$\,mag. Because of the extinction,
faint background stars are much brighter at infrared wavelength
than in the optical. Faint stars in the foreground have typically also red
colours, due to being late-type stars, unless they are white dwarfs or subdwarfs. 
We imaged the object in the J and K-band. This does not only allow us to
determine the colours of any potential background object but it
also helps to distinguish artifacts from real objects.

We detected two previously undetected stars. Star No. 1 is
$\rm 1\farcs94\pm0\farcs02$ W, $\rm 0\farcs30\pm0\farcs03$ S, and has
$\rm J=15.7\pm0.1$, $\rm K=16.2\pm0.1$, and star No. 2 is
$\rm 3\farcs42\pm0\farcs02$ E, $\rm 0\farcs53\pm0\farcs02$ N, and has
$J=17.1\pm0.1$, $K=16.8\pm0.1$. Both stars are not visible in the
images taken with the WISE and IAC-80-cm telescopes, but are identified in the Gaia DR3 catalogue as Gaia DR3 4477340339082969856 (G=18.12), and Gaia DR3 4477340339063864064 (G=19.96) for star No. 1 and 2 respectively. The Gaia parallaxes excluded both stars to be physical companions. Star No. 2 is clearly too faint in the optical to be a FP. Despite star No. 1 is in the optical at the limiting magnitude to cause a FP, we can rule out this scenario as by-product of our spectroscopic transit observations (see Sect.~\ref{RM_C36}).

\subsubsection{Analysis of the CoRoT light curve of CoRoT--36}

The star was continuously observed with the CoRoT satellite over a period of 81.2 days from April $\rm 8^{th}$ to June $\rm 28^{th}$ 2011. In total 218\,441 intensity measurements were obtained in three colors with a time-sampling of 32 seconds. Fig.\,\ref{mask36} shows an DSS image with the CoRoT mask 
superimposed. The observation were taken with a roll angle of $23.967^o$ and an image scale of 2.32 arcsec/pixel. The images of the stars are elongated, because of the bi-prism which provides the three colour photometry. In Total 15 transits can be seen in the raw-data. We divided the CoRoT light curve by a polynomial model to remove any long term variability of the star. Fig.\,\ref{307_trans} shows the phase-folded light curve, as well as, the best fitting transit model with the orbital eccentricity set to zero. The derived transit parameters are given in Table~\ref{tab:sub_obs}.

\subsubsection{Spectroscopic observations of CoRoT--36}\label{RM_C36}

Radial velocity measurements of CoRoT--36 were taken for two purposes: first to constrain the mass of the planet, and second to independently confirm the planet using time resolved transit spectroscopy.

\begin{figure}
   \centering
   \includegraphics[width=\hsize]{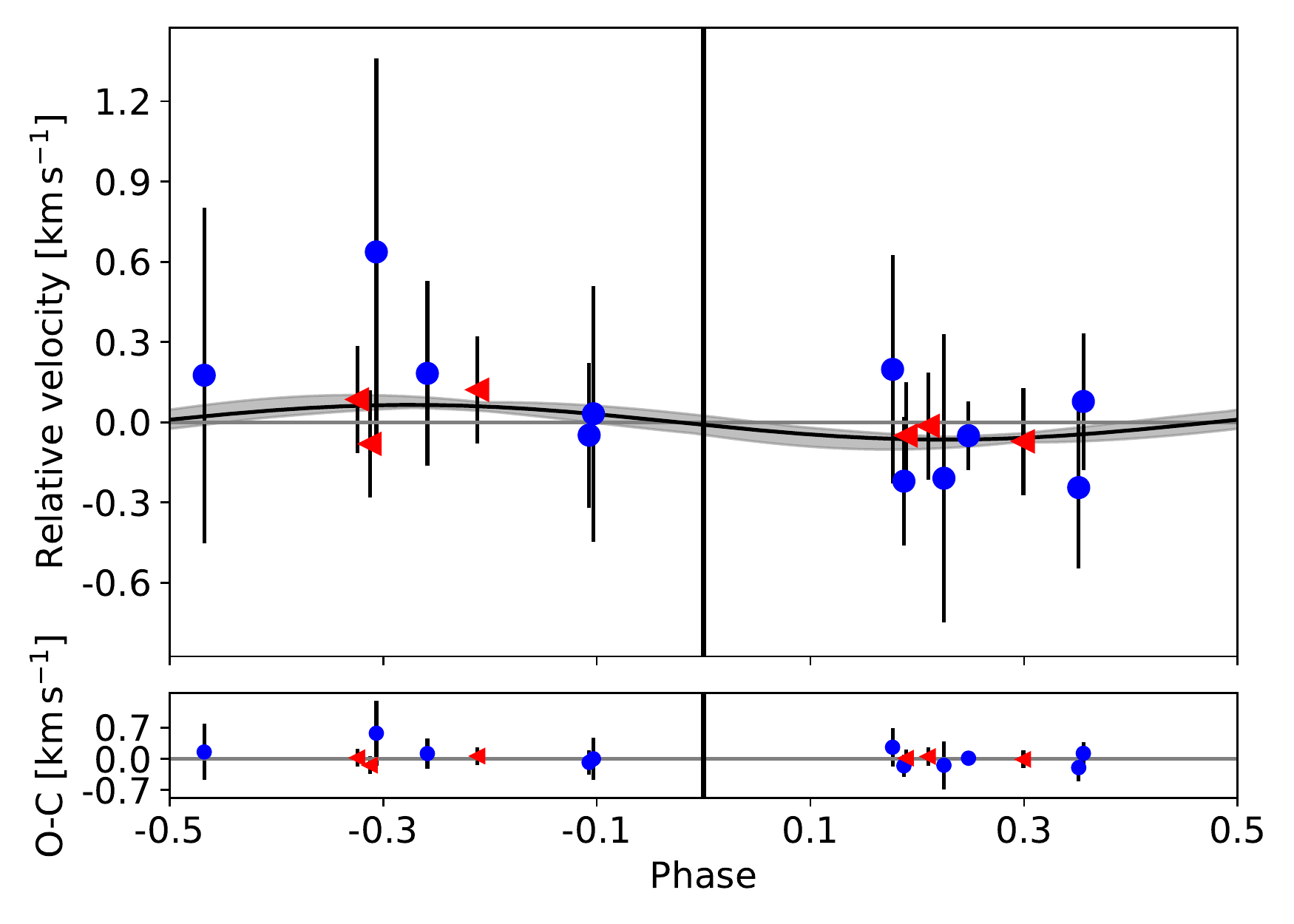}
   
   \caption{Phase folded radial-velocity measurements of CoRoT--36b obtained out of transit with HARPS (red triangles) and FIES (blue points). The photometric epoch of the transit is at phase\,=\,0. The black curve represents the best fitting orbit model, with the grey shaded region depicting the uncertainty of the fit. The lower panel shows the residuals from the orbital fit.}
    \label{fig:307RV_out}
\end{figure}

We obtained 9 measurements with 30\,min exposure time, reaching a SNR of about 80 with the TWIN spectrograph in June 2014 and 3 measurements with the SANDIFORD spectrograph. We observed CoRoT--36 with 30\,min exposures, resulting in an average SNR of about 80. The instrumental stability does not allowed us to use these spectra to exclude a stellar companion for the orbital period. Therefore, we used these data to determine the spectral type listed in Table\,\ref{tab.061}. 
The accuracy of the SANDIFORD spectra is, however, sufficient to rule out a massive binary companion. 
To finally narrow down the planet mass, we used two different instruments. Using the fibre-fed Echelle spectrograph FIES, we obtained 11 spectra of CoRoT--36 in four observing runs between June 2012 and June 2013. The data were reduced as described in Section.~\ref{sec_cand}. Additionally, six spectra of CoRoT--36 were obtained with HARPS in EGGS mode between June$\rm 15^{th}$ and July $\rm 10^{th}$ 2012 under ESO program 188.C-0779. Due to the lack of instrumental stability, we excluded the RV measurements obtained with the TWIN and SANDIFORD spectrographs, but used our measurements obtained with HARPS and FIES. Because of the line broadening due to stellar rotation, it is very difficult to determine the radial velocity variation caused by a Jupiter-mass companion. Nevertheless, we fitted all the different runs, by using instrumental offsets as free parameters, which allowed us to narrow down the semi amplitude of the orbit. The best fit parameters are listed in Table\,\ref{tab:sub_obs}, the orbital fit is shown in Fig\,\ref{fig:307RV_out}. Given the large uncertainties in the data, this measurement is rather an upper limit and confirms CoRoT--36b to be a Jupiter mass planet. All RV measurements, covering the orbital phase are listed in the Appendix, Table~\ref{tab:rv_corot36}.

While the rotation is a problem for precise RV observations, it opens up the possibility to confirm the planet by Doppler imaging (e.g. \citealt{bouchy08}).
Comparing out-of-transit spectra with spectroscopic time series taken during the transit allows us to confirm the presence of a transiting object, to determine its relative size, and its
orbital inclination relative to the spin-axis of the star.
 
We obtained two independent data sets of the transit. The first with 17 FIES spectra on July $\rm 25^{th}$ 2013. We used as templates the average spectrum of the target, by excluding the telluric lines.

Another data-set of 11 spectra was obtained with UVES in service mode (on August $\rm 6^{th}$ 2013). Each in-transit spectrum was exposed for 970\,s, the out-of-transit spectrum was exposed for 1450\,s. The in-transit observations were obtained for $\rm 2^h55^m$, and the out-of-transit spectrum was taken $\rm 2^h9^m$ after the end of in-transit observations in the same night. By taking the out-of-transit spectrum in the same night, we minimised the risk for any variations of the point-spread function, and we ensured that the out-of transit observations were certainly taken after the transit had ended. The achieved SNR at 600\,nm was on average 240 per resolution element.

\begin{figure}
   \centering
   \includegraphics[width=\hsize]{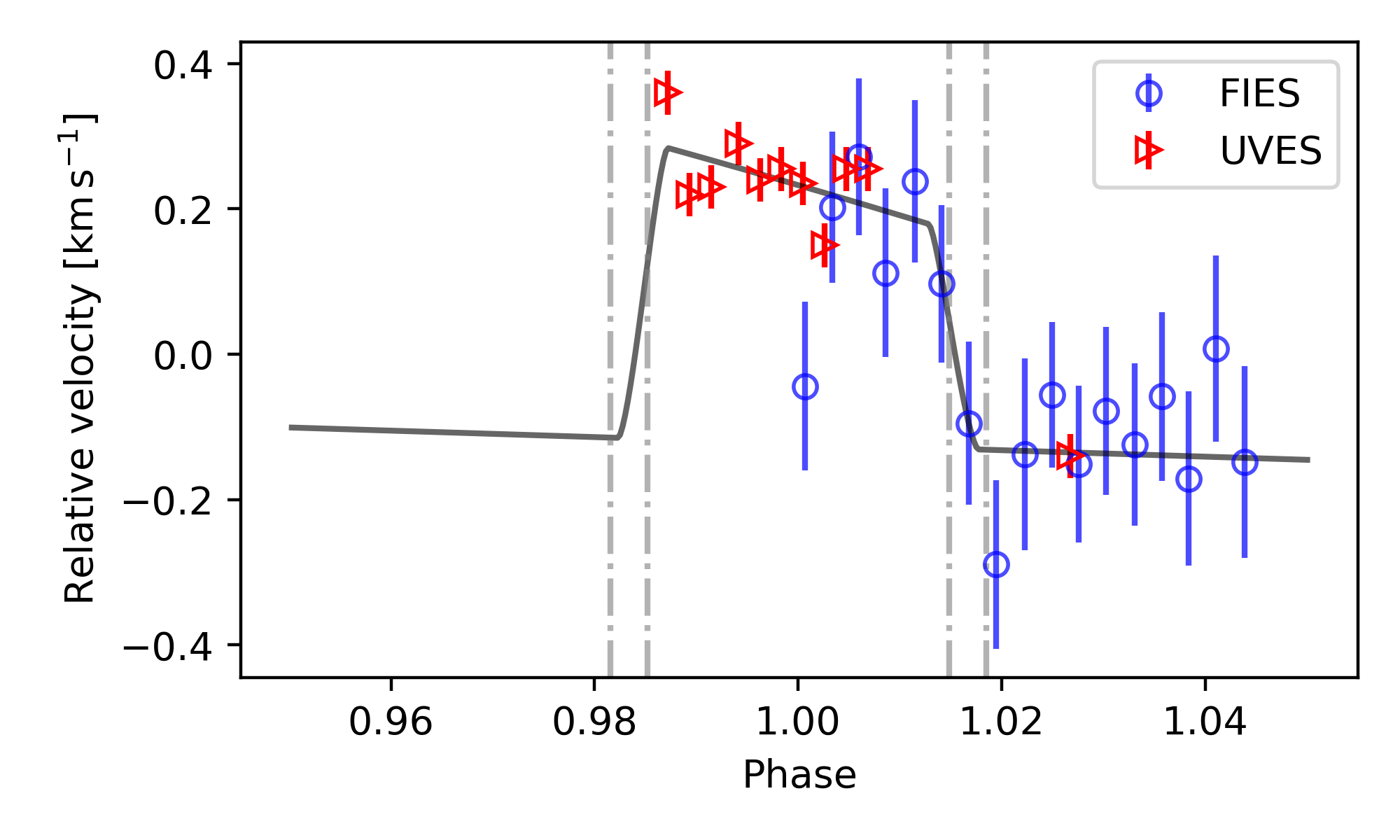}
   
   \caption{RV measurements obtained for CoRoT--36
   during the transit with UVES (red points)
   and FIES (blue triangles). The best fitting RM model is shown as black line. The dashed lines indicate the position of the transit.}
         \label{307RV}
\end{figure}

The RV measurements of both data sets are listed in the Appendix, Table~\ref{tab:RV3}. The measurements obtained with FIES are relative velocities in respect to the template used. We applied an RV offset of 22.79\,$\rm km\,s\rm ^{-1}$ to best match both data sets during the overlapping phase, This is possible since both data sets have been observed in- and out-of-transit. We find that both data sets agree very well and show an almost constant drift between the measurements taken out and in transit, due to the Rossiter-McLaughlin (RM) anomaly, a change in the stellar line profile due to the 'shadow' of the planet. We used the RV model of \texttt{ellc} \citep{maxted16} and the Markov chain Monte Carlo (MCMC) code \texttt{EMCEE} \citep{Foreman-Mackey13} to sample the model parameters, using the sampled parameters from the light curve fit as priors. The resulting RV data and best fitting RV model is shown in Fig.~\ref{307RV}. We concluded that the planet has to be in a nearly polar orbit with a projected angle between the orbital axis of the planet and the spin axis of the star $\rm \lambda_b = 275.8 \pm 11.3$.

We used high-resolution spectra obtained with UVES and HARPS to determine the atmospheric parameters $\rm T_{eff}$, $\rm log(g)_{sp}$, as well as the rotational velocity $\rm v\,sin(i_{\star})$. The stellar parameters, radius and mass, derived from evolutionary models confirm it to be an early F-type star with a mass of $\rm 1.49^{+0.20}_{-0.18}\,M_\odot$. We used this mass as prior for the light curve fit to optimise the surface gravity $\rm log(g)_{lc}$, which we used to optimise the stellar mass and radius using evolutionary models. All derived stellar parameters are listed in Table\,\ref{tab:sub_obs} and are, within the errors, consistent with the values reported in \cite{boufleur18}.

We used the fact that both stars No. 1 and 2 (see Sect.~\ref{Sec:onoff}) are well resolved in the UVES acquisition images, which allowed us to monitor their brightness during and out-of-transit. None of these stars show a notable drop in brightness, which rules out both to cause a potential FP.

Thus, all the evidence speaks for a close-in giant planet in 
polar orbit. All parameters derived for this object are given 
in Table~\ref{tab:sub_obs}.

\section{Discarded candidates}\label{sec_candidates}

In the previous sections, we have presented three cases for 
transiting objects, discovered in this survey. Two are giant planets, and one is an object that is at the border between a brown dwarf and a low-mass M-star. 
In this section, we will discuss CoRoT\,659668516, and show that the transit does not originate from the star itself. We, furthermore, summarise the outcome for the remaining candidates, showing that these are not planets orbiting IMSs.

Seven of our candidates have been identified as eclipsing binaries and are listed in Table~\ref{tab.061}. CoRoT\,632279463 and CoRoT\,631423419 have been identified as binaries by analysing the CoRoT light curves and have not been followed up with high-resolution spectroscopy (see Section \ref{sec_cand}). The candidate CoRoT\,110660135, which has been identified as binary, from intensive high-resolution spectroscopic follow-up, will be discussed in more detail in the next section. The four candidates CoRoT\,102806520, CoRoT\,102627709, CoRoT\,110853363, and CoRoT\,659719532 have been identified due to large drifts of several km\,s$\rm ^{-1}$ in consecutive spectra. Due to the sparse coverage of their orbits, we did not derive their orbital solutions. Nevertheless, we list the RV measurements of all candidates in the supplementary material.

\subsection{CoRoT 659668516}
\begin{figure}
   \centering
   \includegraphics[width=\hsize]{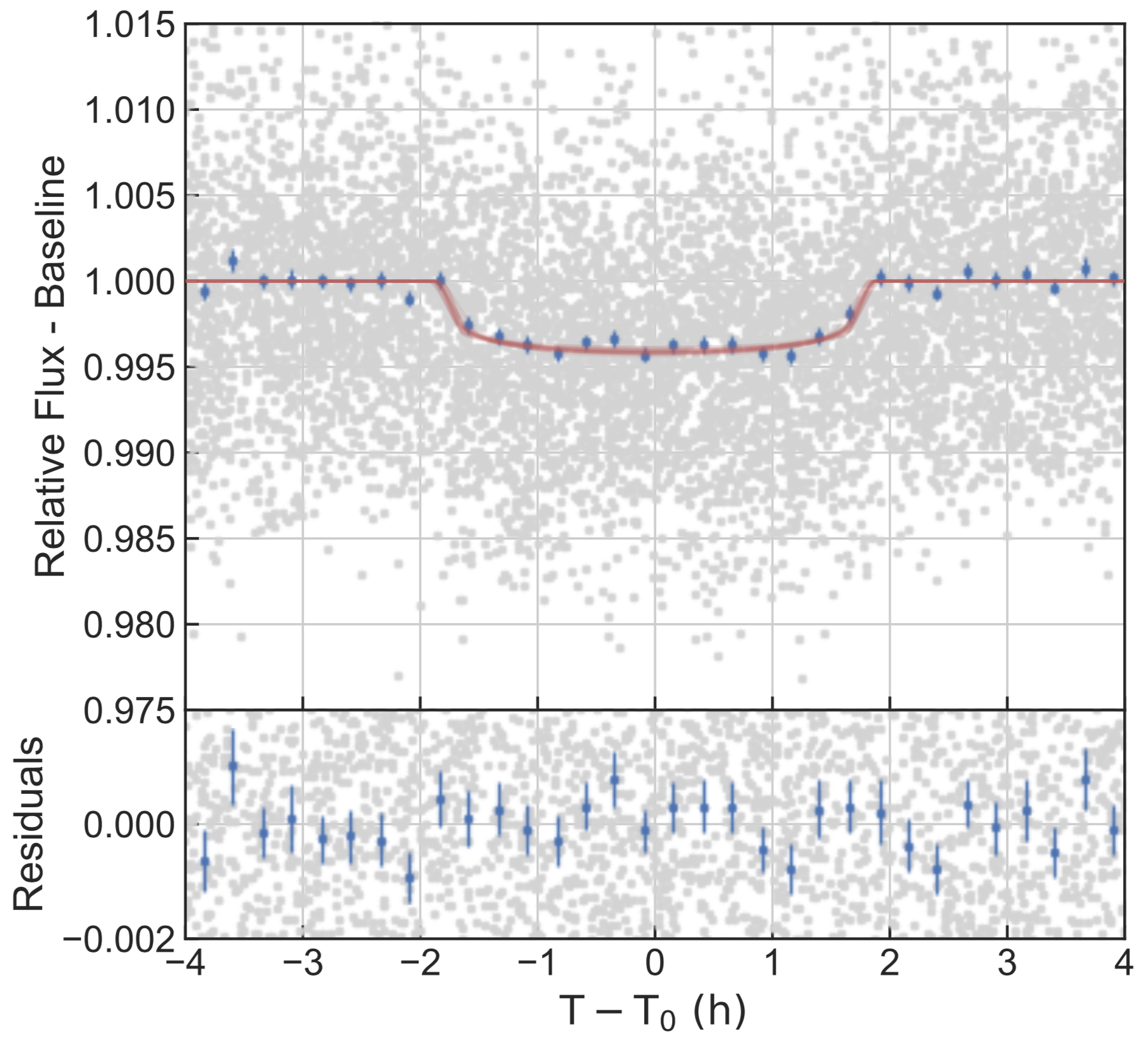}
   
   \caption{Phase folded, white light curve of CoRoT 659668516. The red line is the best fitting model. The lower panel shows the residuals from the transit fit.}
         \label{Fig_lc_4203}
\end{figure}

CoRoT 659668516 is a F3V star that has been observed by CoRoT between July $\rm 7^{th}$ 2011 and September $\rm 30^{th}$ 2011 in monochromatic mode. The light curve shows transits with a period of $\rm 3.287022_{-1.3e-05}^{+1.5e-05}$d. Two contaminating sources (Gaia DR3 4284693601105758592, G=19.1mag and Gaia DR3 4284693601087059328, G=17.7mag), hereafter S1 and S2, have been identified within 3\,arcsec of the star. Since all of them are in the mask of CoRoT, we derived a dilution factor as the sum of the contributing flux, relative to the main source (G=15.2mag) $\rm F_{blend} / F_{source} = 0.13$. Fig.~\ref{Fig_lc_4203} shows the phase-folded light curve. Despite the ephemeris are well constraint, the errors are too large to allow additional photometric follow-up from the ground, additionally this target has not been observed by TESS. Six intermediate resolution spectra were taken with the TWIN spectrograph mounted at 3.5m telescope at Calar Alto observatory which showed a possible BD solution for the companion.
In order to find out what the nature of these eclipses are, we obtained four spectra with UVES and analysed two HARPS spectra, taken 2.1 years before the UVES observations. We determined the stellar parameters with $\rm T_{eff}=7080\pm200\,K$, $\rm log(g)=3.37\pm0.30$, $\rm [Fe/H]=0.0\pm0.2$, and a projected rotational velocity of $\rm v\,sin(i_{\star})=64\pm2\,$km\,s$\rm ^{-1}$. For these parameters, we obtained its mass and radius with $\rm M_{\star}=2.27\pm 0.27\,M_{\odot}$ and $\rm R_{\star} = 5.06\pm1.3\,R_{\odot}$. 
The UVES spectra also show that the spectrum is composite with a slowly rotating component. Due to the blended spectrum, deriving accurate stellar parameters for this component is challenging. We found $\rm T_{eff}=6340\pm300\,K$, $\rm v\,sin(i_{\star})=5\pm3\,$km\,s$\rm ^{-1}$, and obtained $\rm M_{\star} = 1.24\pm0.1\,M_{\odot}$ and $\rm R_{\star} = 1.45\pm0.24\,R_{\odot}$ if a surface gravity of 4.2\,dex is assumed for a star close to the zero-age main-sequence. The blend is unlikely to originate from the visually resolved stars, since S1 is too faint to generate the blend and S2 has been excluded during the UVES and HARPS observations. From these stellar parameters, it is very likely that the transit does not originate from the brighter, fast rotating component since the measured transit duration of $\rm 3.75\pm0.07$\,h is 2 to 3 times shorter than expected for any companion orbiting a star with $\rm 5.06\,R_{\odot}$. Since a large impact parameter is not supported by the transit shape, this scenario would require an eccentric orbit.
Since our least squares fit method cannot be used to measure the relative velocity of a composite spectrum, we measured the slowly rotating component using the cross-correlation with a numerical G2 mask and the fast rotating component by measuring the LSD profiles. The RV measurements are listed in the Appendix, Table~\ref{tab:rv_4203}. Both components do not show any relative or absolute change in RV over the 2.1\,yrs, which excludes a spectroscopic binary with the 3.29\,d period.
In sum, this transit signal originates either from an object orbiting the slow rotating IMS, or from a BEB residing within the photometric mask of CoRoT.

\subsection{Other discarded candidates}

CoRoT 310204242 has been observed in chromatic mode by CoRoT. We identified it to show a planet-like transit with an orbital period of 5.15d. Despite being an evolved star, it drew our attention as the light curve in different colors did not show a clear depth difference and TWIN spectra showed RV variations, comparable with those expected for a BD companion. The two HARPS spectra of this star do not have sufficient SNR to measure the RV of this rapidly rotating star. We obtained and analysed four UVES spectra. The star is an A-giant with $\rm R_{\star} = 3.43\pm0.6\,R_{\sun}$, $\rm T_{eff}=7670\pm160\,K$, $\rm log(g)=3.25\pm0.21$, and a rotational velocity of $\rm v\,sin(i_{\star})=126\pm1\,$km\,s$\rm ^{-1}$. Despite that no close contaminant has been found, using seeing limited observations, the UVES spectra are clearly composite with a slow rotating $\rm v\,sin(i_{\star})=3.6\,$km\,s$\rm ^{-1}$ component. The RV measurements do not show a large variation, but match to an orbital solution with $\rm K \approx 1.5$\,km\,s$\rm ^{-1}$ and an orbital period of 10.3d which is twice the period, found from the light curve. 
We re-analysed the CoRoT light curve using the new orbital period and found that the target has to be a bright object with about $\rm 5\,R_{Jup}$ with a secondary eclipse that is about half as deep as the primary eclipse. Albeit, the blue and red CoRoT light curves do not show any change in eclipse depth, the eclipse is twice as deep in the green light curve. This color effect can only be noted by phase folding the light curve with the right orbital period. The data fit best to a BEB, which contaminates part of the CoRoT mask and is bright enough to also contaminate the RV measurements.

CoRoT 110660135 is a B-star that shows pulsations of 1.7d with 12\,mmag amplitude and transit-like signals with a period of $\rm 8.17041\pm0.00008\,d$. The prospect of a possible discovery of a planet orbiting a B-star initiated our detailed RV follow-up on this target, which finally confirmed that this object is a triple system with a close binary component.
We covered the orbital phase spectroscopically with spectra from SANDIFORD, CAFE, FIES, and HARPS. Using HARPS spectra, we derived the stellar parameters and found it to be a very young B-type star with $\rm R_{\star} = 2.26\pm0.8\,R_{\sun}$, $\rm T_{eff}=15900\pm400\,$K, and a high surface gravity ($\rm log(g)=4.44\pm0.11$). The latter places the star
close to the zero age main-sequence (ZAMS) in the HRD and, thus, it is probably a very young star
with an age of less than 4\,Myr. Due to its grazing eclipse configuration, the size of the companion is not well constrained by the best fit to the CoRoT light curve. Nevertheless, using the stellar parameters, the radius of the companion has to be $\rm 2.6\pm1.0\,R_{Jup}$, which indicates a possible late-type stellar companion. A detailed inspection of the LSD line profiles shows, that, CoRoT 110660135 is indeed a SB2. The stellar rotation with $\rm v\,sin(i_{\star})=33.1\,$km\,s$\rm ^{-1}$ makes it difficult to measure the reflex motion of the orbit. To combine the radial-velocity measurements from all observations, we had to correct for a linear trend of
$\delta v_{rad}=-7.3\,$km\,s$\rm ^{-1}$\,yr$\rm ^{-1}$, which implies the presence of another unresolved stellar companion, orbiting CoRoT 110660135 in a long period orbit.

Other candidates have been discarded by more detailed analysis of their light curves. The transit signal of CoRoT 102850921 has been mimicked by stellar pulsations and the transit of CoRoT 102605773 could not be confirmed after the star has been re-observed with CoRoT in LRa06. 
Additionally to these, there are three candidates that turned out to be background eclipsing binaries (BEB). (i)
CoRoT 102584409, is an SB2, but with a much longer period, as found with CoRoT. (ii) For CoRoT 659721996, we do not see any RV variation, and derive an upper limit of 3.8km\,s$\rm ^{-1}$ ($\rm M_{Pl}<30\,M_{Jup}$), corresponding to a low-mass BD. Nevertheless, the light curve shows a periastron brightening which has to be caused by a very luminous companion or most likely a BEB. (iii) For CoRoT 110858446, available spectra point at a blend with another star that turned out to be a BEB.

\section{The frequency of IMSs with close-in Giant Planets}\label{sec_occurence} 

With the completion of the CoRoT survey we can now determine the frequency of IMSs to harbour close-in gas giant planets (GPs) and compare it to that of solar-mass stars. As close-in GPs we count transiting companions with $\rm a \lesssim$ 0.1\,AU and $\rm R_{Pl} > 0.8\,R_{Jup}$ in the planetary regime which is consistent to \cite{deleuil18b}.
Thirtysix planets and BDs (34 of them transiting) in 33 planetary systems\footnote{The Extrasolar Planets Encyclopedia \href{http://exoplanet.eu/}{ \citep{schneider11}} (Aug. 2021)} have been found with CoRoT after observing
163,665 stars of which 101,083 are dwarfs (luminosity class IV and V, \citealt{deleuil18}). The stellar population of this sample has been studied in detail (see Sect.~\ref{sec_stellar}). \cite{deleuil18b} derived a frequency for this sample of $\rm 0.98\pm0.26\%$ to host GPs with orbital periods of less than 10 days.

Four of the published systems meet our mass-criterion for IMSs ($\rm M_{\star}=1.3-3.2 M_{\rm \odot}$). These are the two brown dwarfs CoRoT--3b
($\rm M_{\star}=1.37\pm0.09\,M_{\odot}$; \citealt{deleuil08}), and CoRoT--15b
($\rm M_{\star}=1.32\pm0.12\,M_{\odot}$; \citealt{bouchy11}), as well as the GPs CoRoT--11b  ($\rm M_{\star}=1.56\pm0.10\,M_{\odot}$; \citealt{gandolfi10,tsantaki14}) and CoRoT--21b ($\rm M_{\star}=1.29\pm0.09\,M_{\odot}$; \citet{paetzold12}). The latter is within its errors just at the border of being an IMS. 
In this article we presented the cases for CoRoT--34b and CoRoT--36b, and gave evidence that first is a brown dwarf, orbiting an A-star of $\rm 1.66\,M_{\odot}$ and the latter is GP orbiting an F-star of $\rm 1.32\,M_{\odot}$.
Despite that we can show CoRoT--35b being a GP orbiting an F-star, the mass of its host star ($\rm M_{\star} = 1.0\,M_{\odot}$) falls well below our mass-criterion for IMSs.
We also discussed the remaining candidate for a sub-stellar companion of CoRoT 659668516 and found it likely to orbit an $\rm M = 1.24\pm0.1\,M_{\odot}$ star residing within the photometric mask, which is at the border of being an IMS.

That means that with the CoRoT survey three GPs of IMSs with spectral type F have been found, but none with a host mass larger than 1.6$\,M_{\odot}$. On the other hand, CoRoT detected three brown dwarfs orbiting IMS with one of them orbiting an A-type star. Given these discoveries, we can derive the frequency for IMSs to host close-in GPs and, thus, further extend the results from \cite{deleuil18} to early F-type stars with 1.6\,$\rm M_{\odot}$.

Applying our before-mentioned limits for close-in giant planets, we excluded in this statistic (i) close-in brown dwarfs (CoRoT--3b, CoRoT--15b, CoRoT--33b, and CoRoT--34b), (ii) planets with sizes from the terrestrial to Neptune regime (CoRoT--7b, CoRoT--22b, CoRoT--24b, and CoRoT--32b), (iii) planets with orbital periods longer than 10 days (CoRoT--9b, CoRoT--10b, CoRoT--24c), and (iv) the non transiting CoRoT--7c and CoRoT--20c without direct radius measurement. The resulting sample contains 26 transiting GPs with host star masses between $\rm 0.88\pm0.04\,M_{\odot}$ $\rm 1.56\pm0.1\,M_{\odot}$, taking into account the updated host star masses for CoRoT--3 and CoRoT--11 \citep{tsantaki14}. This selection leads to a sample of 26 GPs detected with CoRoT.

This result enables us to directly compare the number of host stars (with at least one GP detected) to the number of stars analysed by CoRoT, in order to derive the planet frequency. To calculate how many IMSs have been analysed, we cross-matched the stellar population for main-sequence stars (see Sect.~\ref{sec_stellar}) with the total number of stars with CoRoT light curves (without duplication) obtained via EXODAT\footnote{\url{http://cesam.oamp.fr/exodat/}} \citep{deleuil09}. 
Here, we denoted main-sequence stars as classified with IV and V from the Galactic anticentre fields \citep{guenther12} as well as the Galactic centre fields \citep{damiani16} with a magnitude limit of $r' <$ 15.4 (Corresponding to the faintest planetary host stars found by the CoRoT survey). As in Sect.~\ref{sec_stellar}, we select the mass interval of 0.75\,$\rm M_{\odot}$ to 3.2\,$\rm M_{\odot}$ to include the lowest mass GP host star, as well as the whole sample of IMS observed with CoRoT, leading to 70,525 stars with 33,846 IMSs.
We divided the sample of 26 detected GP host stars according to their mass, in the same mass interval we have used for the stellar sample in Sect.~\ref{sec_stellar}. To derive the GP frequency from the number of stars in each mass bin, we assume an average error of 0.1\,$\rm M_{\odot}$ and derived the uncertainty of the GP frequency from the variance of the different resulting solutions. 

Assuming arbitrary inclinations, we can correct for the geometric transit probability $\rm p=R_{\star}/a$, because we can only detect transiting objects. We derive this correction for each of the 26 GPs individually, reaching a mean transit probability of 12.1\,\% This means that the true number of close-in GP host stars in the CoRoT sample is more than 250.

By combining the corrected GPs, as well as the stellar sample from the centre and anticentre fields, we derive a GP frequency of $\rm 0.70 \pm 0.16\,\%$ for FGK stars (0.75\,$\rm M_{\odot}$ to 1.26\,$\rm M_{\odot}$), assuming a detection efficiency of 90\%. This result is only slightly different when including the candidate CoRoT 659668516 into the sample ($\rm 0.71 \pm 0.15\,\%$). 
This is consistent, but slightly smaller compared to \cite{deleuil18b} ($\rm 0.98 \pm 0.23\,\%$), who used a slightly brighter magnitude limit, as well as a smaller sample of stars for the comparison.

\begin{figure}
   \centering
   \includegraphics[width=\columnwidth]{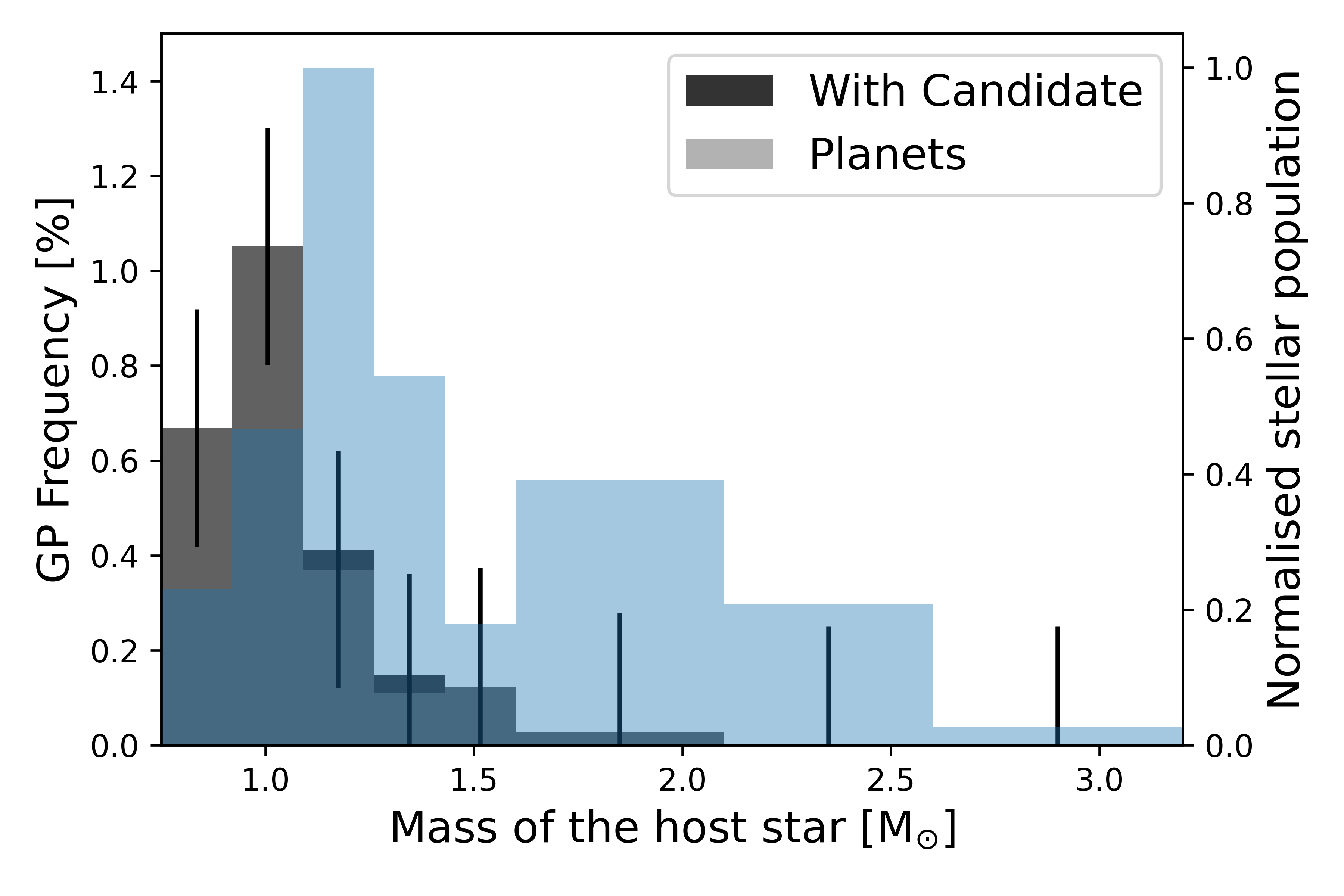}

   \caption{GP Frequency of the CoRoT sample for FGK and IMS stars. Gray, frequency based on confirmed planets, dark gray, including the candidate CoRoT 659668516. Light blue: the stellar sample, observed with CoRoT.}
    \label{Fig:GP_stat}
\end{figure}

We find a GP frequency of $\rm0.12 \pm 0.1\,\%$ for IMSs between 1.26\,$\rm M_{\odot}$ and 1.6\,$M_{\odot}$\footnote{For the stellar sample, we use the same binning as in Sect.~\ref{sec_stellar} with 1.26\,$\rm M_{\odot}$ being the lower border of the bin that includes the 1.3\,$\rm M_{\odot}$ limit for IMSs.} ($\rm 0.10 \pm 0.1\,\%$, when adding CoRoT 659668516 into the sample.)
The GP frequency vs. stellar mass is shown in Fig.~\ref{Fig:GP_stat}. For comparison we added the normalised stellar sample and like to point out that the majority of planet host stars detected with CoRoT are of solar mass, while the majority of stars, surveyed with CoRoT are early F-type stars with masses between 1.1 and 1.2\,$M_{\odot}$. The absence of many F-type planets in the sample, thus, leads to a steep decline of the planet frequency for stars more massive than 1.1\,$M_{\odot}$.

About 5,900 IMS with masses $\rm >1.6\,M_{\odot}$ have been observed with CoRoT. The absence of planet detections in this sample does not allow to derive a GP frequency directly. Nevertheless, we can derive an upper limit by deriving the GP frequency, in the case at least one GP host star would have been detected. We assumed an detection probability of 12\% for a 10\,d orbit to derive the upper limit of $\rm 0.25 \pm 0.16\,\%$ for the mass range between 1.6\,$\rm M_{\odot}$ and 3.2\,$\rm M_{\odot}$. 

\section{Summary and Conclusions}\label{sec_concls}

The CoRoT survey is one of the few photometric surveys that allow us to determine the frequency of close-in planets of IMSs on the main-sequence. As IMSs, we denote main-sequence stars with $\rm M_{\star}=1.3-3.2 M_{\rm \odot}$. Roughly one third of the main-sequence stars that have been observed with CoRoT are stars in this mass range. The sensitivity of the mission is large enough to detect all objects with $\rm R_{planet} > 0.8\,R_{Jup}$ orbiting at $\rm a \lesssim$ 0.1\,AU with a completeness rate of better than 90\%. That means CoRoT has the sensitivity to detect basically all transiting, close-in gas-giant (GP) planets, brown dwarfs and stellar companions. 
In our survey, we analysed all IMSs, observed by CoRoT and studied 17 promising candidates in detail. Only three of them could be confirmed as close-in sub-stellar companions:

CoRoT--34b, is a brown dwarf with $\rm M_{BD}= 71.4_{-8.6}^{+8.9}\,M_{Jup}$ and, thus, just below the hydrogen burning limit which marks the border towards low-mass stars. Only $\approx 30$ transiting BDs, that allow to directly measure their mass and radius, are known to date. Since CoRoT--34b is fully transiting, the transit parameters are well defined, which make it to an important test bench for testing models of high mass BDs. Its mass and radius lead to a surface gravity $\rm log(g)_{BD} = 5.14 \pm 0.10$, which is slightly lower than predicted from models at an stellar age of about 2.1\,Gyr \citep[e.g.][]{baraffe03,saumon08,phillips20}. This can possibly be explained by the inflated radius of $\rm 21\pm13\,\%$ compared to the \texttt{COND03} model \citep{baraffe03}.
This apparent discrepancy would be resolved if the BD were inflated by the irradiation from its host star. \cite{Acton21} showed in their Fig.~7, the relationship between equilibrium temperature and radius for known transiting BDs. Notably all known BDs with $\rm T_{equ} > 2000\,$K have radii larger than predicted by evolutionary models of BDs at the age of their host stars. These are HATSs-70b \citep{zhou19a}, KELT-1b \citep{siverd12}, and TOI-503b \citep{subjak2020}, which span, together with CoRoT--34b, through the whole mass-range of BDs from deuterium burning to the hydrogen burning limit. The fact that it turned out that CoRoT--34b is an inflated 2.1\,Gyr old BD at the hydrogen burning limit, may allow us to study possible heating mechanisms by stellar irradiation and their dependence from the mass of the BD. 
Since high-mass BDs thought to be formed in situ, it might also be possible that further, non-transiting sub-stellar companions with orbital periods <100\,d exist around CoRoT--34 \citep{batygin16}. 
We find that CoRoT--34 pulsates with 0.71\,d, which is only a bit shorter, but very close to one third of the orbital period. This strong synchronisation of the stellar pulsations with the orbit of the brown dwarf might be explained by tidally excited oscillation modes caused by the close-in Brown dwarf as a result of tidal circularisation \citep{Kumar95}.   

CoRoT--35b and CoRoT--36b are both giant planets which show anomalously large radii, only consistent with the largest known hot Jupiters. Those anomalous radii are probably best explained by heating processes related to the strong irradiation from their host stars, which lead to equilibrium temperatures larger than 1700K and 1500K, respectively \citep{sarkis21}. 
The discovery of CoRoT--35b plays a particularly interesting role, as it orbits a metal poor star. The CoRoT centre fields are located at low Galactic latitudes, which are dominated by stars of solar metallicity in the galactic thin disc, which makes CoRoT--35 a rather exceptional object in the CoRoT sample \citep{gazzano13}. Only very few metal-poor stars ($\rm [Fe/H] < -0.3 dex$) have been confirmed to host hot Jupiters and up to date less than ten of them have GPs with inflated radii larger than $\rm 1.4\,R_{Jup}$\footnote{\href{https://exoplanetarchive.ipac.caltech.edu/}{NASA exoplanet archive}, Aug. 2021}. Examples of such objects are WTS-1b \citep{cappetta12} or KELT-21b \citep{johnson18}. 
We found transit timing variations for CoRoT--35b, hinting at another massive planet in the system. Only 26 transits have been observed with CoRoT, thus, it will be difficult to model the object causing these variations \citep[e.g.][]{Veras09}. In future research, this can be done by adding more data points, for instance, using ground-  or space based photometry, which will allow to measure the mass of this object. Along the same line, many published CoRoT planets have recently been re-observed by TESS, allowing to optimise the transit periods and ephemeris \citep{Klagyivik21}. 
The possibility to find a planetary system together with the rarity of such inflated GPs orbiting stars with sub-solar metallicity, makes the detection of CoRoT--35b a rare benchmark object for the planet formation of such stars \citep[e.g.][]{Adibekyan19}.

With time-resolved transit spectroscopy, we confirm that CoRoT--36b is orbiting in a polar orbit with a projected obliquity of $\rm275.8 \pm 11.3\,^{\circ}$.
This discovery adds CoRoT--36b to the growing list of misaligned giant planets, orbiting IMSs. It has been confirmed that more massive stars tend to have commonly misaligned planets in nearly circular orbits \citep[e.g.][]{winn10}. This misalignment is generally thought to be triggered by the Kozai mechanism due to an perturbing object, like an outer planet \citep[e.g.][]{nagasawa08, winn10, triaud10}. Other mechanisms might play a role to explain the observed distribution of planet obliquities (see \cite{albrecht22} for a review).

One of our candidates, CoRoT 659668516, still might be a sub-stellar object of an ISM, but is very diluted by other sources, and thus, we cannot conclude on its nature. All other candidates were identified as false signals or binary stars, and therefore, we can extend the frequency of close-in Giant planets of IMSs, and compare it to the frequency of solar-type stars.

We determine the frequency of solar-type stars in the CoRoT sample to host at least one close-in giant planet  to $\rm 0.70 \pm 0.16\,\%$. This result is fully consistent to the findings from other surveys for close-in planets from solar-type stars like \cite{naef05} ($0.7\pm 0.5\,\%$), \cite{cumming08} ($0.64\pm 0.4\,\%$), \cite{mayor11} ($0.89 \pm 0.36\,\%$), and \cite{zhou19b} ($0.71 \pm 0.31\,\%$).

For stars with $\rm M_{\star}=1.26 - 1.6\,M_\odot$, we derive an average GP frequency of $\rm 0.12 \pm 0.10\,\%$ and find a steep drop in this mass range. The upper mass limit of $\rm 1.6\,M_\odot$ is selected to include the highest-mass planet host stars, identified with CoRoT.

Due to the large number of early F-type stars, surveyed during the CoRoT mission, and its high detection efficiency this is a significant result, which cannot be explained by selection effects within the mission design.
We can, thus, exclude any increase of the close-in GP-frequency in this mass range within the CoRoT survey. For stars more massive than IMSs, we derive an upper limit of 0.25\%. 

This GP frequency for IMS is well in agreement with the upper limits derived from the Kepler space mission (<0.75\%, \citealt{sabotta19}) but slightly lower than the preliminary results obtained from the TESS space mission for F-type stars (0.43 $\pm$ 0.15\%, \citealt{zhou19b}). This difference can be explained by the wider mass-range of our survey which includes a steep decline of the GP frequency for stars more massive than $\rm 1.4\,M_\odot$ which was used as upper border in their study. Our upper limit for the GP frequency of stars with masses between $\rm 1.6 and 3.2\,M_\odot$ is fully  consistent with the frequency for more massive IMS, found in TESS.

This is in contrast to the GP frequency at larger separations, from RV surveys of giant stars, that show an increase with stellar mass \citep[e.g.][]{johnson10,johnson10b,reffert15}.
Assuming that this higher planet frequency for IMSs is not overestimated by stellar activity, we can conclude that the efficient migration scenario, which would lead to a higher frequency for close-in planets, that are then engulfed during stellar evolution \citep[e.g.][]{hasegawa13}, cannot play a large role to explain the absence of short period GPs in RV surveys. Nevertheless, it cannot fully be ruled out since such close-in GPs do exist \citep[e.g.][]{zhou19b}. 

\section*{Acknowledgements}

The CoRoT space mission, launched on December $\rm 27^{th}$ 2006, has been 
developed and is operated by CNES, with the contribution of Austria, 
Belgium, Brazil, ESA (RSSD and Science Programme), Germany and Spain.
This work was support by the Deutsche Forschungsgemeinschaft (DFG) 
through grants GU 464/15- 1,GU 464/16-1, GU 464/17-1, GU 464/18-1 and 
Ts~17/2-1. EG and DS also acknowledge the generous support by the Th\"uringer Ministerium f\"ur Wirtschaft, Wissenschaft und Digitale Gesellschaft. CH, UH, AI, and MD acknowledge funding by the Deutsche Forschungsgemeinschaft (DFG) through grants HE1356/62-1, HE1356/71-1, and IR190/1-1. This research is also supported work funded from the European Research Council (ERC) the European Union’s Horizon 2020 research and innovation programme (grant agreement n$^{\circ}$ 803193/BEBOP). HJD acknowledges support from the Spanish Research Agency of the Ministry of Science and Innovation (AEI-MICINN) under grant PID2019-107061GB-C66, DOI: 10.13039/501100011033. JK gratefully acknowledges the support of the Swedish National Space Agency (DNR 2020-00104).
This research has made use of the SIMBAD database, operated at CDS, Strasbourg, France.
We made also use of data from the European Space Agency (ESA)
mission {\it Gaia} (\url{https://www.cosmos.esa.int/gaia}), processed
by the {\it Gaia} Data Processing and Analysis Consortium
(DPAC,\url{https://www.cosmos.esa.int/web/gaia/dpac/consortium}).
The ground- based observations were obtained with the VLT at the European 
Southern Observatory at Paranal, Chile (291.C-5028(A), 092.C- 0222(A), 
and 093.C-0027(A), with the 2.2-m and 3.5-m telescope at German-Spanish 
Astronomical Center at Calar Alto (H12-2.2-037, H14-2.2-043), with the 
Nordic Optical Telescope at La Palma (NOT runs P45-109, P45-206, P46-113, 
P46-215, P47-105, P47-211, P48-217, P49-205; funding from the European 
Union Seventh Framework Programme FP7/2013-2016 under grant agreement No. 312430, OPTICON), 
the 2.1-m at McDonald observatory, Texas, the 80-cm telescope of the 
Instituto de Astrofisica de Canarias at the observatorio del Teide, the Tel-Aviv University 1-m telescope at WISE Observatory, 
the Alfred-Jensch telescope in Tautenburg, and the Large Binocular Telescope. 
The NOT is operated on the island of La Palma jointly by Denmark, Finland, 
Iceland, Norway, and Sweden, in the Spanish Observatorio del Roque de los Muchachos 
of the Instituto de Astrofisica de Canarias. 
The 2.2-m and 3.5-m telescopes of the Calar Alto observatory is located in the 
Sierra de Los Filabres (Andalucıa, Southern Spain) north of Almeria. It is operated 
jointly by the Max-Planck-Institut for Astronomie (MPIA) in Heidelberg, Germany, 
and the Instituto de Astrofisica de Andalucıa (CSIC) in 
Granada/Spain. The LBT is an international collaboration. LBT Corporation partners are: 
The University of Arizona on behalf of the Arizona university system; Istituto Nazionale 
di Astrofisica, Italy; LBT Beteiligungsgesellschaft, Germany, representing the Max-Planck 
Society, the Astrophysical Institute Potsdam, 
and Heidelberg University; The Ohio State University, and The Research Corporation, 
on behalf of the University of Notre Dame, University of Minnesota and University of Virginia. 
This publication makes use of data products from the Two Micron All Sky Survey, which is a 
joint project of the University of Massachusetts and the Infrared Processing and Analysis 
Center/California Institute of Technology, funded by the National Aeronautics and Space 
Administration and the National Science Foundation.
The authors thank the referee for the
constructive remarks.

\section*{Data Availability}
Most data underlying this article is available online as indicated in the specific section or reference. Data, obtained with ESO telescopes are available in the ESO Science Archive Facility, at \url{http://archive.eso.org/cms.html}. Other data underlying this article will be shared on reasonable request to the corresponding author.




\bibliographystyle{mnras}
\bibliography{bibliothek} 

\begin{thebibliography}{}
\makeatletter
\relax
\def\mn@urlcharsother{\let\do\@makeother \do\$\do\&\do\#\do\^\do\_\do\%\do\~}
\def\mn@doi{\begingroup\mn@urlcharsother \@ifnextchar [ {\mn@doi@}
  {\mn@doi@[]}}
\def\mn@doi@[#1]#2{\def\@tempa{#1}\ifx\@tempa\@empty \href
  {http://dx.doi.org/#2} {doi:#2}\else \href {http://dx.doi.org/#2} {#1}\fi
  \endgroup}
\def\mn@eprint#1#2{\mn@eprint@#1:#2::\@nil}
\def\mn@eprint@arXiv#1{\href {http://arxiv.org/abs/#1} {{\tt arXiv:#1}}}
\def\mn@eprint@dblp#1{\href {http://dblp.uni-trier.de/rec/bibtex/#1.xml}
  {dblp:#1}}
\def\mn@eprint@#1:#2:#3:#4\@nil{\def\@tempa {#1}\def\@tempb {#2}\def\@tempc
  {#3}\ifx \@tempc \@empty \let \@tempc \@tempb \let \@tempb \@tempa \fi \ifx
  \@tempb \@empty \def\@tempb {arXiv}\fi \@ifundefined
  {mn@eprint@\@tempb}{\@tempb:\@tempc}{\expandafter \expandafter \csname
  mn@eprint@\@tempb\endcsname \expandafter{\@tempc}}}

\bibitem[\protect\citeauthoryear{{Aceituno} et~al.,}{{Aceituno}
  et~al.}{2013}]{aceituno13}
{Aceituno} J.,  et~al., 2013, \mn@doi [\aap] {10.1051/0004-6361/201220361},
  \href {http://adsabs.harvard.edu/abs/2013A%26A...552A..31A} {552, A31}

\bibitem[\protect\citeauthoryear{{Acton} et~al.,}{{Acton}
  et~al.}{2021}]{Acton21}
{Acton} J.~S.,  et~al., 2021, \mn@doi [\mnras] {10.1093/mnras/stab1459}, \href
  {https://ui.adsabs.harvard.edu/abs/2021MNRAS.505.2741A} {505, 2741}

\bibitem[\protect\citeauthoryear{{Adibekyan}}{{Adibekyan}}{2019}]{Adibekyan19}
{Adibekyan} V.,  2019, \mn@doi [Geosciences] {10.3390/geosciences9030105},
  \href {https://ui.adsabs.harvard.edu/abs/2019Geosc...9..105A} {9, 105}

\bibitem[\protect\citeauthoryear{{Albrecht}, {Dawson}  \& {Winn}}{{Albrecht}
  et~al.}{2022}]{albrecht22}
{Albrecht} S.~H.,  {Dawson} R.~I.,   {Winn} J.~N.,  2022, arXiv e-prints, \href
  {https://ui.adsabs.harvard.edu/abs/2022arXiv220305460A} {p. arXiv:2203.05460}

\bibitem[\protect\citeauthoryear{{Almenara} et~al.,}{{Almenara}
  et~al.}{2009}]{almenara09}
{Almenara} J.~M.,  et~al., 2009, \mn@doi [\aap] {10.1051/0004-6361/200911926},
  \href {https://ui.adsabs.harvard.edu/abs/2009A&A...506..337A} {506, 337}

\bibitem[\protect\citeauthoryear{{Ammler-von Eiff}, {Sebastian}, {Guenther},
  {Stecklum}  \& {Cabrera}}{{Ammler-von Eiff} et~al.}{2015}]{ammler15}
{Ammler-von Eiff} M.,  {Sebastian} D.,  {Guenther} E.~W.,  {Stecklum} B.,
  {Cabrera} J.,  2015, \mn@doi [Astronomische Nachrichten]
  {10.1002/asna.201412153}, \href
  {http://adsabs.harvard.edu/abs/2015AN....336..134A} {336, 134}

\bibitem[\protect\citeauthoryear{{Anderson} et~al.,}{{Anderson}
  et~al.}{2018}]{anderson18}
{Anderson} D.~R.,  et~al., 2018, arXiv e-prints, \href
  {https://ui.adsabs.harvard.edu/abs/2018arXiv180904897A} {p. arXiv:1809.04897}

\bibitem[\protect\citeauthoryear{{Baglin} et~al.,}{{Baglin}
  et~al.}{2006}]{baglin06}
{Baglin} A.,  et~al., 2006, in 36th COSPAR Scientific Assembly. p.~3749

\bibitem[\protect\citeauthoryear{{Balona}}{{Balona}}{2014}]{balona14}
{Balona} L.~A.,  2014, \mn@doi [\mnras] {10.1093/mnras/stu822}, \href
  {https://ui.adsabs.harvard.edu/abs/2014MNRAS.441.3543B} {441, 3543}

\bibitem[\protect\citeauthoryear{{Baraffe}, {Chabrier}, {Barman}, {Allard}  \&
  {Hauschildt}}{{Baraffe} et~al.}{2003}]{baraffe03}
{Baraffe} I.,  {Chabrier} G.,  {Barman} T.~S.,  {Allard} F.,   {Hauschildt}
  P.~H.,  2003, \mn@doi [\aap] {10.1051/0004-6361:20030252}, \href
  {https://ui.adsabs.harvard.edu/abs/2003A&A...402..701B} {402, 701}

\bibitem[\protect\citeauthoryear{{Baranne} et~al.,}{{Baranne}
  et~al.}{1996}]{baranne96}
{Baranne} A.,  et~al., 1996, \aaps, \href
  {http://adsabs.harvard.edu/abs/1996A%26AS..119..373B} {119, 373}

\bibitem[\protect\citeauthoryear{{Batygin}, {Bodenheimer}  \&
  {Laughlin}}{{Batygin} et~al.}{2016}]{batygin16}
{Batygin} K.,  {Bodenheimer} P.~H.,   {Laughlin} G.~P.,  2016, \mn@doi [\apj]
  {10.3847/0004-637X/829/2/114}, \href
  {https://ui.adsabs.harvard.edu/abs/2016ApJ...829..114B} {829, 114}

\bibitem[\protect\citeauthoryear{{Borgniet} et~al.,}{{Borgniet}
  et~al.}{2014}]{borgniet14}
{Borgniet} S.,  et~al., 2014, \mn@doi [\aap] {10.1051/0004-6361/201321783},
  \href {https://ui.adsabs.harvard.edu/abs/2014A&A...561A..65B} {561, A65}

\bibitem[\protect\citeauthoryear{{Borgniet} et~al.,}{{Borgniet}
  et~al.}{2019}]{borgniet19}
{Borgniet} S.,  et~al., 2019, \mn@doi [\aap] {10.1051/0004-6361/201833431},
  \href {https://ui.adsabs.harvard.edu/abs/2019A&A...621A..87B} {621, A87}

\bibitem[\protect\citeauthoryear{{Bouchy} et~al.,}{{Bouchy}
  et~al.}{2008}]{bouchy08}
{Bouchy} F.,  et~al., 2008, \mn@doi [\aap] {10.1051/0004-6361:200809433}, \href
  {https://ui.adsabs.harvard.edu/abs/2008A&A...482L..25B} {482, L25}

\bibitem[\protect\citeauthoryear{{Bouchy} et~al.,}{{Bouchy}
  et~al.}{2011}]{bouchy11}
{Bouchy} F.,  et~al., 2011, \mn@doi [\aap] {10.1051/0004-6361/201015276}, \href
  {http://adsabs.harvard.edu/abs/2011A%26A...525A..68B} {525, A68}

\bibitem[\protect\citeauthoryear{{Boufleur}, {Emilio}, {Janot-Pacheco},
  {Andrade}, {Ferraz-Mello}, {do Nascimento}  \& {de La Reza}}{{Boufleur}
  et~al.}{2018}]{boufleur18}
{Boufleur} R.~C.,  {Emilio} M.,  {Janot-Pacheco} E.,  {Andrade} L.,
  {Ferraz-Mello} S.,  {do Nascimento} Jr. J.-D.,   {de La Reza} R.,  2018,
  \mn@doi [\mnras] {10.1093/mnras/stx2187}, \href
  {http://adsabs.harvard.edu/abs/2018MNRAS.473..710B} {473, 710}

\bibitem[\protect\citeauthoryear{{Boulade} et~al.,}{{Boulade}
  et~al.}{2003}]{boulade03}
{Boulade} O.,  et~al., 2003, in {Iye} M.,  {Moorwood} A.~F.~M.,  eds,  Society
  of Photo-Optical Instrumentation Engineers (SPIE) Conference Series Vol.
  4841, Instrument Design and Performance for Optical/Infrared Ground-based
  Telescopes. pp 72--81, \mn@doi{10.1117/12.459890}

\bibitem[\protect\citeauthoryear{{Bourrier} et~al.,}{{Bourrier}
  et~al.}{2015}]{bourrier15}
{Bourrier} V.,  et~al., 2015, \mn@doi [\aap] {10.1051/0004-6361/201525750},
  \href {https://ui.adsabs.harvard.edu/abs/2015A&A...579A..55B} {579, A55}

\bibitem[\protect\citeauthoryear{{Cameron} et~al.,}{{Cameron}
  et~al.}{2010}]{cameron10}
{Cameron} A.~C.,  et~al., 2010, \mn@doi [\mnras]
  {10.1111/j.1365-2966.2010.16922.x}, \href
  {http://adsabs.harvard.edu/abs/2010MNRAS.407..507C} {407, 507}

\bibitem[\protect\citeauthoryear{{Cappetta} et~al.,}{{Cappetta}
  et~al.}{2012}]{cappetta12}
{Cappetta} M.,  et~al., 2012, \mn@doi [\mnras]
  {10.1111/j.1365-2966.2012.21937.x}, \href
  {https://ui.adsabs.harvard.edu/abs/2012MNRAS.427.1877C} {427, 1877}

\bibitem[\protect\citeauthoryear{{Chaintreuil}, {Deru}, {Baudin}, {Ferrigno},
  {Grolleau}, {Romagnan}  \& {CoRot Team}}{{Chaintreuil}
  et~al.}{2016}]{2016cole.book...61C}
{Chaintreuil} S.,  {Deru} A.,  {Baudin} F.,  {Ferrigno} A.,  {Grolleau} E.,
  {Romagnan} R.,   {CoRot Team} 2016, {II.4 The ``ready to use'' CoRoT data}.
p.~61, \mn@doi{10.1051/978-2-7598-1876-1.c024}

\bibitem[\protect\citeauthoryear{{Chambers} et~al.,}{{Chambers}
  et~al.}{2017}]{2017yCat.2349....0C}
{Chambers} K.~C.,  et~al., 2017, VizieR Online Data Catalog, \href
  {http://adsabs.harvard.edu/abs/2017yCat.2349....0C} {2349}

\bibitem[\protect\citeauthoryear{{Choi}, {Dotter}, {Conroy}, {Cantiello},
  {Paxton}  \& {Johnson}}{{Choi} et~al.}{2016}]{choi2016}
{Choi} J.,  {Dotter} A.,  {Conroy} C.,  {Cantiello} M.,  {Paxton} B.,
  {Johnson} B.~D.,  2016, \mn@doi [\apj] {10.3847/0004-637X/823/2/102}, \href
  {https://ui.adsabs.harvard.edu/abs/2016ApJ...823..102C} {823, 102}

\bibitem[\protect\citeauthoryear{{Cumming}, {Butler}, {Marcy}, {Vogt}, {Wright}
   \& {Fischer}}{{Cumming} et~al.}{2008}]{cumming08}
{Cumming} A.,  {Butler} R.~P.,  {Marcy} G.~W.,  {Vogt} S.~S.,  {Wright} J.~T.,
   {Fischer} D.~A.,  2008, \mn@doi [\pasp] {10.1086/588487}, \href
  {http://adsabs.harvard.edu/abs/2008PASP..120..531C} {120, 531}

\bibitem[\protect\citeauthoryear{{Currie}}{{Currie}}{2009}]{currie09}
{Currie} T.,  2009, \mn@doi [\apjl] {10.1088/0004-637X/694/2/L171}, \href
  {http://adsabs.harvard.edu/abs/2009ApJ...694L.171C} {694, L171}

\bibitem[\protect\citeauthoryear{{Damiani}, {Meunier}, {Moutou}, {Deleuil},
  {Ysard}, {Baudin}  \& {Deeg}}{{Damiani} et~al.}{2016}]{damiani16}
{Damiani} C.,  {Meunier} J.-C.,  {Moutou} C.,  {Deleuil} M.,  {Ysard} N.,
  {Baudin} F.,   {Deeg} H.,  2016, \mn@doi [\aap]
  {10.1051/0004-6361/201628627}, \href
  {http://adsabs.harvard.edu/abs/2016A%26A...595A..95D} {595, A95}

\bibitem[\protect\citeauthoryear{{Deeg} et~al.,}{{Deeg}
  et~al.}{2009}]{2009A&A...506..343D}
{Deeg} H.~J.,  et~al., 2009, \mn@doi [\aap] {10.1051/0004-6361/200912011},
  \href {https://ui.adsabs.harvard.edu/abs/2009A&A...506..343D} {506, 343}

\bibitem[\protect\citeauthoryear{{Deeg} et~al.,}{{Deeg}
  et~al.}{2020}]{2020JAVSO..48..201D}
{Deeg} H.~J.,  et~al., 2020, JAAVSO, \href
  {https://ui.adsabs.harvard.edu/abs/2020JAVSO..48..201D} {48, 201}

\bibitem[\protect\citeauthoryear{{Dekker}, {D'Odorico}, {Kaufer}, {Delabre}  \&
  {Kotzlowski}}{{Dekker} et~al.}{2000}]{dekker00}
{Dekker} H.,  {D'Odorico} S.,  {Kaufer} A.,  {Delabre} B.,   {Kotzlowski} H.,
  2000, in {Iye} M.,  {Moorwood} A.~F.,  eds,  Society of Photo-Optical
  Instrumentation Engineers (SPIE) Conference Series Vol. 4008, Optical and IR
  Telescope Instrumentation and Detectors. pp 534--545

\bibitem[\protect\citeauthoryear{{Deleuil} \& {Fridlund}}{{Deleuil} \&
  {Fridlund}}{2018}]{deleuil18}
{Deleuil} M.,  {Fridlund} M.,  2018, {CoRoT: The First Space-Based Transit
  Survey to Explore the Close-in Planet Population}.
p.~79, \mn@doi{10.1007/978-3-319-55333-7_79}

\bibitem[\protect\citeauthoryear{{Deleuil} et~al.,}{{Deleuil}
  et~al.}{2008}]{deleuil08}
{Deleuil} M.,  et~al., 2008, \mn@doi [\aap] {10.1051/0004-6361:200810625},
  \href {http://adsabs.harvard.edu/abs/2008A%26A...491..889D} {491, 889}

\bibitem[\protect\citeauthoryear{{Deleuil} et~al.,}{{Deleuil}
  et~al.}{2009}]{deleuil09}
{Deleuil} M.,  et~al., 2009, \mn@doi [\aj] {10.1088/0004-6256/138/2/649}, \href
  {http://adsabs.harvard.edu/abs/2009AJ....138..649D} {138, 649}

\bibitem[\protect\citeauthoryear{{Deleuil} et~al.,}{{Deleuil}
  et~al.}{2018}]{deleuil18b}
{Deleuil} M.,  et~al., 2018, \mn@doi [\aap] {10.1051/0004-6361/201731068},
  \href {https://ui.adsabs.harvard.edu/abs/2018A&A...619A..97D} {619, A97}

\bibitem[\protect\citeauthoryear{{Delgado Mena} et~al.,}{{Delgado Mena}
  et~al.}{2018}]{Delgado18}
{Delgado Mena} E.,  et~al., 2018, \mn@doi [\aap] {10.1051/0004-6361/201833152},
  \href {https://ui.adsabs.harvard.edu/abs/2018A&A...619A...2D} {619, A2}

\bibitem[\protect\citeauthoryear{{Desort}, {Lagrange}, {Galland}, {Beust},
  {Udry}, {Mayor}  \& {Lo Curto}}{{Desort} et~al.}{2009a}]{desort09a}
{Desort} M.,  {Lagrange} A.~M.,  {Galland} F.,  {Beust} H.,  {Udry} S.,
  {Mayor} M.,   {Lo Curto} G.,  2009a, \mn@doi [\aap]
  {10.1051/0004-6361/200810241e}, \href
  {https://ui.adsabs.harvard.edu/abs/2009A&A...499..623D} {499, 623}

\bibitem[\protect\citeauthoryear{{Desort} et~al.,}{{Desort}
  et~al.}{2009b}]{desort09b}
{Desort} M.,  et~al., 2009b, \mn@doi [\aap] {10.1051/0004-6361/200911731},
  \href {https://ui.adsabs.harvard.edu/abs/2009A&A...506.1469D} {506, 1469}

\bibitem[\protect\citeauthoryear{{Donati}, {Semel}, {Carter}, {Rees}  \&
  {Collier Cameron}}{{Donati} et~al.}{1997}]{donati97}
{Donati} J.-F.,  {Semel} M.,  {Carter} B.~D.,  {Rees} D.~E.,   {Collier
  Cameron} A.,  1997, \mnras, \href
  {http://adsabs.harvard.edu/abs/1997MNRAS.291..658D} {291, 658}

\bibitem[\protect\citeauthoryear{{Dong} \& {Zhu}}{{Dong} \&
  {Zhu}}{2013}]{dong13}
{Dong} S.,  {Zhu} Z.,  2013, \mn@doi [\apj] {10.1088/0004-637X/778/1/53}, \href
  {https://ui.adsabs.harvard.edu/abs/2013ApJ...778...53D} {778, 53}

\bibitem[\protect\citeauthoryear{{Dorval} et~al.,}{{Dorval}
  et~al.}{2019}]{dorval19}
{Dorval} P.,  et~al., 2019, arXiv e-prints, \href
  {https://ui.adsabs.harvard.edu/abs/2019arXiv190402733D} {p. arXiv:1904.02733}

\bibitem[\protect\citeauthoryear{{Eastman}, {Siverd}  \& {Gaudi}}{{Eastman}
  et~al.}{2010}]{2010PASP..122..935E}
{Eastman} J.,  {Siverd} R.,   {Gaudi} B.~S.,  2010, \mn@doi [\pasp]
  {10.1086/655938}, \href
  {https://ui.adsabs.harvard.edu/abs/2010PASP..122..935E} {122, 935}

\bibitem[\protect\citeauthoryear{{El-Badry}, {Rix}  \& {Heintz}}{{El-Badry}
  et~al.}{2021}]{2021MNRAS.tmp..394E}
{El-Badry} K.,  {Rix} H.-W.,   {Heintz} T.~M.,  2021, \mn@doi [\mnras]
  {10.1093/mnras/stab323}, \href
  {https://ui.adsabs.harvard.edu/abs/2021MNRAS.tmp..394E} {}

\bibitem[\protect\citeauthoryear{{Esposito} et~al.,}{{Esposito}
  et~al.}{2010}]{esposito10}
{Esposito} S.,  et~al., 2010, \mn@doi [\ao] {10.1364/AO.49.00G174}, \href
  {https://ui.adsabs.harvard.edu/abs/2010ApOpt..49G.174E} {49, G174}

\bibitem[\protect\citeauthoryear{{\swap{ Essen}{von}}, {Mallonn}, {Albrecht},
  {Antoci}, {Smith}, {Dreizler}  \& {Strassmeier}}{{\swap{ Essen}{von}}
  et~al.}{2015}]{vonessen15}
{\swap{ Essen}{von}} C.,  {Mallonn} M.,  {Albrecht} S.,  {Antoci} V.,  {Smith}
  A.~M.~S.,  {Dreizler} S.,   {Strassmeier} K.~G.,  2015, preprint, \href
  {http://adsabs.harvard.edu/abs/2015arXiv150705963V} {} (\mn@eprint {arXiv}
  {1507.05963})

\bibitem[\protect\citeauthoryear{{Fitzpatrick}, {Massa}, {Gordon}, {Bohlin}  \&
  {Clayton}}{{Fitzpatrick} et~al.}{2019}]{2019ApJ...886..108F}
{Fitzpatrick} E.~L.,  {Massa} D.,  {Gordon} K.~D.,  {Bohlin} R.,   {Clayton}
  G.~C.,  2019, \mn@doi [\apj] {10.3847/1538-4357/ab4c3a}, \href
  {https://ui.adsabs.harvard.edu/abs/2019ApJ...886..108F} {886, 108}

\bibitem[\protect\citeauthoryear{{Foreman-Mackey}, {Hogg}, {Lang}  \&
  {Goodman}}{{Foreman-Mackey} et~al.}{2013}]{Foreman-Mackey13}
{Foreman-Mackey} D.,  {Hogg} D.~W.,  {Lang} D.,   {Goodman} J.,  2013, \mn@doi
  [\pasp] {10.1086/670067}, \href
  {https://ui.adsabs.harvard.edu/abs/2013PASP..125..306F} {125, 306}

\bibitem[\protect\citeauthoryear{{Gaia Collaboration} et~al.,}{{Gaia
  Collaboration} et~al.}{2021}]{2021A&A...649A...1G}
{Gaia Collaboration} et~al., 2021, \mn@doi [\aap]
  {10.1051/0004-6361/202039657}, \href
  {https://ui.adsabs.harvard.edu/abs/2021A&A...649A...1G} {649, A1}

\bibitem[\protect\citeauthoryear{{Galland}, {Lagrange}, {Udry}, {Chelli},
  {Pepe}, {Beuzit}  \& {Mayor}}{{Galland} et~al.}{2006a}]{galland06a}
{Galland} F.,  {Lagrange} A.~M.,  {Udry} S.,  {Chelli} A.,  {Pepe} F.,
  {Beuzit} J.~L.,   {Mayor} M.,  2006a, \mn@doi [\aap]
  {10.1051/0004-6361:20054080}, \href
  {https://ui.adsabs.harvard.edu/abs/2006A&A...447..355G} {447, 355}

\bibitem[\protect\citeauthoryear{{Galland}, {Lagrange}, {Udry}, {Beuzit},
  {Pepe}  \& {Mayor}}{{Galland} et~al.}{2006b}]{galland06b}
{Galland} F.,  {Lagrange} A.~M.,  {Udry} S.,  {Beuzit} J.~L.,  {Pepe} F.,
  {Mayor} M.,  2006b, \mn@doi [\aap] {10.1051/0004-6361:20054079}, \href
  {https://ui.adsabs.harvard.edu/abs/2006A&A...452..709G} {452, 709}

\bibitem[\protect\citeauthoryear{{Gandolfi} et~al.,}{{Gandolfi}
  et~al.}{2010}]{gandolfi10}
{Gandolfi} D.,  et~al., 2010, \mn@doi [\aap] {10.1051/0004-6361/201015132},
  \href {http://adsabs.harvard.edu/abs/2010A%26A...524A..55G} {524, A55}

\bibitem[\protect\citeauthoryear{{Gaudi} et~al.,}{{Gaudi}
  et~al.}{2017}]{gaudi17}
{Gaudi} B.~S.,  et~al., 2017, \mn@doi [\nat] {10.1038/nature22392}, \href
  {https://ui.adsabs.harvard.edu/abs/2017Natur.546..514G} {546, 514}

\bibitem[\protect\citeauthoryear{{Gazzano} et~al.,}{{Gazzano}
  et~al.}{2010}]{gazzano10}
{Gazzano} J.-C.,  et~al., 2010, \mn@doi [\aap] {10.1051/0004-6361/201014708},
  \href {http://adsabs.harvard.edu/abs/2010A%26A...523A..91G} {523, A91}

\bibitem[\protect\citeauthoryear{{Gazzano}, {Kordopatis}, {Deleuil}, {de
  Laverny}, {Recio-Blanco}  \& {Hill}}{{Gazzano} et~al.}{2013}]{gazzano13}
{Gazzano} J.-C.,  {Kordopatis} G.,  {Deleuil} M.,  {de Laverny} P.,
  {Recio-Blanco} A.,   {Hill} V.,  2013, \mn@doi [\aap]
  {10.1051/0004-6361/201117747}, \href
  {http://adsabs.harvard.edu/abs/2013A%26A...550A.125G} {550, A125}

\bibitem[\protect\citeauthoryear{{Giddings}}{{Giddings}}{1981}]{Giddings1981}
{Giddings} J.~R.,  1981, PhD thesis, -

\bibitem[\protect\citeauthoryear{{Gray}}{{Gray}}{2005}]{gray05}
{Gray} D.~F.,  2005, {The Observation and Analysis of Stellar Photospheres}

\bibitem[\protect\citeauthoryear{{Grziwa}, {P{\"a}tzold}  \& {Carone}}{{Grziwa}
  et~al.}{2012}]{grziwa12}
{Grziwa} S.,  {P{\"a}tzold} M.,   {Carone} L.,  2012, \mn@doi [\mnras]
  {10.1111/j.1365-2966.2011.19970.x}, \href
  {http://adsabs.harvard.edu/abs/2012MNRAS.420.1045G} {420, 1045}

\bibitem[\protect\citeauthoryear{{Guenther}, {Hartmann}, {Esposito}, {Hatzes},
  {Cusano}  \& {Gandolfi}}{{Guenther} et~al.}{2009}]{guenther09}
{Guenther} E.~W.,  {Hartmann} M.,  {Esposito} M.,  {Hatzes} A.~P.,  {Cusano}
  F.,   {Gandolfi} D.,  2009, \mn@doi [\aap] {10.1051/0004-6361/200912112},
  \href {https://ui.adsabs.harvard.edu/abs/2009A&A...507.1659G} {507, 1659}

\bibitem[\protect\citeauthoryear{{Guenther}, {Gandolfi}, {Sebastian},
  {Deleuil}, {Moutou}  \& {Cusano}}{{Guenther} et~al.}{2012}]{guenther12}
{Guenther} E.~W.,  {Gandolfi} D.,  {Sebastian} D.,  {Deleuil} M.,  {Moutou} C.,
    {Cusano} F.,  2012, \mn@doi [\aap] {10.1051/0004-6361/201219121}, \href
  {http://adsabs.harvard.edu/abs/2012A%26A...543A.125G} {543, A125}

\bibitem[\protect\citeauthoryear{{Guenther} et~al.,}{{Guenther}
  et~al.}{2013}]{guenther13}
{Guenther} E.~W.,  et~al., 2013, \mn@doi [\aap] {10.1051/0004-6361/201220902},
  \href {http://adsabs.harvard.edu/abs/2013A%26A...556A..75G} {556, A75}

\bibitem[\protect\citeauthoryear{{Guenther} et~al.,}{{Guenther}
  et~al.}{2016}]{guenther16}
{Guenther} E.~W.,  et~al., 2016, {III.7 Planets orbiting stars more massive
  than the Sun}.
p.~149, \mn@doi{10.1051/978-2-7598-1876-1.c037}

\bibitem[\protect\citeauthoryear{Guerra, Boutsia, Rakich, Green, Mccarthy  \&
  Steward}{Guerra et~al.}{2011}]{Guerra11}
Guerra J.~C.,  Boutsia K.,  Rakich A.,  Green R.,  Mccarthy D.,   Steward C.,
  2011.

\bibitem[\protect\citeauthoryear{{G{\"u}nther} \& {Daylan}}{{G{\"u}nther} \&
  {Daylan}}{2019}]{guenther19}
{G{\"u}nther} M.~N.,  {Daylan} T.,  2019, {allesfitter: Flexible star and
  exoplanet inference from photometry and radial velocity}, Astrophysics Source
  Code Library (\mn@eprint {ascl} {1903.003})

\bibitem[\protect\citeauthoryear{{Hasegawa} \& {Pudritz}}{{Hasegawa} \&
  {Pudritz}}{2013}]{hasegawa13}
{Hasegawa} Y.,  {Pudritz} R.~E.,  2013, \mn@doi [\apj]
  {10.1088/0004-637X/778/1/78}, \href
  {http://adsabs.harvard.edu/abs/2013ApJ...778...78H} {778, 78}

\bibitem[\protect\citeauthoryear{{Hatzes} et~al.,}{{Hatzes}
  et~al.}{2015}]{hatzes15b}
{Hatzes} A.~P.,  et~al., 2015, \mn@doi [\aap] {10.1051/0004-6361/201425519},
  \href {https://ui.adsabs.harvard.edu/abs/2015A&A...580A..31H} {580, A31}

\bibitem[\protect\citeauthoryear{{Hatzes} et~al.,}{{Hatzes}
  et~al.}{2018}]{Hatzes18}
{Hatzes} A.~P.,  et~al., 2018, \mn@doi [\aj] {10.3847/1538-3881/aaa8e1}, \href
  {https://ui.adsabs.harvard.edu/abs/2018AJ....155..120H} {155, 120}

\bibitem[\protect\citeauthoryear{{Heber}, {Irrgang}  \& {Schaffenroth}}{{Heber}
  et~al.}{2018}]{Heber2018}
{Heber} U.,  {Irrgang} A.,   {Schaffenroth} J.,  2018, \mn@doi [Open Astronomy]
  {10.1515/astro-2018-0008}, \href
  {http://adsabs.harvard.edu/abs/2018OAst...27...35H} {27, 35}

\bibitem[\protect\citeauthoryear{{Henden}, {Levine}, {Terrell}  \&
  {Welch}}{{Henden} et~al.}{2015}]{2015AAS...22533616H}
{Henden} A.~A.,  {Levine} S.,  {Terrell} D.,   {Welch} D.~L.,  2015, in
  American Astronomical Society Meeting Abstracts. p. 336.16

\bibitem[\protect\citeauthoryear{{Heuser}}{{Heuser}}{2018}]{heuser18}
{Heuser} C.,  2018, PhD thesis, Dr. Karl Remeis-Observatory \&amp; ECAP,
  Astronomical Institute, Friedrich-Alexander University Erlangen-N{\"u}rnberg,
  Sternwartstr. 7, D-96049 Bamberg, Germany

\bibitem[\protect\citeauthoryear{{Hidas} et~al.,}{{Hidas}
  et~al.}{2005}]{hidas05}
{Hidas} M.~G.,  et~al., 2005, \mn@doi [\mnras]
  {10.1111/j.1365-2966.2005.09061.x}, \href
  {http://adsabs.harvard.edu/abs/2005MNRAS.360..703H} {360, 703}

\bibitem[\protect\citeauthoryear{{Irrgang}, {Przybilla}, {Heber}, {B{\"o}ck},
  {Hanke}, {Nieva}  \& {Butler}}{{Irrgang} et~al.}{2014}]{2014A&A...565A..63I}
{Irrgang} A.,  {Przybilla} N.,  {Heber} U.,  {B{\"o}ck} M.,  {Hanke} M.,
  {Nieva} M.~F.,   {Butler} K.,  2014, \mn@doi [\aap]
  {10.1051/0004-6361/201323167}, \href
  {https://ui.adsabs.harvard.edu/abs/2014A&A...565A..63I} {565, A63}

\bibitem[\protect\citeauthoryear{{Irrgang}, {Desphande}, {Moehler}, {Mugrauer}
  \& {Janousch}}{{Irrgang} et~al.}{2016}]{2016A&A...591L...6I}
{Irrgang} A.,  {Desphande} A.,  {Moehler} S.,  {Mugrauer} M.,   {Janousch} D.,
  2016, \mn@doi [\aap] {10.1051/0004-6361/201628844}, \href
  {https://ui.adsabs.harvard.edu/abs/2016A&A...591L...6I} {591, L6}

\bibitem[\protect\citeauthoryear{{Irrgang}, {Geier}, {Heber}, {Kupfer},
  {El-Badry}  \& {Bloemen}}{{Irrgang} et~al.}{2020}]{2020arXiv200703350I}
{Irrgang} A.,  {Geier} S.,  {Heber} U.,  {Kupfer} T.,  {El-Badry} K.,
  {Bloemen} S.,  2020, arXiv e-prints, \href
  {https://ui.adsabs.harvard.edu/abs/2020arXiv200703350I} {p. arXiv:2007.03350}

\bibitem[\protect\citeauthoryear{{Johnson}, {Aller}, {Howard}  \&
  {Crepp}}{{Johnson} et~al.}{2010a}]{johnson10}
{Johnson} J.~A.,  {Aller} K.~M.,  {Howard} A.~W.,   {Crepp} J.~R.,  2010a,
  \mn@doi [\pasp] {10.1086/655775}, \href
  {http://adsabs.harvard.edu/abs/2010PASP..122..905J} {122, 905}

\bibitem[\protect\citeauthoryear{{Johnson} et~al.,}{{Johnson}
  et~al.}{2010b}]{johnson10b}
{Johnson} J.~A.,  et~al., 2010b, \mn@doi [\apjl]
  {10.1088/2041-8205/721/2/L153}, \href
  {http://adsabs.harvard.edu/abs/2010ApJ...721L.153J} {721, L153}

\bibitem[\protect\citeauthoryear{{Johnson} et~al.,}{{Johnson}
  et~al.}{2018}]{johnson18}
{Johnson} M.~C.,  et~al., 2018, \mn@doi [\aj] {10.3847/1538-3881/aaa5af}, \href
  {https://ui.adsabs.harvard.edu/abs/2018AJ....155..100J} {155, 100}

\bibitem[\protect\citeauthoryear{{Kipping}}{{Kipping}}{2013}]{Kipping13}
{Kipping} D.~M.,  2013, \mn@doi [\mnras] {10.1093/mnras/stt1435}, \href
  {https://ui.adsabs.harvard.edu/abs/2013MNRAS.435.2152K} {435, 2152}

\bibitem[\protect\citeauthoryear{{Klagyivik}, {Deeg}, {Csizmadia}, {Cabrera}
  \& {Nowak}}{{Klagyivik} et~al.}{2021}]{Klagyivik21}
{Klagyivik} P.,  {Deeg} H.~J.,  {Csizmadia} S.,  {Cabrera} J.,   {Nowak} G.,
  2021, \mn@doi [Frontiers in Astronomy and Space Sciences]
  {10.3389/fspas.2021.792823}, \href
  {https://ui.adsabs.harvard.edu/abs/2021FrASS...8..210K} {8, 210}

\bibitem[\protect\citeauthoryear{{Kumar}, {Ao}  \& {Quataert}}{{Kumar}
  et~al.}{1995}]{Kumar95}
{Kumar} P.,  {Ao} C.~O.,   {Quataert} E.~J.,  1995, \mn@doi [\apj]
  {10.1086/176055}, \href
  {https://ui.adsabs.harvard.edu/abs/1995ApJ...449..294K} {449, 294}

\bibitem[\protect\citeauthoryear{{Kupka}, {Ryabchikova}, {Piskunov}, {Stempels}
   \& {Weiss}}{{Kupka} et~al.}{2000}]{kupka00}
{Kupka} F.~G.,  {Ryabchikova} T.~A.,  {Piskunov} N.~E.,  {Stempels} H.~C.,
  {Weiss} W.~W.,  2000, Baltic Astronomy, \href
  {http://adsabs.harvard.edu/abs/2000BaltA...9..590K} {9, 590}

\bibitem[\protect\citeauthoryear{{Kurucz}}{{Kurucz}}{1993}]{Kurucz1993}
{Kurucz} R.~L.,  1993, {SYNTHE spectrum synthesis programs and line data}

\bibitem[\protect\citeauthoryear{{Kurucz}}{{Kurucz}}{1996}]{Kurucz1996}
{Kurucz} R.~L.,  1996, in {Adelman} S.~J.,  {Kupka} F.,   {Weiss} W.~W.,  eds,
  Astronomical Society of the Pacific Conference Series Vol. 108, M.A.S.S.,
  Model Atmospheres and Spectrum Synthesis. p.~160

\bibitem[\protect\citeauthoryear{{Lawrence} et~al.,}{{Lawrence}
  et~al.}{2007}]{2007MNRAS.379.1599L}
{Lawrence} A.,  et~al., 2007, \mn@doi [\mnras]
  {10.1111/j.1365-2966.2007.12040.x}, \href
  {https://ui.adsabs.harvard.edu/abs/2007MNRAS.379.1599L} {379, 1599}

\bibitem[\protect\citeauthoryear{{L{\'e}ger} et~al.,}{{L{\'e}ger}
  et~al.}{2009}]{leger09}
{L{\'e}ger} A.,  et~al., 2009, \mn@doi [\aap] {10.1051/0004-6361/200911933},
  \href {http://adsabs.harvard.edu/abs/2009A\%26A...506..287L} {506, 287}

\bibitem[\protect\citeauthoryear{{Lehmann} et~al.,}{{Lehmann}
  et~al.}{2011}]{lehmann11}
{Lehmann} H.,  et~al., 2011, \mn@doi [\aap] {10.1051/0004-6361/201015769},
  \href {http://adsabs.harvard.edu/abs/2011A%26A...526A.124L} {526, A124}

\bibitem[\protect\citeauthoryear{{Lehmann}, {Guenther}, {Sebastian},
  {D{\"o}llinger}, {Hartmann}  \& {Mkrtichian}}{{Lehmann}
  et~al.}{2015}]{lehm15}
{Lehmann} H.,  {Guenther} E.,  {Sebastian} D.,  {D{\"o}llinger} M.,  {Hartmann}
  M.,   {Mkrtichian} D.~E.,  2015, \mn@doi [\aap]
  {10.1051/0004-6361/201526176}, \href
  {http://adsabs.harvard.edu/abs/2015A%26A...578L...4L} {578, L4}

\bibitem[\protect\citeauthoryear{{Lindegren}}{{Lindegren}}{2018}]{RUWE}
{Lindegren} L.,  2018, Re-normalising the astrometric chi-square in Gaia DR2,
  GAIA-C3-TN-LU-LL-124, \url{www.rssd.esa.int/doc_fetch.php?id=3757412}

\bibitem[\protect\citeauthoryear{{Lindegren} et~al.,}{{Lindegren}
  et~al.}{2021a}]{2021A&A...649A...2L}
{Lindegren} L.,  et~al., 2021a, \mn@doi [\aap] {10.1051/0004-6361/202039709},
  \href {https://ui.adsabs.harvard.edu/abs/2021A&A...649A...2L} {649, A2}

\bibitem[\protect\citeauthoryear{{Lindegren} et~al.,}{{Lindegren}
  et~al.}{2021b}]{Lindegren2021}
{Lindegren} L.,  et~al., 2021b, \mn@doi [\aap] {10.1051/0004-6361/202039653},
  \href {https://ui.adsabs.harvard.edu/abs/2021A&A...649A...4L} {649, A4}

\bibitem[\protect\citeauthoryear{{Lovis} \& {Mayor}}{{Lovis} \&
  {Mayor}}{2007}]{Lovis07}
{Lovis} C.,  {Mayor} M.,  2007, \mn@doi [\aap] {10.1051/0004-6361:20077375},
  \href {https://ui.adsabs.harvard.edu/abs/2007A&A...472..657L} {472, 657}

\bibitem[\protect\citeauthoryear{{Mamajek}}{{Mamajek}}{2009}]{mamajek09}
{Mamajek} E.~E.,  2009, in {Usuda} T.,  {Tamura} M.,   {Ishii} M.,  eds,
  American Institute of Physics Conference Series Vol. 1158, American Institute
  of Physics Conference Series. pp 3--10 (\mn@eprint {arXiv} {0906.5011}),
  \mn@doi{10.1063/1.3215910}

\bibitem[\protect\citeauthoryear{{Maxted}}{{Maxted}}{2016}]{maxted16}
{Maxted} P.~F.~L.,  2016, \mn@doi [\aap] {10.1051/0004-6361/201628579}, \href
  {https://ui.adsabs.harvard.edu/abs/2016A&A...591A.111M} {591, A111}

\bibitem[\protect\citeauthoryear{{Maxted} et~al.,}{{Maxted}
  et~al.}{2021}]{maxted21}
{Maxted} P.~F.~L.,  et~al., 2021, \mn@doi [\mnras] {10.1093/mnras/stab3371},
  \href {https://ui.adsabs.harvard.edu/abs/2021MNRAS.tmp.3057M} {}

\bibitem[\protect\citeauthoryear{{Mayor} et~al.,}{{Mayor}
  et~al.}{2003}]{mayor03}
{Mayor} M.,  et~al., 2003, The Messenger, \href
  {http://adsabs.harvard.edu/abs/2003Msngr.114...20M} {114, 20}

\bibitem[\protect\citeauthoryear{{Mayor} et~al.,}{{Mayor}
  et~al.}{2011}]{mayor11}
{Mayor} M.,  et~al., 2011, preprint, \href
  {http://adsabs.harvard.edu/abs/2011arXiv1109.2497M} {} (\mn@eprint {arXiv}
  {1109.2497})

\bibitem[\protect\citeauthoryear{{Mazeh}, {Nachmani}, {Sokol}, {Faigler}  \&
  {Zucker}}{{Mazeh} et~al.}{2012}]{mazeh12}
{Mazeh} T.,  {Nachmani} G.,  {Sokol} G.,  {Faigler} S.,   {Zucker} S.,  2012,
  \mn@doi [\aap] {10.1051/0004-6361/201117908}, \href
  {http://adsabs.harvard.edu/abs/2012A%26A...541A..56M} {541, A56}

\bibitem[\protect\citeauthoryear{{McCarthy}, {Sandiford}, {Boyd}  \&
  {Booth}}{{McCarthy} et~al.}{1993}]{mccarthy93}
{McCarthy} J.~K.,  {Sandiford} B.~A.,  {Boyd} D.,   {Booth} J.,  1993, \mn@doi
  [\pasp] {10.1086/133250}, \href
  {http://adsabs.harvard.edu/abs/1993PASP..105..881M} {105, 881}

\bibitem[\protect\citeauthoryear{{Moutou} et~al.,}{{Moutou}
  et~al.}{2013}]{moutou13}
{Moutou} C.,  et~al., 2013, \mn@doi [\icarus] {10.1016/j.icarus.2013.03.022},
  \href {http://adsabs.harvard.edu/abs/2013Icar..226.1625M} {226, 1625}

\bibitem[\protect\citeauthoryear{{Naef}, {Mayor}, {Beuzit}, {Perrier},
  {Queloz}, {Sivan}  \& {Udry}}{{Naef} et~al.}{2005}]{naef05}
{Naef} D.,  {Mayor} M.,  {Beuzit} J.-L.,  {Perrier} C.,  {Queloz} D.,  {Sivan}
  J.-P.,   {Udry} S.,  2005, in {Favata} F.,  {Hussain} G.~A.~J.,   {Battrick}
  B.,  eds,  ESA Special Publication Vol. 560, 13th Cambridge Workshop on Cool
  Stars, Stellar Systems and the Sun. p.~833 (\mn@eprint {} {astro-ph/0409230})

\bibitem[\protect\citeauthoryear{{Nagasawa}, {Ida}  \& {Bessho}}{{Nagasawa}
  et~al.}{2008}]{nagasawa08}
{Nagasawa} M.,  {Ida} S.,   {Bessho} T.,  2008, \mn@doi [\apj]
  {10.1086/529369}, \href {http://adsabs.harvard.edu/abs/2008ApJ...678..498N}
  {678, 498}

\bibitem[\protect\citeauthoryear{{Onken} et~al.,}{{Onken}
  et~al.}{2019}]{2019PASA...36...33O}
{Onken} C.~A.,  et~al., 2019, \mn@doi [\pasa] {10.1017/pasa.2019.27}, \href
  {https://ui.adsabs.harvard.edu/abs/2019PASA...36...33O} {36, e033}

\bibitem[\protect\citeauthoryear{{Oshagh}, {Santos}, {Ehrenreich},
  {Haghighipour}, {Figueira}, {Santerne}  \& {Montalto}}{{Oshagh}
  et~al.}{2014}]{Oshagh14}
{Oshagh} M.,  {Santos} N.~C.,  {Ehrenreich} D.,  {Haghighipour} N.,  {Figueira}
  P.,  {Santerne} A.,   {Montalto} M.,  2014, \mn@doi [\aap]
  {10.1051/0004-6361/201424059}, \href
  {https://ui.adsabs.harvard.edu/abs/2014A&A...568A..99O} {568, A99}

\bibitem[\protect\citeauthoryear{{P{\"a}tzold} et~al.,}{{P{\"a}tzold}
  et~al.}{2012}]{paetzold12}
{P{\"a}tzold} M.,  et~al., 2012, \mn@doi [\aap] {10.1051/0004-6361/201118425},
  \href {http://adsabs.harvard.edu/abs/2012A%26A...545A...6P} {545, A6}

\bibitem[\protect\citeauthoryear{{Pepe} et~al.,}{{Pepe} et~al.}{2002}]{pepe02}
{Pepe} F.,  et~al., 2002, The Messenger, \href
  {http://adsabs.harvard.edu/abs/2002Msngr.110....9P} {110, 9}

\bibitem[\protect\citeauthoryear{{Phillips} et~al.,}{{Phillips}
  et~al.}{2020}]{phillips20}
{Phillips} M.~W.,  et~al., 2020, \mn@doi [\aap] {10.1051/0004-6361/201937381},
  \href {https://ui.adsabs.harvard.edu/abs/2020A&A...637A..38P} {637, A38}

\bibitem[\protect\citeauthoryear{{Pribulla} et~al.,}{{Pribulla}
  et~al.}{2014}]{pribulla14}
{Pribulla} T.,  et~al., 2014, \mn@doi [\mnras] {10.1093/mnras/stu1333}, \href
  {http://adsabs.harvard.edu/abs/2014MNRAS.443.2815P} {443, 2815}

\bibitem[\protect\citeauthoryear{{Reffert}, {Bergmann}, {Quirrenbach},
  {Trifonov}  \& {K{\"u}nstler}}{{Reffert} et~al.}{2015}]{reffert15}
{Reffert} S.,  {Bergmann} C.,  {Quirrenbach} A.,  {Trifonov} T.,
  {K{\"u}nstler} A.,  2015, \mn@doi [\aap] {10.1051/0004-6361/201322360}, \href
  {http://adsabs.harvard.edu/abs/2015A%26A...574A.116R} {574, A116}

\bibitem[\protect\citeauthoryear{{Reichert}, {Reffert}, {Stock}, {Trifonov}  \&
  {Quirrenbach}}{{Reichert} et~al.}{2019}]{reichert19}
{Reichert} K.,  {Reffert} S.,  {Stock} S.,  {Trifonov} T.,   {Quirrenbach} A.,
  2019, \mn@doi [\aap] {10.1051/0004-6361/201834028}, \href
  {https://ui.adsabs.harvard.edu/abs/2019A&A...625A..22R} {625, A22}

\bibitem[\protect\citeauthoryear{{Ricker} et~al.,}{{Ricker}
  et~al.}{2015}]{ricker15}
{Ricker} G.~R.,  et~al., 2015, \mn@doi [Journal of Astronomical Telescopes,
  Instruments, and Systems] {10.1117/1.JATIS.1.1.014003}, \href
  {http://adsabs.harvard.edu/abs/2015JATIS...1a4003R} {1, 014003}

\bibitem[\protect\citeauthoryear{{Riello} et~al.,}{{Riello}
  et~al.}{2021}]{2021A&A...649A...3R}
{Riello} M.,  et~al., 2021, \mn@doi [\aap] {10.1051/0004-6361/202039587}, \href
  {https://ui.adsabs.harvard.edu/abs/2021A&A...649A...3R} {649, A3}

\bibitem[\protect\citeauthoryear{{Royer}, {Zorec}  \& {G{\'o}mez}}{{Royer}
  et~al.}{2007}]{royer07}
{Royer} F.,  {Zorec} J.,   {G{\'o}mez} A.~E.,  2007, \mn@doi [\aap]
  {10.1051/0004-6361:20065224}, \href
  {https://ui.adsabs.harvard.edu/abs/2007A&A...463..671R} {463, 671}

\bibitem[\protect\citeauthoryear{Sabotta, Kabath, Korth, Guenther, Dupkala,
  Grziwa, Klocova  \& Skarka}{Sabotta et~al.}{2019}]{sabotta19}
Sabotta S.,  Kabath P.,  Korth J.,  Guenther E.~W.,  Dupkala D.,  Grziwa S.,
  Klocova T.,   Skarka M.,  2019, \mn@doi [Monthly Notices of the Royal
  Astronomical Society] {10.1093/mnras/stz2232}, 489, 2069

\bibitem[\protect\citeauthoryear{{Sarkis}, {Mordasini}, {Henning}, {Marleau}
  \& {Molli{\`e}re}}{{Sarkis} et~al.}{2021}]{sarkis21}
{Sarkis} P.,  {Mordasini} C.,  {Henning} T.,  {Marleau} G.~D.,   {Molli{\`e}re}
  P.,  2021, \mn@doi [\aap] {10.1051/0004-6361/202038361}, \href
  {https://ui.adsabs.harvard.edu/abs/2021A&A...645A..79S} {645, A79}

\bibitem[\protect\citeauthoryear{{Sarro} et~al.,}{{Sarro}
  et~al.}{2013}]{sarro13}
{Sarro} L.~M.,  et~al., 2013, \mn@doi [\aap] {10.1051/0004-6361/201220184},
  \href {http://cdsads.u-strasbg.fr/abs/2013A%26A...550A.120S} {550, A120}

\bibitem[\protect\citeauthoryear{{Saumon} \& {Marley}}{{Saumon} \&
  {Marley}}{2008}]{saumon08}
{Saumon} D.,  {Marley} M.~S.,  2008, \mn@doi [\apj] {10.1086/592734}, \href
  {https://ui.adsabs.harvard.edu/abs/2008ApJ...689.1327S} {689, 1327}

\bibitem[\protect\citeauthoryear{{Schlafly}, {Meisner}  \& {Green}}{{Schlafly}
  et~al.}{2020}]{2020yCat.2363....0S}
{Schlafly} E.~F.,  {Meisner} A.~M.,   {Green} G.~M.,  2020, VizieR Online Data
  Catalog, \href {https://ui.adsabs.harvard.edu/abs/2020yCat.2363....0S} {p.
  II/363}

\bibitem[\protect\citeauthoryear{{Schneider}, {Dedieu}, {Le Sidaner}, {Savalle}
   \& {Zolotukhin}}{{Schneider} et~al.}{2011}]{schneider11}
{Schneider} J.,  {Dedieu} C.,  {Le Sidaner} P.,  {Savalle} R.,   {Zolotukhin}
  I.,  2011, \mn@doi [\aap] {10.1051/0004-6361/201116713}, \href
  {http://adsabs.harvard.edu/abs/2011A%26A...532A..79S} {532, A79}

\bibitem[\protect\citeauthoryear{{Seager} \& {Mall{\'e}n-Ornelas}}{{Seager} \&
  {Mall{\'e}n-Ornelas}}{2003}]{seager03}
{Seager} S.,  {Mall{\'e}n-Ornelas} G.,  2003, \mn@doi [\apj] {10.1086/346105},
  \href {http://adsabs.harvard.edu/abs/2003ApJ...585.1038S} {585, 1038}

\bibitem[\protect\citeauthoryear{{Sebastian}}{{Sebastian}}{2017}]{sebastian17}
{Sebastian} D.,  2017, PhD thesis, Friedrich-Schiller-Universitaet Jena
  Physikalisch-Astronomische Fakultaet Thueringer Landessternwarte Tautenburg

\bibitem[\protect\citeauthoryear{{Sebastian}, {Guenther}, {Schaffenroth},
  {Gandolfi}, {Geier}, {Heber}, {Deleuil}  \& {Moutou}}{{Sebastian}
  et~al.}{2012}]{sebastian12}
{Sebastian} D.,  {Guenther} E.~W.,  {Schaffenroth} V.,  {Gandolfi} D.,  {Geier}
  S.,  {Heber} U.,  {Deleuil} M.,   {Moutou} C.,  2012, \mn@doi [\aap]
  {10.1051/0004-6361/201118032}, \href
  {http://adsabs.harvard.edu/abs/2012A%26A...541A..34S} {541, A34}

\bibitem[\protect\citeauthoryear{{Shporer} et~al.,}{{Shporer}
  et~al.}{2011}]{shporer11}
{Shporer} A.,  et~al., 2011, \mn@doi [\aj] {10.1088/0004-6256/142/6/195}, \href
  {http://adsabs.harvard.edu/abs/2011AJ....142..195S} {142, 195}

\bibitem[\protect\citeauthoryear{{\swap{Silva}{da }}, {Maceroni}, {Gandolfi},
  {Lehmann}  \& {Hatzes}}{{\swap{Silva}{da }} et~al.}{2014}]{silva14}
{\swap{Silva}{da }} R.,  {Maceroni} C.,  {Gandolfi} D.,  {Lehmann} H.,
  {Hatzes} A.~P.,  2014, \mn@doi [\aap] {10.1051/0004-6361/201322917}, \href
  {http://adsabs.harvard.edu/abs/2014A%26A...565A..55D} {565, A55}

\bibitem[\protect\citeauthoryear{{Siverd} et~al.,}{{Siverd}
  et~al.}{2012}]{siverd12}
{Siverd} R.~J.,  et~al., 2012, \mn@doi [\apj] {10.1088/0004-637X/761/2/123},
  \href {https://ui.adsabs.harvard.edu/abs/2012ApJ...761..123S} {761, 123}

\bibitem[\protect\citeauthoryear{{Skrutskie} et~al.,}{{Skrutskie}
  et~al.}{2006}]{2006AJ....131.1163S}
{Skrutskie} M.~F.,  et~al., 2006, \mn@doi [\aj] {10.1086/498708}, \href
  {http://adsabs.harvard.edu/abs/2006AJ....131.1163S} {131, 1163}

\bibitem[\protect\citeauthoryear{{Speagle}}{{Speagle}}{2019}]{speagle19}
{Speagle} J.~S.,  2019, arXiv e-prints, \href
  {http://adsabs.harvard.edu/abs/2019arXiv190402180S} {}

\bibitem[\protect\citeauthoryear{{\v{S}}ubjak et~al.,}{{\v{S}}ubjak
  et~al.}{2020}]{subjak2020}
{\v{S}}ubjak J.,  et~al., 2020, \mn@doi [The Astronomical Journal]
  {10.3847/1538-3881/ab7245}, 159, 151

\bibitem[\protect\citeauthoryear{{Telting} et~al.,}{{Telting}
  et~al.}{2014}]{telting14}
{Telting} J.~H.,  et~al., 2014, \mn@doi [Astronomische Nachrichten]
  {10.1002/asna.201312007}, \href
  {http://adsabs.harvard.edu/abs/2014AN....335...41T} {335, 41}

\bibitem[\protect\citeauthoryear{{Tody}}{{Tody}}{1993}]{tody93}
{Tody} D.,  1993, in {Hanisch} R.~J.,  {Brissenden} R.~J.~V.,   {Barnes} J.,
  eds,  Astronomical Society of the Pacific Conference Series Vol. 52,
  Astronomical Data Analysis Software and Systems II. p.~173

\bibitem[\protect\citeauthoryear{{Triaud} et~al.,}{{Triaud}
  et~al.}{2010}]{triaud10}
{Triaud} A.~H.~M.~J.,  et~al., 2010, \mn@doi [\aap]
  {10.1051/0004-6361/201014525}, \href
  {https://ui.adsabs.harvard.edu/abs/2010A&A...524A..25T} {524, A25}

\bibitem[\protect\citeauthoryear{{Tsantaki}, {Sousa}, {Santos}, {Montalto},
  {Delgado-Mena}, {Mortier}, {Adibekyan}  \& {Israelian}}{{Tsantaki}
  et~al.}{2014}]{tsantaki14}
{Tsantaki} M.,  {Sousa} S.~G.,  {Santos} N.~C.,  {Montalto} M.,  {Delgado-Mena}
  E.,  {Mortier} A.,  {Adibekyan} V.,   {Israelian} G.,  2014, \mn@doi [\aap]
  {10.1051/0004-6361/201424257}, \href
  {http://adsabs.harvard.edu/abs/2014A%26A...570A..80T} {570, A80}

\bibitem[\protect\citeauthoryear{{Valdes}, {Gupta}, {Rose}, {Singh}  \&
  {Bell}}{{Valdes} et~al.}{2004}]{valdes04}
{Valdes} F.,  {Gupta} R.,  {Rose} J.~A.,  {Singh} H.~P.,   {Bell} D.~J.,  2004,
  \mn@doi [\apjs] {10.1086/386343}, \href
  {http://adsabs.harvard.edu/abs/2004ApJS..152..251V} {152, 251}

\bibitem[\protect\citeauthoryear{{Veras} \& {Ford}}{{Veras} \&
  {Ford}}{2009}]{Veras09}
{Veras} D.,  {Ford} E.~B.,  2009, in {Pont} F.,  {Sasselov} D.,   {Holman}
  M.~J.,  eds,  Proceedings of the International Astronomical Union Vol. 253,
  Transiting Planets. pp 486--489, \mn@doi{10.1017/S1743921308027002}

\bibitem[\protect\citeauthoryear{{Vogt} et~al.,}{{Vogt} et~al.}{1994}]{vogt94}
{Vogt} S.~S.,  et~al., 1994, in {Crawford} D.~L.,  {Craine} E.~R.,  eds,
  Society of Photo-Optical Instrumentation Engineers (SPIE) Conference Series
  Vol. 2198, Instrumentation in Astronomy VIII. p.~362

\bibitem[\protect\citeauthoryear{{Winn}, {Fabrycky}, {Albrecht}  \&
  {Johnson}}{{Winn} et~al.}{2010}]{winn10}
{Winn} J.~N.,  {Fabrycky} D.,  {Albrecht} S.,   {Johnson} J.~A.,  2010, \mn@doi
  [\apjl] {10.1088/2041-8205/718/2/L145}, \href
  {http://adsabs.harvard.edu/abs/2010ApJ...718L.145W} {718, L145}

\bibitem[\protect\citeauthoryear{{Wittenmyer} et~al.,}{{Wittenmyer}
  et~al.}{2020a}]{wittenmyer20a}
{Wittenmyer} R.~A.,  et~al., 2020a, \mn@doi [\mnras] {10.1093/mnras/stz3378},
  \href {https://ui.adsabs.harvard.edu/abs/2020MNRAS.491.5248W} {491, 5248}

\bibitem[\protect\citeauthoryear{{Wittenmyer} et~al.,}{{Wittenmyer}
  et~al.}{2020b}]{wittenmyer20b}
{Wittenmyer} R.~A.,  et~al., 2020b, \mn@doi [\mnras] {10.1093/mnras/stz3436},
  \href {https://ui.adsabs.harvard.edu/abs/2020MNRAS.492..377W} {492, 377}

\bibitem[\protect\citeauthoryear{{Zhou} et~al.,}{{Zhou}
  et~al.}{2019a}]{zhou19a}
{Zhou} G.,  et~al., 2019a, \mn@doi [\aj] {10.3847/1538-3881/aaf1bb}, \href
  {https://ui.adsabs.harvard.edu/abs/2019AJ....157...31Z} {157, 31}

\bibitem[\protect\citeauthoryear{{Zhou} et~al.,}{{Zhou}
  et~al.}{2019b}]{zhou19b}
{Zhou} G.,  et~al., 2019b, \mn@doi [\aj] {10.3847/1538-3881/ab36b5}, \href
  {https://ui.adsabs.harvard.edu/abs/2019AJ....158..141Z} {158, 141}

\makeatother
\end{thebibliography}



\newpage
\appendix

\section{Spectrophotometric temperature, angular diameter and interstellar reddening}\label{sect:sed_fit}

\begin{figure}
\centering
\includegraphics[width=1\columnwidth]{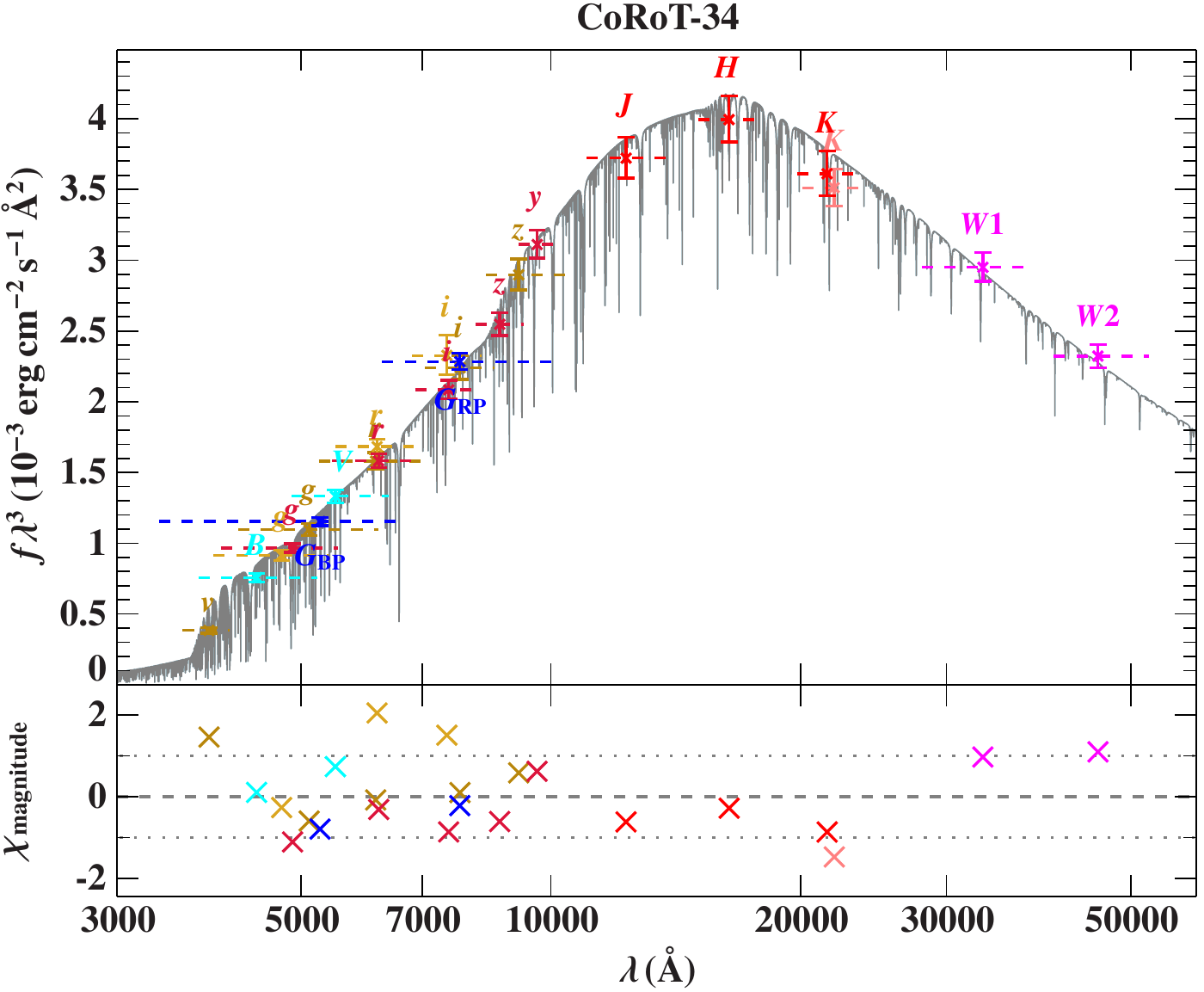}

\caption{Comparison of synthetic and observed photometry for CoRoT--34. 
\textit{Top panel:} SED with filter-averaged fluxes (times wavelength to the power of three) converted from observed magnitudes. The approximate width of the respective filters (widths at 10\% of maximum) are shown by dashed lines. The  best-fitting model SED, smoothed to a spectral resolution of 6\,{\AA}, is shown in grey. 
\textit{Bottom panel:} Residuals $\chi$, i.e., the difference between synthetic and observed magnitudes divided by the corresponding uncertainties.
The different photometric systems are displayed in the following colours: 
APASS-griz \citep[golden,][]{2015AAS...22533616H}; 
PAN-STARRS \citep[red,][]{2017yCat.2349....0C}; 
SkyMapper \citep[dark yellow,][]{2019PASA...36...33O};
APASS-Johnson \citep[cyan,][]{2015AAS...22533616H}; 
{\it Gaia } \citep[blue,][]{2021A&A...649A...3R}; 
2MASS \citep[red,][]{2006AJ....131.1163S}; 
UKIDDS \citep[pink][]{2007MNRAS.379.1599L}; 
UNWISE \citep[magenta,][]{2020yCat.2363....0S}.
}
\label{fig:photometry_sed_corot34} 
\end{figure}

The spectroscopic analysis provided us with the atmospheric parameters. The spectral energy distribution (SED) provides us with an independent estimate of the effective temperature (see Table~\ref{tab:sed_parameter}).
Making use of the geometric flux dilution we derive the angular diameter ($\Theta = 2R/d$) as a scaling factor from the observed flux $f(\lambda)$ and the synthetic stellar surface flux $F(\lambda)$: $\Theta^2=f(\lambda)/F(\lambda)/4$.
Because the CoRoT stars are
 located at low Galactic latitude interstellar reddening is important to be accounted for. We fit the interstellar colour excess and the R$_V$ parameter simultaneously with the angular diameter and the effective temperature, using the reddening law of \citet{2019ApJ...886..108F}. 
For CoRoT--34 the SED is well covered (see Fig.~\ref{fig:photometry_sed_corot34}) allowing us to determine the R$_V$ parameter of the reddening law to be close to the standard (R$_V$=3.0).
The synthetic flux distributions are interpolated from a grid of model SEDs calculated with the ATLAS12 code and matched to the observed magnitudes by $\chi^2$ minimization \citep[see][for details]{Heber2018,2020arXiv200703350I}. 
The surface gravity and metallicities of the stars are fixed. 
The results are summarized in Table~\ref{tab:sed_parameter}. Effective temperatures agree well with the spectroscopic ones listed in Table~\ref{tab:sub_obs}.

\begin{table}
\caption{\label{tab:sed_parameter} Resulting parameters of the fit of the spectral energy distributions.}
\begin{center}
\renewcommand{\arraystretch}{1.2}
\begin{tabular}{lr}
\hline
\multicolumn{2}{l}{CoRoT--34:}\\
\hline
Color excess $E44-55)$ & $0.323^{+0.040}_{-0.029}$\,mag \\
Extinction parameter $R(55)$ & $3.67^{+0.23}_{-0.21}$ \\
Angular diameter $\log(\Theta\,\mathrm{(rad)})$ & $-10.259^{+0.007}_{-0.016}$ \\
Effective temperature $T_{\mathrm{eff}}$ & $7820^{+320}_{-220}$\,K \\
\hline
\multicolumn{2}{l}{CoRoT\,35:}\\
\hline
Color excess $E(44-55)$ & $0.566^{+0.022}_{-0.049}$\,mag \\
Angular diameter $\log(\Theta\,\mathrm{(rad)})$ & $-10.254 \pm 0.010$ \\
Effective temperature $T_{\mathrm{eff}}$ & $6400^{+120}_{-260}$\,K \\
\hline
\multicolumn{2}{l}{CoRoT\,36:}\\
\hline
Color excess $E(44-55)$ & $0.206^{+0.020}_{-0.097}$\,mag \\
Angular diameter $\log(\Theta\,\mathrm{(rad)})$ & $-10.030^{+0.016}_{-0.020}$ \\
Effective temperature $T_{\mathrm{eff}}$ & $6800^{+100}_{-520}$\,K \\
\hline
\end{tabular}
\end{center}

\end{table}

\section{Radial velocity measurements}\label{RV_measures}

The following tables list the RV measurements for the confirmed Sub-stellar companions CoRoT--34b, 35b, \& 36b, as well as for the candidate CoRoT 659668516. All other RV measurements are available in machine readable form as supplementary material.

\begin{table}
\caption{Radial velocity measurements of CoRoT--36 obtained during several orbital phases out of transit. Horizontal lines mark data sets obtained with different instruments or different observing runs.}
\label{tab:rv_corot36}
\begin{tabular}{crrl}
\hline
\noalign{\smallskip}
HJD$^1$     & RV     & $\pm \sigma$ & Instrument \\
-2 450 000  & [km\,s$\rm ^{-1}$] & [km\,s$\rm ^{-1}$]       & \\
\hline
6093.80675	&	10.18	&	0.20	&	HARPS	\\
6096.67636	&	9.99	&	0.20	&	HARPS	\\
6098.79155	&	10.15	&	0.20	&	HARPS	\\
6101.67704	&	10.01	&	0.20	&	HARPS	\\
6115.70754	&	9.98	&	0.20	&	HARPS	\\
6118.64215	&	10.05	&	0.20	&	HARPS	\\
\hline
6101.60540	&	9.68	&	0.43	&	FIES	\\
6102.58413	&	9.24	&	0.30	&	FIES	\\
6104.50742	&	10.12	&	0.72	&	FIES	\\
6107.49199	&	9.27	&	0.54	&	FIES	\\
\hline			
6118.51460	&	9.28	&	0.24	&	FIES	\\
6119.45816	&	9.58	&	0.26	&	FIES	\\
6120.45199	&	9.68	&	0.63	&	FIES	\\
6121.62526	&	9.68	&	0.34	&	FIES	\\
6122.49860	&	9.53	&	0.48	&	FIES	\\
\hline		
6459.46791	&	9.65	&	0.27	&	FIES	\\
6461.46262	&	9.65	&	0.13	&	FIES	\\
\hline
6115.69276	&	19.95	&	0.21	&	SANDIFORD	\\
6116.85210	&	19.02	&	0.25	&	SANDIFORD	\\
6117.80003	&	18.09	&	0.26	&	SANDIFORD	\\
\hline

\end{tabular}
\\
$^1$ The Heliocentric Julian date is calculated directly from the UTC.\\
\end{table}

\begin{table}
\caption{Radial velocity measurements  of CoRoT--36 obtained during and close to the transit.}
\begin{tabular}{c r r  l}
\hline
\noalign{\smallskip}
HJD$^1$     & RV     & $\pm \sigma$ & Instrument \\
-2 450 000  & [km\,s$\rm ^{-1}$] & [km\,s$\rm ^{-1}$]       & \\
\hline
6499.3910 & -0.044 & 0.116 & FIES$^2$ \\
6499.4059 &  0.202 & 0.104 & FIES \\
6499.4207 &  0.272 & 0.108 & FIES \\
6499.4356 &  0.112 & 0.116 & FIES \\
6499.4517 &  0.238 & 0.112 & FIES \\
6499.4665 &  0.097 & 0.108 & FIES \\
6499.4814 & -0.095 & 0.112 & FIES \\
6499.4963 & -0.289 & 0.116 & FIES \\
6499.5123 & -0.138 & 0.132 & FIES \\
6499.5272 & -0.056 & 0.100 & FIES \\
6499.5420 & -0.151 & 0.108 & FIES \\
6499.5569 & -0.078 & 0.116 & FIES \\
6499.5729 & -0.124 & 0.112 & FIES \\
6499.5878 & -0.058 & 0.116 & FIES \\
6499.6026 & -0.171 & 0.120 & FIES \\
6499.6176 &  0.008 & 0.128 & FIES \\
6499.6336 & -0.148 & 0.132 & FIES \\

\hline
6510.5485 & 23.150 & 0.030 & UVES$^3$  \\
6510.5604 & 23.010 & 0.030 & UVES \\
6510.5722 & 23.020 & 0.030 & UVES \\
6510.5873 & 23.080 & 0.030 & UVES \\
6510.5992 & 23.030 & 0.030 & UVES \\
6510.6110 & 23.045 & 0.030 & UVES \\
6510.6231 & 23.025 & 0.030 & UVES \\
6510.6349 & 22.940 & 0.030 & UVES \\
6510.6468 & 23.045 & 0.030 & UVES \\
6510.6589 & 23.045 & 0.030 & UVES \\
6510.7703 & 22.650 & 0.030 & UVES \\
\hline
\end{tabular}
\label{tab:RV3}
\\
$^1$ The Heliocentric Julian date is calculated directly from the UTC.\\
$^2$ The RVs obtained with FIES are measured relative to the
template. \\
$^3$ The RVs obtained with UVES are heliocentric. \\
\end{table}

\begin{table}
\caption{Radial velocity measurements of CoRoT--34.}
\label{tab:rv_1475}
\begin{tabular}{c r r  l}
\hline
\noalign{\smallskip}
HJD$^1$     & RV     & $\pm \sigma$ & Instrument \\
-2 450 000  & [km\,s$\rm ^{-1}$] & [km\,s$\rm ^{-1}$]       & \\
\hline
5962.96299	&	43.34	&	6.77    &   HIRES	\\
5964.87595	&	42.44	&	1.47    &   HIRES	\\
6315.83576	&	27.45	&	5.43    &   HIRES	\\
6315.95414	&	33.80	&	14.34   &   HIRES	\\
6317.94208	&	29.15	&	4.56    &   HIRES	\\\hline
5546.71708	&	27.60	&	1.97    &   HARPS	\\
5547.71536	&	44.16	&	1.75    &   HARPS	\\
6307.66380	&	36.34	&	1.89    &   HARPS	\\
6308.66913	&	45.29	&	2.35    &   HARPS	\\
6312.68506	&	42.43	&	5.72    &   HARPS	\\
6332.60448	&	28.60	&	0.49    &   HARPS	\\
6334.72000	&	30.49	&	6.54    &   HARPS	\\
6335.63909	&	46.67	&	5.63    &   HARPS	\\
6336.56863	&	33.64	&	0.53    &   HARPS	\\
6353.66197	&	25.77	&	1.00    &   HARPS	\\
6354.65832	&	44.04	&	9.29    &   HARPS	\\
6355.64480	&	34.42	&	2.19    &   HARPS	\\
6360.66932	&	27.65	&	4.49    &   HARPS	\\
6361.53093	&	41.77	&	11.33   &   HARPS	\\
6362.58672	&	30.77	&	0.98    &   HARPS	\\
6364.58864	&	31.13	&	1.77    &   HARPS	\\
6365.58175	&	45.39	&	2.99    &   HARPS	\\\hline
6587.72768	&	41.12	&	0.78    &   UVES	\\
6587.74101	&	43.10	&	1.75    &   UVES	\\
6629.67304	&	38.23	&	2.02    &   UVES	\\
6629.68407	&	35.59	&	1.67    &   UVES	\\
6656.56692	&	33.19	&	3.16    &   UVES	\\
6656.57796	&	34.63	&	5.17    &   UVES	\\
6667.75962	&	31.70	&	6.46    &   UVES	\\
6667.77063	&	31.11	&	1.02    &   UVES	\\
6669.79442	&	34.24	&	0.84    &   UVES	\\
6669.80543	&	36.40	&	2.77    &   UVES	\\
6675.63855	&	31.59	&	3.15    &   UVES	\\
6675.64955	&	31.15	&	2.56    &   UVES	\\
6676.79285	&	40.28	&	2.32    &   UVES	\\
6676.80386	&	45.35	&	0.68    &   UVES	\\
6694.55314	&	36.78	&	3.18    &   UVES	\\
\hline
	\end{tabular}
\\
$^1$ The Heliocentric Julian date is calculated directly from the UTC.\\
\end{table}

\begin{table}
\caption{RV measurements CoRoT--35.}
\label{tab:RV2}
\begin{tabular}{c r r  l}
\hline
\noalign{\smallskip}
BJD$^1$     & RV     & $\pm \sigma$ & Instrument \\
-2 450 000  & [km\,s$\rm ^{-1}$] & [km\,s$\rm ^{-1}$]       & \\
\hline
6158.68514 & 3.9054 & 0.0719 & HARPS\\
6160.63150 & 4.1314 & 0.0713 & HARPS\\
6452.86950 & 3.8436 & 0.0639 & HARPS\\
6534.63480 & 4.1474 & 0.0652 & HARPS\\
6536.52832 & 3.8552 & 0.0670 & HARPS\\
6544.54782 & 4.2670 & 0.1212 & HARPS\\
6564.56242 & 3.9744 & 0.0844 & HARPS\\
6565.50257 & 3.9698 & 0.0866 & HARPS\\
\hline
\end{tabular}
\\
$^1$ The Barycentric Julian date is calculated directly from the UTC.\\
\end{table}

\begin{table}
\caption{Radial velocity measurements of CoRoT 659668516. frc and src denote "fast-" and "slow rotating component", respectively.}
\label{tab:rv_4203}
\begin{tabular}{c r r  l}
\hline
\noalign{\smallskip}
HJD$^1$     & RV     & $\pm \sigma$ & Instrument \\
-2 450 000  & [km\,s$\rm ^{-1}$] & [km\,s$\rm ^{-1}$]       & \\
\hline
6818.48244	&	15.95	&	1.38    &   TWIN	\\
6818.57451	&	17.64	&	0.24    &   TWIN	\\
6820.48877	&	23.69	&	3.24    &   TWIN	\\
6821.46328	&	15.68	&	1.42    &   TWIN	\\
6821.60504	&	21.30	&	1.67    &   TWIN	\\
6855.44570	&	26.23	&	1.61    &   TWIN	\\
\hline			
6097.77494	&	16.34	&	0.26    &   HARPS, src	\\
6099.67744$^2$	&	16.08	&	0.19    &   HARPS, src	\\
6097.77494	&	18.14	&	3.00    &   HARPS, frc	\\
\hline			
6861.54386	&	14.80	&	0.66    &   UVES, frc	\\
6863.56748	&	14.96	&	0.51    &   UVES, frc	\\
6866.73130	&	14.93	&	0.76    &   UVES, frc	\\
6868.52861	&	15.01	&	0.56    &   UVES, frc	\\
6861.54386	&	16.01	&	0.55    &   UVES, src	\\
6863.56748	&	16.30	&	0.82    &   UVES, src	\\
6866.73130	&	15.52	&	0.84    &   UVES, src	\\
6868.52861	&	16.76	&	0.73    &   UVES, src	\\
\hline
	\end{tabular}
\\
$^1$ The Heliocentric Julian date is calculated directly from the UTC.\\
$^2$ Only the slow rotating component could be measured for this spectrum due to low SNR.\\
\end{table}

\section{Candidates of this survey}

\begin{table}
\caption{Observational parameters of the 17 candidates analysed.}
\label{tab:obs_param}
\begin{tabular}{l|lll}
 CoRoT ID &  Right Ascension & Declination & apparent magnitude$^1$ \\\hline\hline
 102806520 & 06:46:10.244  & -01:42:23.630  &  B:14.48; 		R:13.21 \\
 102850921 & 06:47:23.865  & -03:08:32.377  &  R:12.98  \\ 
 102584409 & 06:41:00.051  & -01:29:29.738  &  B:11.91; V:11.51; 	R:11.56 \\
 102605773 & 06:41:34.477  & -00:53:57.746  &  B:14.80; 		R:14.15  \\ 
 102627709 & 06:42:05.924  & -00:31:31.390  &  B:14.73; 		R:13.57  \\ 
 110853363 & 06:52:36.485  & -03:07:30.151  &  B:13.53; 		R:13.42  \\
 110756834 & 06:51:29.005  & -03:49:03.486  &  B:13.99; 		R:13.41  \\
 110858446 & 06:53:25.072  & -05:42:47.027  &  R:13.35  \\
 110660135 & 06:50:01.371  & -05:12:07.448  &  B:10.62; V:10.55; 	R:10.82  \\
 310204242 & 18:31:19.816  & -06:21:23.875  &  B:14.65; 		R:13.14  \\
 659719532 & 18:33:35.740  & +07:42:19.278  &  B:11.65; V:11.15; 	R:11.44  \\
 652345526 & 18:31:00.241  & +07:11:00.125  &  B:13.72; V:13.06; 	R:12.72  \\
 659668516 & 18:33:27.345  & +05:20:01.334  &  B:16.04; 		R:14.64 \\ 
 659460079 & 19 17 15.43   & -02 46 28.82   &  B:16.55;           R:15.28 \\
 659721996 & 18:35:59.542  & +07:49:08.436  &  B:14.93; 		R:14.03  \\
 632279463 & 18:30:9.370   & +07:23:45.478  &  V:12.54;       R:12.56 \\
 631423419 & 18:34:0.905   & +06:50:22.567  &  B:13.02;       R:12.1  \\
 \hline         
\end{tabular}  
\\
$^1$ Obtained from the EXODAT database \cite{deleuil09}.\\

\end{table}

\section{TTVs measured for CoRoT--35b.}

\begin{table}
\caption{ $\rm T_{exp}$ are the expected transit mid times from linear ephemeris, TTV is the measured mid-transit time variation from the expected mid-time.}
\label{tab:ttvs}
\begin{tabular}{lcc}
 \hline
& $\rm T_{exp}$ [BJD] & TTV [min]\\\hline
TTV$_\mathrm{b;1} $ &  56032.2843 & $\rm -51.66 \pm 4.35$\\
TTV$_\mathrm{b;2} $ &  56035.5117 & $\rm 52.56 \pm 4.50$\\
TTV$_\mathrm{b;3} $ &  56038.7392 & $\rm 48.11 \pm 4.40$\\
TTV$_\mathrm{b;4} $ &  56041.9667 & $\rm -27.05 \pm 4.50$\\
TTV$_\mathrm{b;5} $ &  56045.1942 & $\rm 5.34 \pm 4.34$\\
TTV$_\mathrm{b;6} $ &  56048.4217 & $\rm 11.63 \pm 4.37$\\
TTV$_\mathrm{b;7} $ &  56051.6491 & $\rm 2.40 \pm 4.31$\\
TTV$_\mathrm{b;8} $ &  56054.8766 & $\rm -2.68 \pm 4.38$\\
TTV$_\mathrm{b;9} $ &  56058.1041 & $\rm -43.35 \pm 4.49$\\
TTV$_\mathrm{b;10} $ &  56061.3316 & $\rm -10.17 \pm 4.43$\\
TTV$_\mathrm{b;11} $ &  56064.5591 & $\rm -13.60 \pm 4.44$\\
TTV$_\mathrm{b;12} $ &  56067.7865 & $\rm 54.18 \pm 4.42$\\
TTV$_\mathrm{b;13} $ &  56071.0140 & $\rm -21.44 \pm 4.32$\\
TTV$_\mathrm{b;14} $ &  56074.2415 & $\rm 11.69 \pm 4.37$\\
TTV$_\mathrm{b;15} $ &  56077.4690 & $\rm -31.82 \pm 4.33$\\
TTV$_\mathrm{b;16} $ &  56080.6965 & $\rm 4.58 \pm 4.35$\\
TTV$_\mathrm{b;17} $ &  56083.9239 & $\rm -6.07 \pm 4.38$\\
TTV$_\mathrm{b;18} $ &  56087.1514 & $\rm -2.06 \pm 4.35$\\
TTV$_\mathrm{b;19} $ &  56090.3789 & $\rm 28.04 \pm 4.41$\\
TTV$_\mathrm{b;20} $ &  56093.6064 & $\rm -9.28 \pm 4.27$\\
TTV$_\mathrm{b;21} $ &  56096.8339 & $\rm -9.27 \pm 4.40$\\
TTV$_\mathrm{b;22} $ &  56100.0613 & $\rm -17.02 \pm 4.43$\\
TTV$_\mathrm{b;23} $ &  56103.2888 & $\rm 17.38 \pm 4.43$\\
TTV$_\mathrm{b;24} $ &  56106.5163 & $\rm -24.02 \pm 4.39$\\
TTV$_\mathrm{b;25} $ &  56109.7438 & $\rm 8.60 \pm 4.38$\\
TTV$_\mathrm{b;26} $ &  56112.9713 & $\rm -31.27 \pm 4.34$\\\hline
\end{tabular}  
\\

\end{table}

\bsp	
\label{lastpage}
\end{document}